\RequirePackage{fix-cm} 
\documentclass[fleqn,usenatbib]{mnras}
\usepackage{newtxtext,newtxmath}
\DeclareFontShape{OML}{cmm}{b}{it}{<->ssub*cmm/m/it}{}
\usepackage[T1]{fontenc}
\usepackage{graphicx}	
\usepackage{amsmath}	
\usepackage{subcaption} 
\usepackage{hyperref}
\usepackage{xcolor}
\usepackage{threeparttable}
\usepackage{enumitem}
\setlist[itemize]{leftmargin=*}
\usepackage{orcidlink}
\hypersetup{
    colorlinks=true,
    linkcolor=blue,
    citecolor=blue,
    urlcolor=blue
}
\graphicspath{{figures/}}

\title[Off-Nuclear AGN Signatures in MaNGA]{Off-Nuclear AGN Signatures in MaNGA: Spatially resolved selection using MEGACUBES}

\author[S. Jaison et al.]{Sandra Jaison\orcidlink{0009-0006-7814-2334},$^{\hyperlink{aff1}{1}}$\thanks{E-mail: sandra.jaison@ufrgs.br}
Rog\'erio Riffel\orcidlink{0000-0002-1321-1320},$^{\hyperlink{aff1}{1}}$
Dominika Wylezalek\orcidlink{0000-0003-2212-6045},$^{\hyperlink{aff2}{2}}$
Sandro B. Rembold\orcidlink{0000-0003-0880-5738},$^{\hyperlink{aff3}{3}}$
Sandra I. Raimundo\orcidlink{0000-0002-6248-398X},$^{\hyperlink{aff4}{4},\hyperlink{aff5}{5}}$\newauthor
Rogemar A. Riffel\orcidlink{0000-0003-0483-3723},$^{\hyperlink{aff3}{3}}$
Marco Albán\orcidlink{0009-0005-2036-9299},$^{\hyperlink{aff6}{6}}$
Mark Armah\orcidlink{0000-0002-9746-3938},$^{\hyperlink{aff7}{7}}$
Theodor Brosi\orcidlink{0009-0003-1508-522X},$^{\hyperlink{aff2}{2}}$
Quinn O. Casey\orcidlink{0009-0005-9002-4800},$^{\hyperlink{aff8}{8}}$\newauthor
Oli L. Dors\orcidlink{0000-0003-4782-1570},$^{\hyperlink{aff7}{7}}$
Ryan C. Hickox\orcidlink{0000-0003-1468-9526},$^{\hyperlink{aff8}{8}}$
Thaisa Storchi-Bergmann\orcidlink{0000-0003-1772-0023},$^{\hyperlink{aff1}{1}}$
and Tiago H. S. Weirich\orcidlink{0009-0005-8886-1298},$^{\hyperlink{aff3}{3}}$
\\
\hypertarget{aff1}{$^{1}$}Departamento de Astronomia, Instituto de Fisica, Universidade Federal do Rio Grande do Sul, CP 15051, 91501-970 Porto Alegre, RS, Brazil\\
\hypertarget{aff2}{$^{2}$}Zentrum für Astronomie der Universität Heidelberg, Astronomisches Rechen-Institut, Mönchhofstr, 12-14, 69120 Heidelberg, Germany\\
\hypertarget{aff3}{$^{3}$}Departamento de Física, CCNE, Universidade Federal de Santa Maria, Av. Roraima 1000, 97105-900 Santa Maria, RS, Brazil\\
\hypertarget{aff4}{$^{4}$}Physics and Astronomy, University of Southampton, Highfield, Southampton, SO17 1BJ, UK\\
\hypertarget{aff5}{$^{5}$}DARK, Niels Bohr Institute, University of Copenhagen, Jagtvej 155, Copenhagen N, 2200, Denmark\\
\hypertarget{aff6}{$^{6}$}Max-Planck-Institut für Radioastronomie, Auf dem Hügel 69, 53121 Bonn, Germany\\
\hypertarget{aff7}{$^{7}$}Universidade do Vale do Paraíba, Instituto de Pesquisa e Desenvolvimento, Av. Shishima Hifumi, 2911, CEP: 12244-000, São José dos Campos, SP, Brazil\\
\hypertarget{aff8}{$^{8}$}Department of Physics and Astronomy, Dartmouth College, Hanover, NH 03755, USA\\
}

\date{Accepted 2026 November 18. Received 2026 November 17; in original form 2026 September 18}

\pubyear{2026}

\begin{document}
\label{firstpage}
\pagerange{\pageref{firstpage}--\pageref{lastpage}}
\maketitle
\begin{abstract}
We present a spatially resolved search for off-nuclear active galactic nucleus (AGN) candidates in the MaNGA survey. Using {\sc megacubes}, we analyze over 10,000 galaxies, separating AGN activity from diffuse ionized gas with refined equivalent-width and surface-brightness thresholds. Combining a spatial radial algorithm with BPT, WHaN, and WHaD diagnostics, we classify 261 nuclear AGNs, 201 off-nuclear AGNs, and 159 `quiet' (composite-spaxel-dominated) sources. We test whether off-nuclear AGNs reside in distinct host galaxies or whether the offset is a local, gas-driven phenomenon. Nuclear and off-nuclear AGN hosts are statistically indistinguishable in stellar mass, age, star formation rate, and gas-phase metallicity. However, the excitation and kinematics of the AGN-ionized gas differ significantly: nuclear AGN spaxels have [\ion{O}{iii}]$\lambda5007$ luminosities $\sim 0.7$ dex higher than off-nuclear spaxels, whose lower [\ion{O}{iii}]/H$\beta$ ratios indicate genuinely lower-excitation gas rather than an obscured view of the central engine. We also find an orientation trend: 72\% of nuclear hosts are face-on, versus 50\% of off-nuclear hosts, so projection effects contribute to, but do not fully explain, off-nuclear detections. The classical WHaN boundary at $\log([\ion{N}{ii}]/\mathrm{H}\alpha) = -0.4$ misclassifies high-metallicity diffuse ionized gas in integral field data; an empirical shift to $-0.25$ removes over 48,000 spurious AGN spaxels. Finally, the off-nuclear fraction implies an episodic AGN phase lifetime of $\sim 5 \times 10^4$ yr, consistent within a factor of a few with independent fading-AGN estimates and supporting short, recurrent accretion bursts.

\end{abstract}


\section{Introduction}
\label{sec:intro}

Most present-day galaxies are inferred to host a supermassive black hole (SMBH) at their gravitational centers \citep{1998AJ....115.2285M, 2000ApJ...539L...9F, 2000ApJ...539L..13G, Greene2020, Caglar2023}. Only a small fraction of SMBHs are observed to be in an actively accreting state at any given cosmic epoch; however, they are expected to experience recurrent episodes of accretion over the course of their host galaxies’ lifetimes \citep{Martini2004, Heckman2014, Hickox2014, Schawinski2015}. During these active phases, SMBHs accrete ambient material and can drive powerful outflows and relativistic jets. Such systems, collectively termed active galactic nuclei (AGNs), are among the most luminous and energetic phenomena in the universe and are frequently classified according to their luminosity—distinguishing lower-luminosity Seyfert galaxies from highly luminous quasars \citep[e.g.,][]{1943ApJ....97...28S, schmidt19633c273,2006AJ....131.2766R}. Additionally, AGNs can be classified using a variety of other multi-wavelength observational criteria \citep[e.g.,][]{padovani2017}. 

Extensive theoretical and observational investigations have been undertaken to characterize the physical nature of AGNs and to quantify their impact on galaxy evolution. Accumulating evidence indicates that SMBHs exert a fundamental influence on the assembly histories of their host galaxies. In particular, AGN feedback—mediated by radiative output, winds, and jets—is widely invoked as a key mechanism capable of suppressing (quenching) star formation and thereby regulating the stellar mass growth of galaxies \citep[e.g.,][and references therein]{2012ARA&A..50..455F, 2017MNRAS.468.4956Z, Harrison2024}. Conversely, alternative scenarios propose that AGN-driven outflows can, under certain conditions, compress the surrounding interstellar medium and thereby induce or enhance star formation \citep[e.g.,][]{Silk2005,2012MNRAS.420L...8H, 2019MNRAS.485.3409G}.

A central challenge in understanding SMBH-galaxy co-evolution is obtaining an unbiased census of AGN across the full range of luminosities, host morphologies, and spatial configurations. Traditional single-fiber optical spectroscopic surveys, such as the Sloan Digital Sky Survey \citep[SDSS;][]{2000AJ....120.1579Y, 2006AJ....131.2332G}, are limited to the central region of each galaxy. Studies employing these surveys and standard narrow emission-line diagnostic diagrams \citep{1981PASP...93....5B, Kewley2001, Kauffmann2003} are inherently biased toward more luminous AGNs \citep[e.g.,][]{Trump_2015, Jones2016}. Furthermore, single-fiber Baldwin-Phillips-Terlevich \citep[BPT;][]{1981PASP...93....5B} classification can fail to find AGNs for a variety of reasons: active star formation \citep{ 2020MNRAS.492.2268B}; strong host galaxy dust, or nuclear torus obscuration that dilutes the AGN signatures \citep{2002ApJ...579L..71M, Almeida2017}; contamination by low-ionization nuclear emission-line regions (LINER)-like emission from hot low-mass evolved stars; and low-metallicity or low-mass AGN hosts \citep{Groves2006}.

The major drawback of these single-fiber surveys is that they miss some unique AGN sources, such as off-nuclear and extended AGN signatures, among others. For instance, the integrated spectrum of a galaxy can hide AGN signatures when the central region is dominated by star formation or is heavily obscured by dust \citep{Wylezalek2017}, pushing any detectable AGN emission to the outskirts, beyond the fiber aperture. Single-fiber surveys further miss AGNs that are spatially offset from the galaxy center due to mergers \citep{Greene2011-merger, Comerford_2014,2018ApJ...869..154B}, or that were recently switched off while relic ionization is still visible in the outskirts \citep{Keel_2012,2015AJ....149..155K}. Studying these types of sources is important to shed light on the full picture of SMBH-galaxy co-evolution.

Integral field spectroscopy (IFS) surveys, such as Mapping Nearby Galaxies at Apache Point Observatory \citep[MaNGA;][]{Bundy2015}, overcome these limitations by providing spatially resolved spectroscopy across the entire galaxy, enabling detection of ionization signatures offset from the galaxy center.

Although integral field unit (IFU) data are powerful, they also introduce new challenges. A variety of diagnostic diagrams have been proposed to classify emission-line regions, including the BPT diagrams \citep{1981PASP...93....5B, VeilleuxS1987}, $EW_{\mathrm{H}\alpha}$ vs. [\ion{N}{ii}]/H$\alpha$ diagram \citep[WHaN;][]{CidFernandes2010}, $EW_{\mathrm{H}\alpha}$ vs.$\sigma_{\mathrm{H}\alpha}$ diagram \citep[WHaD;][]{sanchez2024whad}, $EW_{\mathrm{H}\alpha}$ vs. [\ion{O}{iii}]/[\ion{O}{ii}] diagram \citep[WHaO;][]{Sanchez+25whao}, to name a few. Moreover, traditional diagnostic methods were initially calibrated using single-fiber spectra, and more recent diagnostics specifically developed for IFU data still lack the extensive validation studies needed to apply them with full confidence. Each of these is sensitive to different parameters, and degeneracies among them can lead to misclassifications, as they do not always yield consistent classifications \citep{CidFernandes2010}. The problem is further deepened by the fact that standard optical diagnostics depend on galactocentric distance and aperture size \citep{Sanchez+20, Sanchez+21}. In particular, the fraction of fake AGN classifications increases significantly beyond 1 effective radius ($R_{\rm e}$), primarily due to contamination from diffuse ionized gas \citep[DIG;][]{al2023} and extended LINER-like emission \citep[e.g.,][]{Heckman1980, Binette, stasinska2006, CidFernandes2010, Belfiore2016, Lacerda2018}. To address this, we utilize multiple-extension Flexible Image Transport System (FITS) data products known as {\sc megacubes} \citep{Riffel2023}, which provide robust stellar population and spatially resolved emission-line fitting for every MaNGA galaxy. We pair these data with the AGN selection criteria introduced by \citet{Wylezalek2018}, which are aimed at identifying AGN in optical IFS surveys such as MaNGA and have also been shown to uncover off-nuclear AGN signatures. These criteria were developed while the MaNGA survey was still ongoing, based on a sample of $\sim2800$ galaxies.

In this work, we adapt and refine the AGN identification technique of \citet{Wylezalek2018} and apply it, for the first time, to the complete MaNGA DR17 sample. By combining this approach with a three-diagnostic classification scheme (BPT + WHaN + WHaD) and a spatial radial algorithm, we produce the most complete census to date of nuclear, off-nuclear, and `quiet' AGN candidates across the full stellar-mass range in MaNGA, where `quiet' denotes sources that fail both the nuclear and off-nuclear thresholds. This approach aims to uncover off-nuclear AGN candidates by quantifying the fraction of spaxels dominated by AGN ionization as a function of radius, enabling us to separate nuclear AGN from systems exhibiting significant off-nuclear activity. Beyond this source cataloging, we perform a direct spaxel-level comparison of various properties between the nuclear and off-nuclear sources and also investigate the role of host galaxy orientation in the classification. In addition, we conduct a dedicated examination of the quiet AGN population, identifying a DIG-contamination bias in the WHaN \citep{CidFernandes2010} demarcation for IFU data, and we investigate a revised boundary consistent with the BPT maps. Following this, we test the robustness of our classification against the choice of emission-line diagnostic, repeating the analysis with the kinematically calibrated two-dimensional (2D) BPT boundaries of \citet{law2021diagnostics} and the three-dimensional (3D) BPT framework of \citet{ji2020constraining}, to replace the classical BPT diagrams and assess the changes in our classification.

This paper is organized as follows.
Section \ref{sec:data} describes the data used. Section \ref{sec:diagnostics} presents the diagnostic framework utilized in this study. Section \ref{sec:method} details the sample selection and spatial classification pipeline. The results are presented in Section \ref{sec:results_main}, followed by discussion in Section \ref{sec:discussion}. Finally, we summarize our conclusions in Section \ref{sec:conclusions}.

Throughout this work, we assume a flat $\Lambda$CDM cosmology with $H_0 = 72\,\mathrm{km\,s^{-1}\,Mpc^{-1}}$, $\Omega_m = 0.3$.

\section{DATA}
\label{sec:data}
\subsection{The MaNGA Survey}
\label{sec:manga}

We utilize data from the fourth-generation SDSS sub-project, the MaNGA final data release \citep[DR17;][]{2006AJ....131.2332G, 2013AJ....146...32S, Bundy2015, 2015AJ....149...77D, 2015AJ....150...19L, 2016AJ....152...83L, Law+21, 2016AJ....151....8Y, 2017AJ....154...28B, Wake_2017, 2019ApJS..240...23A, 2019AJ....158..160B, 2019AJ....158...74C, 2019AJ....158..231W, 2022ApJS..259...35A}. The survey observed over 10,000 nearby galaxies covering a wavelength range of 3600-10,400 \AA\ at a spectral resolution of $R\sim2000$  and an angular resolution of 2.5 arcsec, targeting galaxies at an average redshift of $z \sim 0.03$.

Unlike previous SDSS surveys, which obtained spectra only at the centers of target galaxies, MaNGA obtained spatially resolved spectral measurements. This was achieved using custom-designed IFUs, hexagonal bundles consisting of tightly packed arrays of optical fibers (19-127 fibers), which cover a field of 12 to 32 arcsec in diameter \citep{2015AJ....149...77D,2015AJ....150...19L, 2016AJ....151....8Y}. Furthermore, MaNGA observations are divided into `primary' and `secondary' targets, providing spatial coverage up to 1.5 and 2.5 $R_{\rm e}$, respectively. The main goal of the MaNGA survey is to understand the evolutionary life history of present-day galaxies, from their early assembly and the ongoing growth via star formation to their eventual quenching.

\subsection{MEGACUBES}
\label{sec:megacubes}
To obtain the required spatially resolved physical properties from the MaNGA observations, we employed the multiple-extension FITS data products known as {\sc megacubes} \citep{Riffel2023}, which were generated for the full MaNGA DR17 sample. These data products, publicly available via the web interface \url{https://manga.linea.org.br/} (alternatively \url{https://manga.if.ufrgs.br/}), provide 2D maps of several physical quantities, as well as the emission-line profiles for each individual spaxel. The {\sc megacubes} were produced with the in-house pipeline {\sc urutau} \citep[\url{https://github.com/ndmallmann/urutau};][]{Mallmann+23}, which performs full spectral fitting of each spaxel using the {\sc starlight} code \citep{2005MNRAS.358..363C, 2018MNRAS.480.4480C}. This choice is motivated by the fact that {\sc starlight}, in its original implementation, is not configured to treat data cubes or multi-extension FITS files directly, whereas {\sc urutau} provides a modular framework for spaxel-by-spaxel fitting and for the handling of input and output data products. 

In short, for each spaxel, the stellar continuum is modeled as a linear combination of simple stellar populations covering a broad range of ages and metallicities using the GRANADA + MILES stellar population models \citep{GonzalesDelgado2005,2010MNRAS.404.1639V, 2013A&A...557A..86C, Riffel2023}. Furthermore, the spectral base includes a power-law component of the form $F_\nu \propto \nu^{-1.5}$ to represent the possible contribution of an AGN featureless continuum. This comprehensive spectral decomposition enables the determination of all the key physical parameters required for our analysis, including light- and mass-weighted mean stellar ages, stellar metallicities, dust extinction, and star formation histories (SFHs). 

Once the stellar continuum has been accurately modeled and subtracted, {\sc urutau} yields emission-line data that are effectively free from underlying stellar absorption features. The profiles of the main optical emission lines (e.g., H$\beta$, [\ion{O}{iii}]$\lambda5007$, H$\alpha$, [\ion{N}{ii}]$\lambda6583$) are then fitted with Gaussian functions using the {\sc ifscube} Python package \citep{2020zndo...3945237R, 2021MNRAS.507...74R}, providing robust measurements of emission-line fluxes, equivalent width (EW), and kinematic parameters. In addition to the measured quantities, all model components and fitting diagnostics are stored, enabling an assessment of the fit quality on a spaxel-by-spaxel basis. Further details are provided in \citet{Riffel2023}.

The final {\sc megacubes} data products employed in this work comprise spatially resolved maps of both stellar populations and gas properties for every spaxel within the galaxy’s field of view (FoV). These data cubes are particularly well suited for spatially resolved ionization diagnostics, including BPT diagrams \citep{1981PASP...93....5B}, WHaN \citep{CidFernandes2010}, and WHaD \citep{sanchez2024whad}. To further ensure reliable emission line measurements, we followed \citet{Riffel2023} and removed all spaxels whose emission-line fluxes fell below the 1$\sigma$ confidence level (see Section \ref{sec:diagnostics}).

\section{Diagnostic Framework for AGN spaxels Selection}
\label{sec:diagnostics}
Identifying the presence of an AGN in spatially resolved IFU observations requires establishing the dominant ionization mechanism in each individual spaxel. Star-forming \ion{H}{ii} regions are powered by comparatively soft ultraviolet radiation emitted by young, massive stars, whereas AGN activity generates a substantially harder ionizing continuum. This difference is reflected in the observed optical emission-line ratios \citep[e.g.,][]{Kewley2006}. 

To construct a reliable ionization map for a given source, it is crucial to robustly separate true emission features from both instrumental artifacts and random noise. Therefore, we apply a strict masking criterion to the data to maintain the reliability of our analysis. For each galaxy, we retain only those spaxels with a median stellar continuum signal-to-noise ratio (S/N) $\gtrsim 1$ across the fitted spectral range. The principal diagnostic emission lines—H$\alpha$, H$\beta$, [\ion{O}{iii}]$\lambda$5007, [\ion{N}{ii}]$\lambda$6583, [\ion{S}{ii}]$\lambda$6717, and [\ion{S}{ii}]$\lambda$6731—are further required to have a local noise level ($\sigma_{\mathrm{local}}$), estimated from the residual spectrum within a wavelength window centered on each line, below the fitted line amplitude ($A_{\mathrm{line}}$).

Crucially, this masking is applied strictly and simultaneously: any spaxel that fails to meet the continuum S/N threshold, or falls below the amplitude-to-noise limit for any of the six principal emission lines, is masked. These invalid spaxels are entirely excluded from the denominator in all subsequent calculations of AGN spaxel fractions.

\subsection{Diagnostic Diagrams}
\label{sec:diagrams}
Each spaxel may contain contributions from multiple ionization sources, including stellar photoionization, AGN, DIG, hot low-mass evolved stars, and shocks \citep{Binette, stasinska2006, CidFernandes2010, Belfiore2016, 2017MNRAS.466.3217Z, Lacerda2018}. To address this complexity and to correctly classify each region of a source, we employ a multi-diagnostic approach described below.

\subsubsection{BPT Diagrams}
\label{sec:bpt}
The most widely employed emission-line diagnostic frameworks are the traditional BPT diagrams \citep[][]{1981PASP...93....5B, VeilleuxS1987}. In this study, we utilize both the [\ion{N}{ii}]-based BPT diagram, defined in the emission-line ratio plane [\ion{O}{iii}]/H$\beta$ vs. [\ion{N}{ii}]/H$\alpha$, and the [\ion{S}{ii}]-based BPT diagram, defined by [\ion{O}{iii}]/H$\beta$ vs. [\ion{S}{ii}]/H$\alpha$.

To differentiate nebular regions ionized by young, massive stars from those powered by harder ionizing spectra using the [\ion{N}{ii}]-BPT, we make use of standard demarcation curves: the theoretical maximum starburst boundary derived by \citet{Kewley2001} and the empirical separation proposed by \citet{Kauffmann2003}. Spaxels located below the Kauffmann curve are classified as star-forming (SF), those situated between the Kauffmann and Kewley curves are designated as composite, and those lying above the Kewley curve are interpreted as being dominated by AGN or LINER-like excitation \citep[e.g.,][]{Law+21, Riffel2023,deMellos2024}. To more effectively discriminate among SF, AGN, and LINER-like regions, we additionally employ the [\ion{S}{ii}]-BPT diagram.

\subsubsection{WHaN Diagram}
\label{sec:whan}

BPT diagrams are the standard diagnostic tool but have key limitations: they require four strong emission lines and thus often miss weak emission-line regions. As a result, using only emission-line ratios can contaminate AGN samples with pseudo-AGN spaxels powered by old, evolved stars.  To address this, we complement the BPT classification with the WHaN diagram \citep{CidFernandes2010}, which uses the [\ion{N}{ii}]/H$\alpha$ ratio and the H$\alpha$ equivalent width, $EW_{\mathrm{H}\alpha}$, making it more sensitive to the ionizing source strength. Because old stellar populations cannot produce large $EW_{\mathrm{H}\alpha}$ values \citep{CidFernandes2010, Davies2014}, this method effectively separates true AGN activity from emission powered by evolved stars.

In the WHaN diagram, regions with enhanced [\ion{N}{ii}]/H$\alpha$ and $EW_{\mathrm{H}\alpha} > 3$ \AA\ \citep{CidFernandes2010, 2011MNRAS.413.1687C} are classified as genuine AGNs. Among them, $EW_{\mathrm{H}\alpha} > 6$ \AA\ defines strong AGNs (sAGN, Seyfert-like), while $3\text{ \AA} \le EW_{\mathrm{H}\alpha} \le 6\text{ \AA}$ defines weak AGNs (wAGN, LINER-like). Regions with lower $EW_{\mathrm{H}\alpha}$ are classified as Retired Galaxies (RG), dominated by photoionization from old stars.

\subsubsection{WHaD Diagram}
\label{sec:whad}
We also use the WHaD diagram, recently introduced by \citet{sanchez2024whad}, as an independent visual check. WHaD plots $EW_{\mathrm{H}\alpha}$ against the H$\alpha$ velocity dispersion ($\sigma_{{\rm H}\alpha}$), relying solely on a single emission line.

Because it requires only H$\alpha$, WHaD remains applicable in faint or dust-attenuated spaxels where H$\beta$ or [\ion{O}{iii}] is below the detection threshold. Following \citet{sanchez2024whad}, we divide spaxels into five regimes:
RG ($EW_{\mathrm{H}\alpha}$ $< 3$ \AA), SF ($EW_{\mathrm{H}\alpha}$ $> 6$ \AA,  $\sigma_{{\rm H}\alpha} < 57$ km\,s$^{-1}$), wAGN ($EW_{\mathrm{H}\alpha}$ $= 3$–$10$ \AA, $\sigma_{{\rm H}\alpha} \geq 57$ km\,s$^{-1}$), sAGN ($EW_{\mathrm{H}\alpha}$ $> 10$ \AA, $\sigma_{{\rm H}\alpha} \geq 57$ km\,s$^{-1}$), and an ambiguous “Unknown-Unk” class for spaxels between these regimes.

\begin{figure*} 
    \centering
    \includegraphics[width=\textwidth]{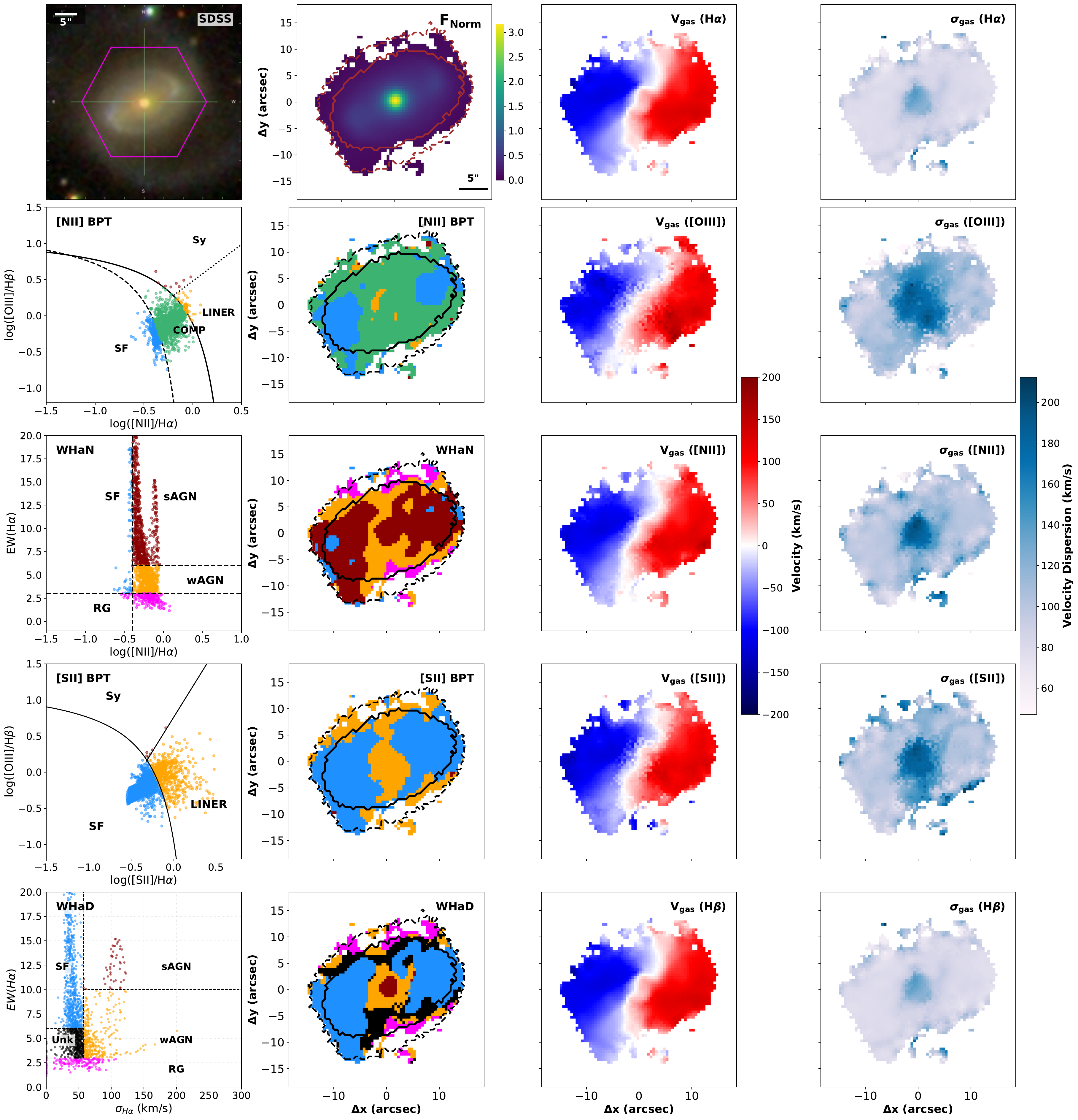}
    \caption{Example diagnostic diagrams and velocity-dispersion maps for MaNGA:1-177624 ($z = 0.031$), with spatial coordinates shown in arcseconds relative to the nucleus. The top row includes the SDSS optical image and the normalized continuum flux map ($F_{Norm}$). Below, we show the spatially resolved [\ion{N}{ii}]-BPT, [\ion{S}{ii}]-BPT, WHaN, and WHaD classification maps, alongside the corresponding diagnostic diagrams. Contours representing continuum S/N $\geq 5$ (dashed) and S/N $\geq 10$ (solid) are included to indicate regions with robust stellar continuum modeling. We also present the velocity and velocity dispersion maps for the main emission lines.}
    \label{fig:diagnostic_maps_example}
\end{figure*}

\subsection{The final AGN candidates}
\label{sec:wylezalek}

To identify AGN candidates from the spatial distribution of ionized gas, we applied the method from \citet{Wylezalek2018}, originally developed for IFU data and first used on the MaNGA Product Launch (MPL-5) sample of 2{,}727 galaxies. This approach quantifies the global extent of AGN-like emission by assigning emission-line classifications to each spaxel from diagnostic diagrams and constructing resolved ionization maps.

Each spaxel is classified with both the [\ion{N}{ii}] and [\ion{S}{ii}] BPT diagrams, which together give a more complete view of the ionization structure. From these maps, they compute the fraction of spaxels classified as AGN/composite in the [\ion{N}{ii}]-BPT and as AGN/LINER-like in the [\ion{S}{ii}]-BPT, using these fractions as AGN activity indicators. This allows the detection of low-luminosity or obscured AGNs that may ionize only a small part of the galaxy and not dominate the integrated emission.

To minimize contamination from RGs or DIG, they applied additional constraints. First, they set a threshold on $EW_{\mathrm{H}\alpha}$. To avoid dilution by outer DIG or SF regions, they computed the mean EW using only the highest percentiles of AGN/LINER-classified spaxels. It is worth noting that while \citet{Wylezalek2018} used a stricter $EW_{\mathrm{H}\alpha}$ limit of 5 \AA, we adopt 3 \AA\ -- the standard WHaN boundary between retired and active galaxies \citep{CidFernandes2010} -- to match the WHaN diagnostic used later in our classification. They also required a minimum H$\alpha$ surface brightness (SB) for AGN/LINER spaxels to exclude DIG-dominated extended regions.

Even after imposing these spaxel-fraction limits, EW filters, and SB cuts, contamination from extreme star-forming spaxels remained an issue. In some low-mass systems with high specific star formation rates (sSFR), shocks or DIG can elevate the emission-line ratios enough to move spaxels across the demarcation curve, especially in the [\ion{S}{ii}]-BPT plane \citep[e.g.,][]{2008A&A...485..657B}. To address this, they defined a new parameter, the BPT distance ($d_{\text{BPT}}$), which quantifies the perpendicular distance of each spaxel classified as AGN or LINER-like from the star-forming demarcation line in the [\ion{S}{ii}]-BPT diagram. By calculating the mean distance of the most extreme spaxels and imposing a stringent cutoff in this metric, they successfully filtered out spurious AGN candidates whose spaxels lie close to the boundary, preserving only those systems whose line ratios unambiguously require a hard ionizing spectrum.

To build our high-purity AGN parent sample, we applied the spaxel-fraction selection discussed above to all the $\sim$ 10\,000 {\sc megacubes}. In this step, the total number of valid spaxels (those simultaneously passing all S/N criteria as defined in Section~\ref{sec:diagnostics}) serves as the denominator for the fraction calculations. To prevent spurious AGN classifications driven by small-number statistics in galaxies with very few high-quality spaxels, we required a global minimum of 10 valid spaxels per galaxy during this initial screening pass. This threshold is subsequently increased to 50 valid spaxels before the spatial classification to ensure robust radial profiling (Section~\ref{sec:method}). More strictly, a source was flagged as an AGN candidate only if it met all of the following: \footnote{$\{A,N\}$ refers to spaxels classified as active (AGN/composite) in the [\ion{N}{ii}]-BPT diagram, while $\{AL,S\}$ refers to spaxels classified as active (AGN/LINER-like) in the [\ion{S}{ii}]-BPT diagram.}

\begin{enumerate}[label=(\roman*), leftmargin=*]
    \item {[\ion{N}{ii}] Criteria:} Fraction of spaxels in the [\ion{N}{ii}] AGN/composite region $> 10\%$ ($f_{A,N} > 0.10$) and mean EW(H$\alpha)_{A,N} > 3.0$ \AA.
    \item {[\ion{S}{ii}] Criteria:} Fraction of spaxels in the [\ion{S}{ii}] AGN/LINER-like region $> 15\%$ ($f_{AL,S} > 0.15$) and mean EW(H$\alpha)_{AL,S} > 3.0$ \AA.
    \item {\raggedright Surface Brightness: $\log(SB(\mathrm{H}\alpha)_{A,N}) > 37.5$ or $\log(SB(\mathrm{H}\alpha)_{AL,S}) > 37.5$\,erg\,s$^{-1}$\,kpc$^{-2}$.\par}
    \item {BPT Distance:} Mean orthogonal distance from the SF demarcation line in the [\ion{S}{ii}]-BPT, $d_{\text{BPT}} > 0.3$.
\end{enumerate}

Applying these cuts to the spaxels that passed the S/N and emission-line detections (discussed above) yielded 776 AGN candidates. For each galaxy in this preliminary sample, we produced side-by-side [\ion{N}{ii}]-BPT, [\ion{S}{ii}]-BPT, WHaN, and WHaD maps, alongside the normalized stellar continuum flux map ($F_{Norm}$) for reference (e.g., Figure \ref{fig:diagnostic_maps_example}). We overlaid continuum S/N $\geq 5$ and S/N $\geq 10$ contours on these maps to indicate regions where the underlying stellar continuum subtraction is most reliable.

We then conducted a visual inspection of all 776 candidates, examining their SDSS images together with these four spatially resolved diagnostic maps to verify the spatial correspondence of the classifications. At this stage, we discarded systems whose spatial ionization patterns could not be robustly interpreted. To ensure the reliability of our visual classifications, the inspection was performed independently in two separate passes, which yielded consistent results for more than 99\% of the objects. We removed two main categories of galaxies:

\begin{enumerate}[label=(\roman*), leftmargin=*, labelwidth=2em]
    \item {Broad-line sources (Type 1 AGN):} Galaxies that display broad Balmer emission lines. These systems are excluded because accurately separating the narrow-line component, which is not included in the {\sc megacubes}, would require specialized broad-line fitting, which lies beyond the scope of our spatially resolved analysis. An example is shown in Figure~\ref{fig:bcomp_example}
    \item {Major mergers and interactions:} Galaxies exhibiting strongly disturbed morphologies or prominent tidal features, in which the nuclear region is ill-defined, and shocks may imitate off-nuclear AGN-like emission. In such cases, no unique photometric center can be assigned in a meaningful way, so the radial zones that underpin our nuclear/off-nuclear classification (Section~\ref{sec:nuclear_class}) cannot be defined. An example is shown in Figure~\ref{fig:merger_example}
\end{enumerate}

\begin{figure} 
    \centering
    \includegraphics[width=0.99\columnwidth]{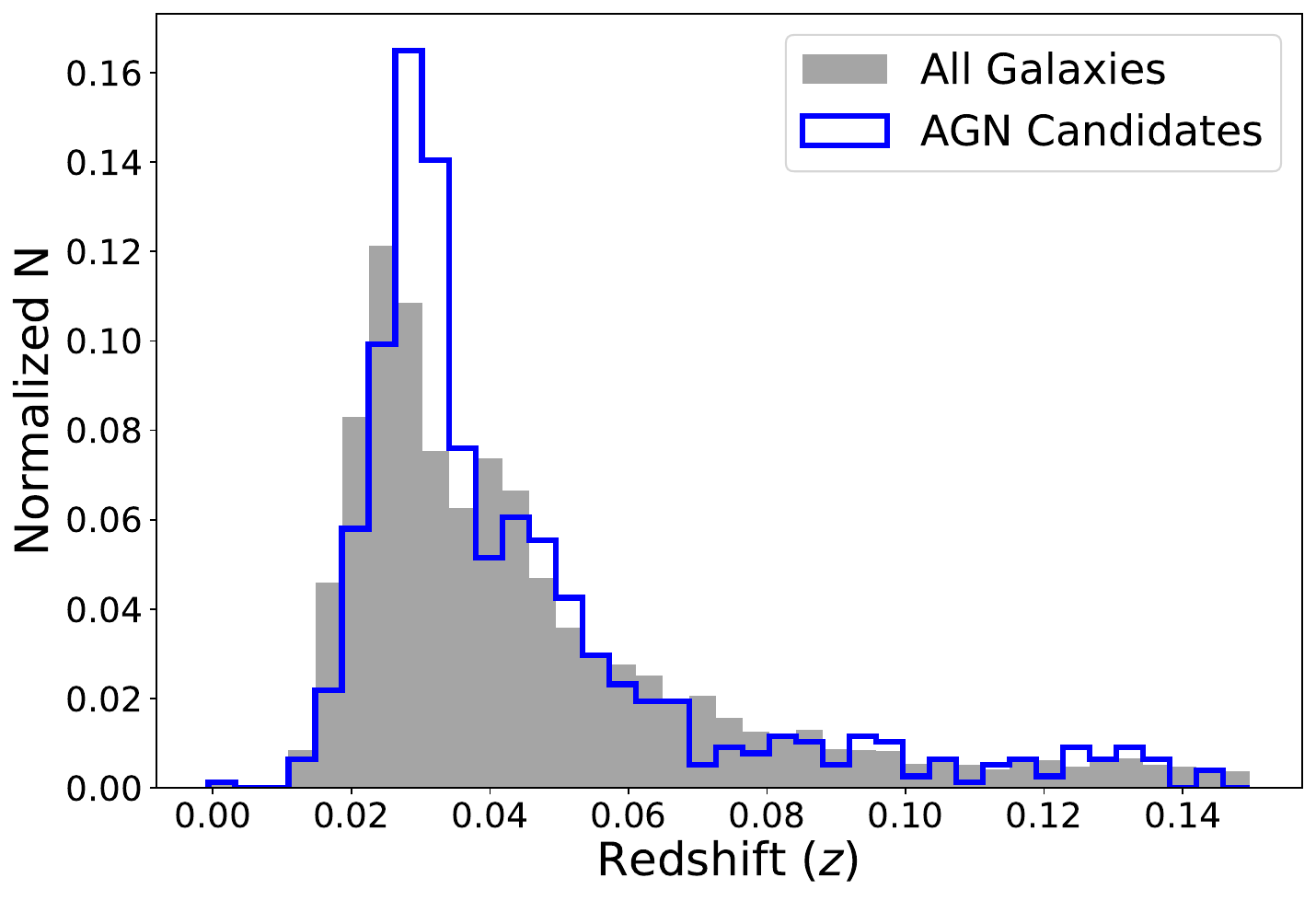}
    \caption{Normalized distribution of the redshift ($z$) for all galaxies in the MaNGA sample (filled, grey histogram) and the AGN candidates (blue line).}
    \label{fig:normalized_z}
\end{figure}

This quality-control step removed 149 galaxies, leaving a final working sample of 627 systems, which we use as input for the spatial classification described below. Comparing their redshift distribution to that of the full MaNGA sample (Figure~\ref{fig:normalized_z}) shows that both peak at $z \sim 0.02$–0.04 with similar overall shapes, indicating our AGN sample is representative of the survey volume.

\section{Nuclear and off-nuclear AGN samples}
\label{sec:method}
Because these diagnostics are defined on a spaxel-by-spaxel basis, they also permit an investigation of the spatial structure of AGN ionization and make it possible to distinguish nuclear from off-nuclear activity. In galaxies hosting nuclear AGNs, AGN-like spaxels are expected to be concentrated toward the center, generally within the inner 1–2 kiloparsec \citep[kpc;][]{al2023, Riffel2023}, consistent with ionization powered by a central SMBH. Off-nuclear AGN candidates, by contrast, should show AGN-like emission displaced from the galaxy center. In combination, the spatially resolved line-ratio classifications, EW thresholds, SB cuts, and the $d_{\text{BPT}}$ distance metric provide a robust framework for identifying and mapping off-nuclear AGN-like ionization in IFS surveys.

\subsection{Spatial Classification: Nuclear and Off-Nuclear AGN}

\subsubsection{Nuclear AGN selection}
\label{sec:nuclear_class}
We defined a nuclear aperture of 2 kpc diameter around the adopted center, which was taken to be the photometric center derived from the $F_{Norm}$ maps. To determine this center, we smoothed the inner 60\% of each flux map using a median filter and identified the flux maximum as the galaxy center. The adopted aperture scale is motivated by the characteristic spatial extent of the narrow-line region (NLR), which can reach up to 1 kpc \citep[e.g.,][]{liu2013,chen2019, Deconto-Machado2022}, though we note that extended NLR in high-luminosity quasars can span several kpc \citep[e.g.,][]{hainline2013, hainline2014}. This 2 kpc diameter is also comparable to the median spatial resolution of MaNGA (1.37 kpc; \citep[][]{Wake_2017}). Furthermore, using this aperture (up to 2 kpc) enables an effective separation between compact nuclear emission and DIG at larger galactocentric radii \citep[e.g.,][]{Lacerda2018,al2023, Riffel2023}.

Within this nuclear aperture, we computed the fraction of valid spaxels classified as active --Seyfert or LINER-like-- in the BPT diagrams or --sAGN or wAGN-- in the WHaN diagram ($f_{\rm nuc}$). As with the global AGN selection, the denominator for $f_{\rm nuc}$ is strictly the number of valid spaxels within the aperture. A galaxy was classified as hosting a nuclear AGN if $ f_{\rm nuc} \geq 0.40$ in either the [\ion{N}{ii}]-BPT or [\ion{S}{ii}]-BPT diagram and, simultaneously, in the WHaN diagram. The threshold of 40\% was determined through a detailed empirical analysis of the diagnostic maps, from which we found that this value robustly identifies genuine nuclear activity while retaining borderline systems. This criterion was further validated by visual inspection of all four diagnostic maps for each candidate. Galaxies that did not satisfy this condition, i.e., those whose centers are dominated by SF or composite emission, were retained as potential off-nuclear AGN candidates. An illustrative example of a nuclear AGN identified with this procedure is presented in Figure~\ref{fig:nucl_example}. We note that the 2\,kpc aperture is comparable to the MaNGA resolution element beyond $z \approx 0.06$ ($\delta \approx 35z$\,kpc; \citealt{Bundy2015}); beam smearing may therefore lower $f_{\rm nuc}$ for the most distant galaxies, and \citet{al2023} found that AGN recovered only in sub-2\,kpc apertures lie at systematically lower redshift.

\subsubsection{Off-Nuclear AGN selection}
\label{sec:offnuclear_class}
Galaxies that did not meet the nuclear threshold ($f_{\rm nuc} < 0.40$) were taken as potential off-nuclear candidates. Since any AGN offset by less than a kpc blends into the nuclear classification, our off-nuclear sample represents a strict lower limit. For these sources, we computed the AGN spaxel fraction in four cumulative integration zones extending beyond the nuclear aperture (i.e., $r > 1.0$\,kpc), where $r$ represents the projected on-sky galactocentric radius:

\begin{enumerate}[leftmargin=*]
    \item {Immediate Off-Nuclear:} $1.0 < r < 2.0$\,kpc
    \item {Inner Disk:} $1.0 < r < 4.0$\,kpc
    \item {Extended Disk:} $1.0 < r < 6.0$\,kpc
    \item {Total Off-Nuclear:} $1.0 < r < R_{max}$\, kpc (integrating out to the edge of the FoV).
\end{enumerate}

Each zone was required to contain at least 30 valid spaxels ($N_{\rm spax} \geq 30$), since a fractional criterion alone can reach high values in sparsely sampled zones; regions lacking sufficient high S/N data were entirely excluded from the denominator. This requirement preferentially removes the innermost zone of the most distant galaxies, where a given physical zone subtends fewer spaxels. A galaxy was classified as an off-nuclear AGN if the fraction of AGN-classified spaxels satisfied $f_{\rm zone} > 0.15$ simultaneously in both the [\ion{N}{ii}]-BPT and [\ion{S}{ii}]-BPT diagrams and in the WHaN diagram, within at least one of the outer zones. Using multiple zones rather than a single fixed aperture allows us to pick up AGN-ionized gas regardless of how far it extends from the center. An example of an off-nuclear AGN identified this way is shown in Figure~\ref{fig:offnucl_example}.

We define these zones in fixed physical units because the extent of AGN-ionized gas is driven by the central source luminosity rather than the host galaxy's size. We validated this choice in two ways. First, we repeated the classification with the outer edge of each zone scaled to fractions of the $R_{\rm e}$ (e.g., $0.5 R_{\rm e}$, $1.0 R_{\rm e}$, $1.5 R_{\rm e}$; where $R_{\rm e} = 1.5$--$9.2$\,kpc across the sample, median $3.7$\,kpc). This approach left the off-nuclear sample essentially unchanged, recovering $99\%$ of the sources, with none of the minor losses caused by the $N_{\rm spax}$ requirement. Second, because a fixed physical zone covers fewer spaxels at larger distances, we verified the exact redshift at which the 30-spaxel requirement becomes restrictive for each zone. The $1$--$2$\,kpc zone drops below the threshold at $z \approx 0.060$, the $1$--$4$\,kpc zone at $z \approx 0.148$, and the $1$--$6$\,kpc zone at $z \approx 0.25$. Given our sample's redshift range, only the innermost zone is affected: it contains sufficient spaxels for $99\%$ of galaxies below $z = 0.06$, while the three outer zones remain well-sampled for every galaxy. Importantly, because the last zone integrates out to the edge of the FoV, every single galaxy is evaluated regardless of distance. Finally, for the sources above $z = 0.06$, we only lose the spatial resolution to localize the emission within that zone of radius $2$\,kpc, not the detection of the off-nuclear activity itself. We therefore retain fixed physical apertures throughout.

\subsection{Quiet AGNs}
\label{sec:quiet_class}
Galaxies classified as AGN but that failed both the nuclear and off-nuclear thresholds were initially labeled as quiet AGNs (Figure~\ref{fig:quiet_example}). When we inspected their diagnostic maps, a clear pattern emerged; many of these sources look composite-dominated in the [\ion{N}{ii}]-BPT but get flagged as AGN by the WHaN diagram. The issue turned out to be the WHaN SF region boundary.

The traditional WHaN diagram places the SF region cutoff at $\log([\ion{N}{ii}]/\text{H}\alpha) = -0.4$ \citep{stasinska2006, CidFernandes2010}, which is the extremely conservative lower limit for pure SF regions. Conversely, the theoretical maximum starburst limit defined by \citet{Kewley2001} establishes an upper extreme at $\log([\ion{N}{ii}]/\text{H}\alpha) = -0.1$ \citep{CidFernandes2010}. In spatially resolved IFU data, the blending of DIG and high-metallicity regions naturally shifts the local line ratios toward this upper limit \citep{2017MNRAS.466.3217Z}. Therefore, maintaining the strict $-0.4$ boundary contradicts the BPT maps and falsely flags extended SF in our sources as AGN activity.

We addressed this by running a secondary diagnostic pass on the quiet sample alone, moving the WHaN boundary to 

\begin{equation}
  \log\!\left(\frac{[\ion{N}{ii}]}{\text{H}\alpha}\right) = -0.25\,,
  \label{eq:new_whan}
\end{equation}
the midpoint between the \citet{stasinska2006} and \citet{Kewley2001} values. Spaxels falling between the original ($-0.4$) and revised ($-0.25$) boundaries are flagged as Transitional (Trans), and are not counted as AGN in subsequent analysis. After applying this shift, the WHaN and BPT classifications come into agreement for the quiet sources (Figure~\ref{fig:quiet_before_after}). The detailed impact of this boundary revision on the quiet sample is discussed in Section~\ref{sec:results_quiet}.

\medskip
\noindent Starting from the 776 AGN candidates, the full selection and classification pipeline produced the final breakdown shown in Table~\ref{tab:sample}.

\begin{table}
\centering
\caption{Summary of the sample selection and classification.}
\label{tab:sample}
\begin{tabular}{lr}
\hline
Classification Step & Number of Galaxies \\
\hline
\textbf{Parent Sample} & \textbf{776} \\
\hline
\textit{Removed: Visual Inspection} & 149 \\
\hline
\textbf{Input to Classification Pipeline} & \textbf{627} \\
\quad -- Rejected: Low valid spaxel count & 6 \\
\quad -- Nuclear AGNs & 261 \\
\quad -- \textbf{Off-nuclear candidates} & \textbf{201} \\
\quad -- Quiet AGNs & 159 \\
\hline
\end{tabular}
\end{table}

\section{Nuclear vs. Off-Nuclear AGN}
\label{sec:results_main}

\subsection{Sample Validation}
\label{sec:validation}

\subsubsection{Comparison with Wylezalek et al. (2018)}
\label{sec:compare_wylezalek}
We validated our selection by comparing it against the sample of \citet{Wylezalek2018}, who ran their criteria on the MPL-5 (2,727 galaxies) with a cut of $EW_{\mathrm{H}\alpha}$ $> 5.0$ \AA\ and found 303 AGN candidates. Cross-matching their catalog with our 776 sources returned 105 in common.

The reasons for this mismatch are: DR17 is a much larger dataset than MPL-5, and our lower EW threshold (3.0 vs.\ 5.0 \AA) brings in weaker candidates that their cut excluded. The more interesting question is why some sources that were flagged as AGN in their list no longer pass the criteria with the current data. To track this down, we recalculated emission-line fluxes for the missed sources using direct integration applied consistently to both the original MPL-5 spectra and the DR17 {\sc megacubes}.

\begin{figure}
    \centering
    \includegraphics[trim=0cm 0cm 0cm 1.5cm, clip,
                     width=\columnwidth]{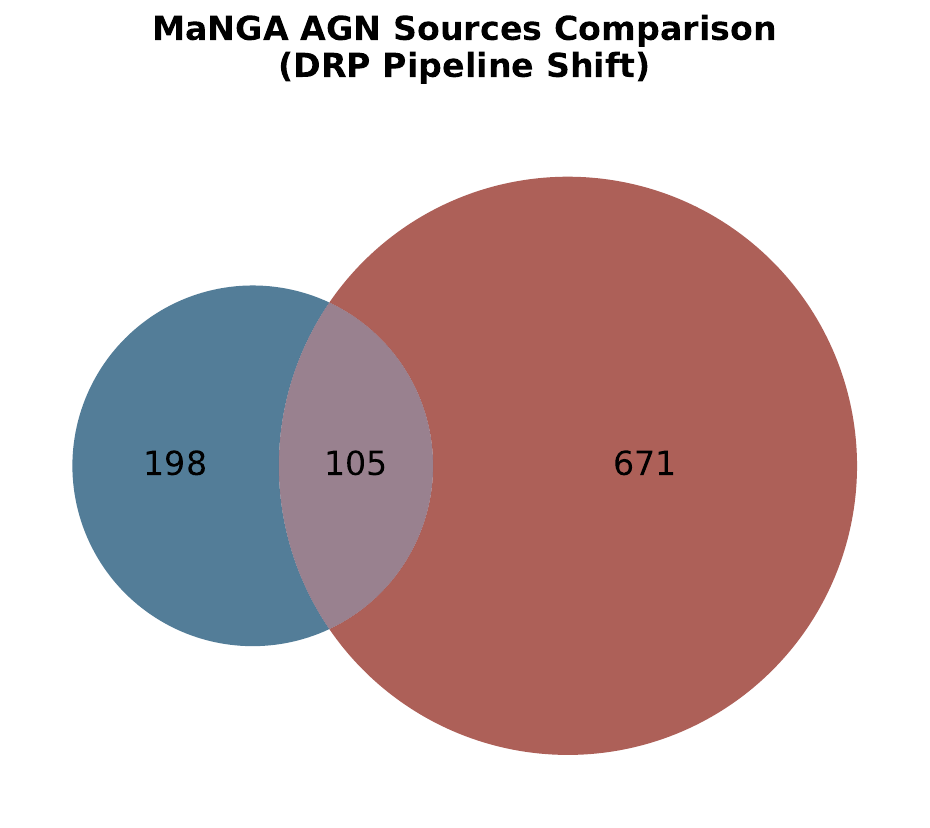}
   
    \caption{Comparison of MaNGA AGN sources between \citet{Wylezalek2018} and this work. left: Wylezalek et al. (2018) only; centre: in common; right: this work only. The 198 missing sources on the left represent targets excluded from our final sample due to updates in the MaNGA data reduction pipeline, as well as borderline targets rejected by the strict valid-spaxel mask we implemented for the emission lines.}
    \label{fig:venn_manga_compare}
\end{figure}

What we found is that the difference comes down to updates in the MaNGA data reduction pipeline (DRP) between the two releases \citep{2016AJ....152...83L, Law+21}. The flux-ratio maps comparison (Figure~\ref{fig:drp_comparison}) shows that in the same spatial regions, the new pipeline measures less [\ion{O}{iii}] flux than the old one (blue in the maps) but more H$\beta$. Dividing a smaller [\ion{O}{iii}] by a larger H$\beta$ pulls the [\ion{O}{iii}]/H$\beta$ ratio down, which on the BPT diagram means those spaxels move downward, often enough to drop below the AGN demarcation line. Specifically, about 60\% of the missing sources now fail the [\ion{S}{ii}] selection criteria described in \ref{sec:wylezalek}, while most of the rest fail the $d_{\text{BPT}}$ cut. This confirms that the combination of calibration-level shifts from the DRP updates and our applied valid-spaxel emission-line masks is sufficient to flip borderline spaxels from AGN to non-AGN, thereby explaining the missing sources shown in Figure~\ref{fig:venn_manga_compare}.

\subsubsection{Comparison with Albán \& Wylezalek (2023)}
\label{sec:compare_alban}
We also cross-matched our 261 nuclear AGN candidates with the catalog of \citet{al2023}. They identified optical AGNs using a 2 kpc aperture centered on each galaxy from the MaNGA Survey, the same nuclear scale we adopt in Section~\ref{sec:nuclear_class}. This returned 190 sources in common.

A number of their nuclear AGN, however, did not survive our visual inspection. Several turned out to be galaxies undergoing major mergers or interactions, where disturbed morphologies and shock fronts can mimic off-nuclear AGN signatures. While the \citet{al2023} approach works well for picking out sources where the AGN dominates the central 2 kpc, our spaxel-fraction method \citep{Wylezalek2018} is better suited to catching activity that sits further away from the nucleus.

\subsubsection{Comparison with Riffel et al. (2023)}
\label{sec:compare_riffel}
As an additional check on our nuclear classifications, we compared our results with \citet{Riffel2023}, who identified 293 optical AGN hosts. Their active sample was defined using integrated nuclear fluxes from SDSS-III single-fiber observations rather than spatially resolved maps.

Of the 293 AGNs from their catalog, our pipeline matched 171 galaxies. The majority of these (126 sources) were classified as nuclear AGNs by our analysis, and 4 were classified as off-nuclear candidates. The remaining galaxies in their sample met our emission-line thresholds but were excluded during visual inspection due to broad-line components or major mergers.

The remaining 122 galaxies from their catalog did not enter our sample. This discrepancy highlights the fundamental physical difference between single-fiber and IFU-based AGN selection. An AGN may be sufficiently luminous to dominate the integrated emission within the central 3 arcseconds of a galaxy yet fail to produce an extended NLR large enough to surpass our requirements (e.g., $f_{A,N} > 10\%$ and $f_{AL,S} > 15\%$).

\subsection{Nuclear vs. Off-Nuclear: Spaxel-Level Physical Properties}
\label{sec:nuc_vs_offnuc}

With the classification established, to investigate the mechanisms underlying the differences between nuclear and off-nuclear AGN hosts, we compared the physical characteristics in the host galaxies on a spaxel-by-spaxel basis. From the {\sc megacubes}, we extracted the following properties for all spaxels with continuum S/N > 10, where the stellar population fits are more robust \citep{Riffel2023}: light- and mass-weighted stellar age, light- and mass-weighted stellar metallicity, stellar velocity dispersion ($\sigma_\star$), dust attenuation ($A_V$), stellar mass, recent star formation rate (SFR), H$\alpha$ velocity dispersion ($\sigma_{\rm gas}$), [\ion{O}{iii}]$\lambda$5007 luminosity and SB.

For each host class, we divided the spaxels into two distinct populations:

\begin{itemize}
    \item \textbf{SF spaxels:} spaxels classified as star-forming in the [\ion{N}{ii}]-BPT diagram. A comparison of these populations probes potential differences in the star-forming disks between the two host classes.
    \item \textbf{AGN spaxels:} spaxels classified as Seyfert- or LINER-like in the [\ion{N}{ii}]-BPT diagram. A comparison of these populations assesses whether the AGN-ionized gas itself exhibits systematic differences.
\end{itemize}

Table~\ref{tab:ks_results} reports the mean values, standard deviations, and two-sample Kolmogorov–Smirnov (KS) statistics for all considered properties. Owing to the very large number of spaxels in both samples, the KS $p$-values are very small in all cases and, by themselves, are not particularly informative. We therefore focus our interpretation on the actual differences between the mean values and on the visual degree of overlap between the corresponding distributions. Using medians instead of means gives the same trends for all properties (e.g.\ $\Delta \log L_{\rm [\ion{O}{iii}]} = 0.71$ vs.\ $0.72$~dex), indicating that our results are not driven by outliers.

\subsubsection{The Star-Forming Environment}
\label{sec:sf_spaxels}

Across most measured quantities, SF spaxels in nuclear and off-nuclear hosts are statistically consistent. Stellar ages ($\langle \log\,t_L \rangle \sim 9.35{-}9.40$; $\langle \log\,t_M \rangle \sim 9.93$), metallicities, stellar mass ($\langle \log\, M_\star \rangle \sim 7.74$), recent SFR, and $\sigma_{\rm gas}$ differ by only a few percent, with strongly overlapping distributions.

\begin{figure}
    \centering
    \includegraphics[trim=0cm 0cm 0cm 1cm, clip, width=\columnwidth]{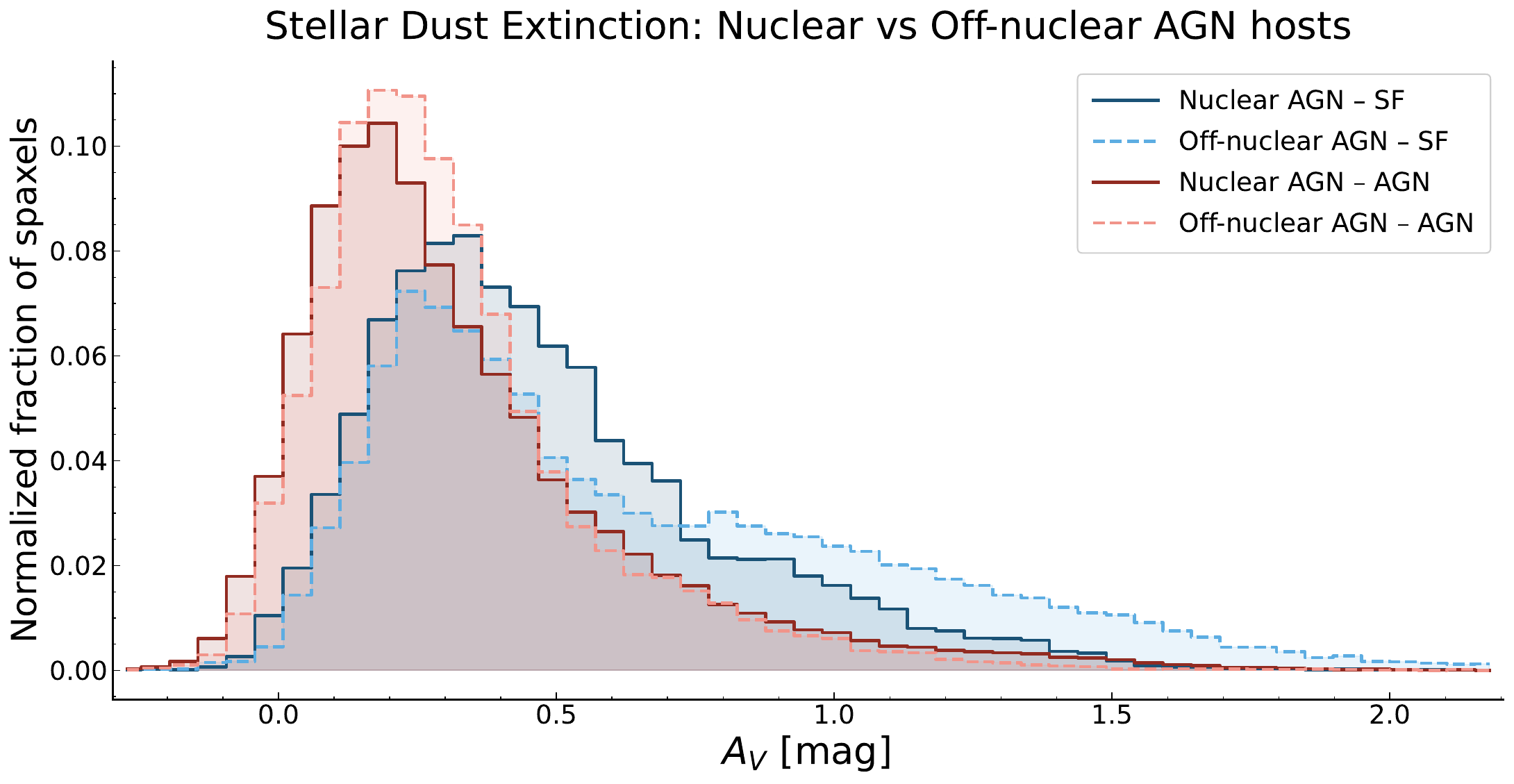}
    \caption{Normalized distribution of the stellar dust attenuation ($A_V$). The SF spaxels (blue) of off-nuclear hosts are systematically dustier than those of nuclear hosts,  while the AGN spaxels (red) show comparable attenuation in both classes.}
    \label{fig:dust_av}
\end{figure}

Dust attenuation is the main exception: SF spaxels in off-nuclear hosts show higher $A_V$ ($0.65$ vs.\ $0.49$, KS~$= 0.172$; Figure~\ref{fig:dust_av}). Most of this gap comes from how the two samples are distributed in inclination rather than from any real difference at a fixed viewing angle—off-nuclear hosts are more inclined (Section~\ref{sec:axial_ratio}), and among face-on galaxies alone the difference shrinks to $0.36$ vs.\ $0.39$~mag (Table~\ref{tab:orientation_full}). We come back to orientation, and to whether dust in the center can explain the off-nuclear classification in Section~\ref{sec:balmer}.

\subsubsection{The AGN-Ionized Environment}
\label{sec:agn_spaxels}

\begin{figure}
    \centering
    \includegraphics[trim=0cm 0cm 0cm 1.3cm, clip, width=\columnwidth]{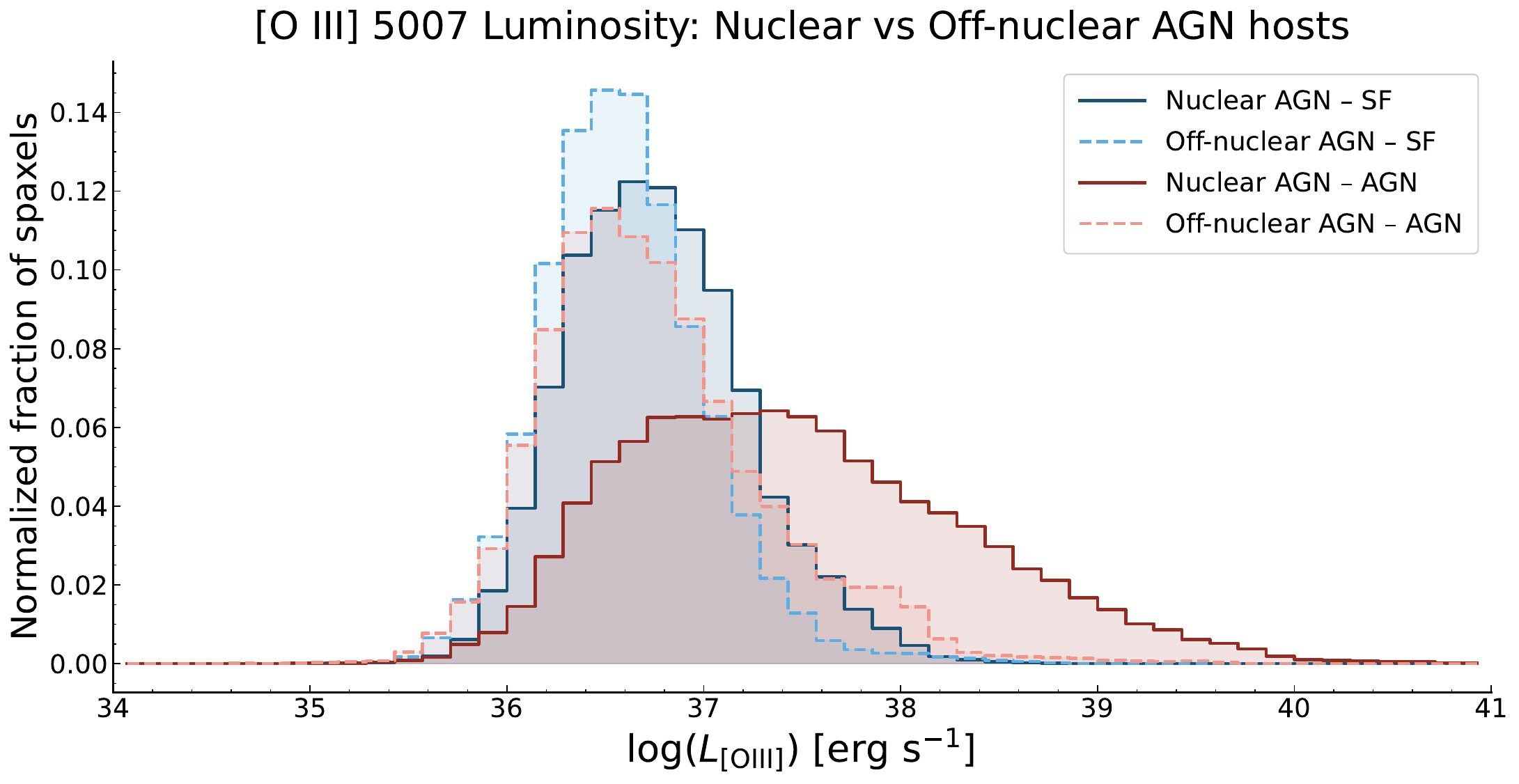}
    \caption{Normalized distribution of the [\ion{O}{iii}]$\lambda$5007 luminosity for nuclear (solid) and off-nuclear (dashed) AGN hosts, split into SF spaxels (blue) and AGN spaxels    
    (red). The nuclear AGN spaxels are $\sim$0.7~dex brighter on average, the largest difference among all properties examined.}
    \label{fig:loiii}
\end{figure}

The stellar populations underlying the AGN spaxels are broadly similar between nuclear and off-nuclear hosts, with comparable stellar mass, age, and metallicity; the primary differences are observed in the gas properties. The strongest contrast is in [\ion{O}{iii}]$\lambda$5007 luminosity (Figure~\ref{fig:loiii}) and is closely followed by the excitation state of the gas. Nuclear AGN spaxels are brighter ($\langle \log\,L_{\rm [\ion{O}{iii}]} \rangle = 37.47$~erg\,s$^{-1}$) than off-nuclear spaxels ($\langle \log\,L_{\rm [\ion{O}{iii}]} \rangle = 36.76$~erg\,s$^{-1}$), corresponding to a $\sim$0.7~dex offset and the most significant separation in our comparison (KS~$= 0.396$). The offset persists when expressed as a SB, $\Sigma = L/A$, with $A$ the proper area of a  $0.5\arcsec$  spaxel at the redshift of each
galaxy ($\langle \log\,\Sigma_{\rm [\ion{O}{iii}]} \rangle = 38.38$ vs. $37.78$~erg\,s$^{-1}$\,kpc$^{-2}$, KS~$= 0.364$); Figure~\ref{fig:sb_oiii}), and is therefore not driven by the differing physical areas subtended by a spaxel across the sample.

\begin{figure}
    \centering
    \includegraphics[trim=0cm 0cm 0cm 1.2cm, clip, width=\columnwidth]{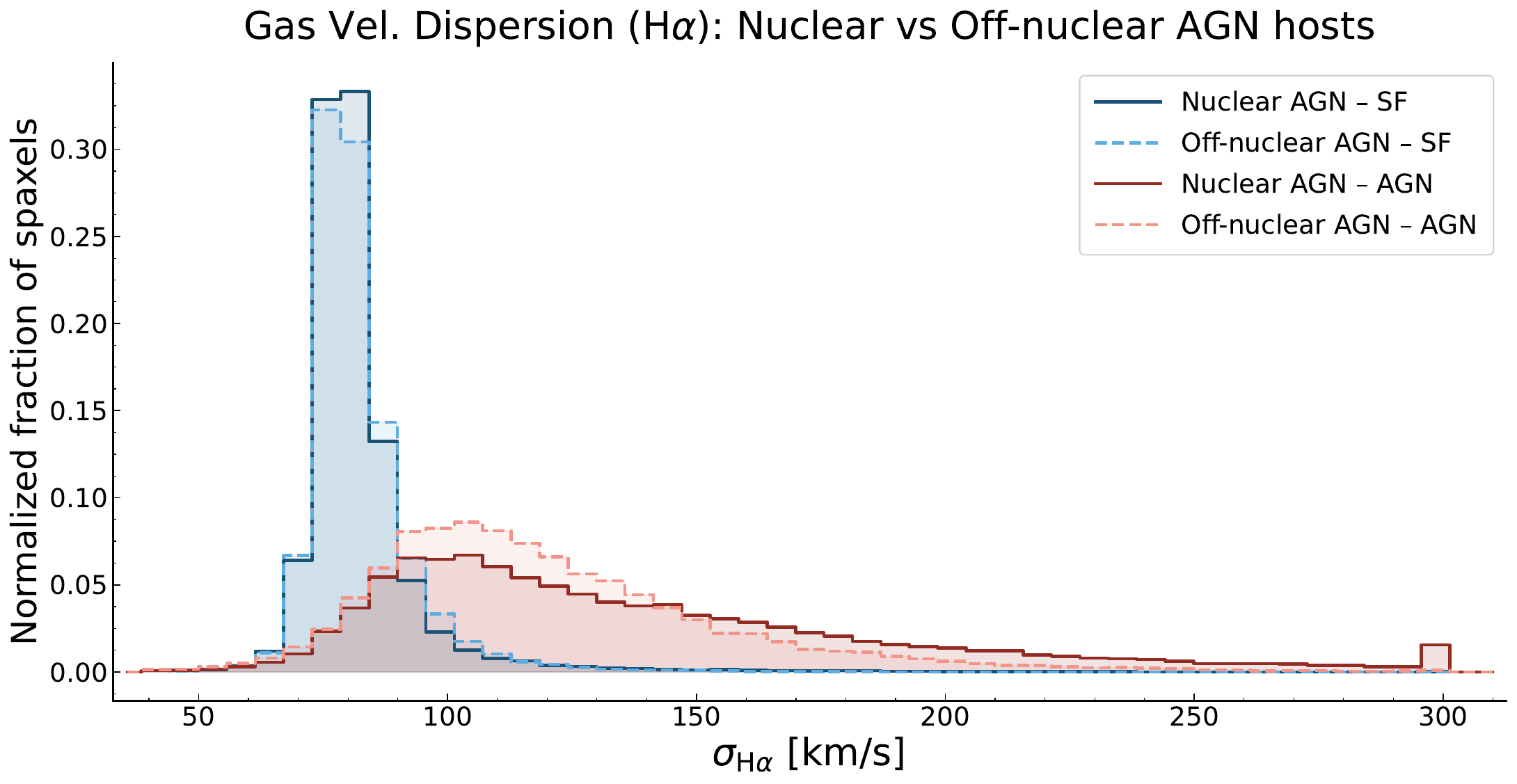}
    \caption{Normalized distribution of the gas velocity dispersion. The SF spaxels (blue) are nearly identical between the two host classes, while the AGN spaxels (red) of nuclear hosts show broader profiles than their off-nuclear counterparts.}
    \label{fig:sig_gas}
\end{figure}

This luminosity deficit is accompanied by a decrease in gas excitation (Figure~\ref{fig:loiii_and_excitation}). The off-nuclear spaxels exhibit systematically lower excitation line ratios, with $\langle \log$([\ion{O}{iii}]/H$\beta)\rangle = 0.29$ compared to $0.50$ in the nuclear AGN spaxels (KS~$= 0.324$). $\sigma_{\rm gas}$ exhibits a consistent trend (Figure~\ref{fig:sig_gas}), with broader line profiles in nuclear AGN spaxels ($\langle \sigma_{\rm gas} \rangle = 138.3$~km\,s$^{-1}$) than in off-nuclear spaxels ($120.5$~km\,s$^{-1}$, KS~$= 0.162$). Recent SFR also differs modestly, with nuclear AGN spaxels showing slightly higher SFR over the last 20~Myr ($\langle \log\,{\rm SFR}_{20} \rangle = -2.60$) relative to off-nuclear spaxels ($-2.79$, KS~$= 0.134$). Distributions for the remaining properties are provided in Appendix~\ref{sec:other_properties}.

\begin{table*}
  \centering
  \begin{threeparttable}
  \caption{Mean values ($\pm$ standard deviation) and KS statistics ($D$-value) for nuclear vs.\ off-nuclear AGN hosts, separated into SF and AGN spaxel populations.}
  \label{tab:ks_results}
  \begin{tabular}{lcccccc}
    \hline
    & \multicolumn{3}{c}{SF Spaxels} &
      \multicolumn{3}{c}{AGN Spaxels} \\
       Property (per spaxel) &
      Nuclear & Off-Nuc & KS &
      Nuclear & Off-Nuc & KS \\
    \hline
    $\log\,t_L$ (age, light-weighted)
      & $9.397 \pm 0.260$ & $9.348 \pm 0.314$ & 0.093
      & $9.699 \pm 0.398$ & $9.714 \pm 0.291$ & 0.033 \\
    $\log\,t_M$ (age, mass-weighted)
      & $9.932 \pm 0.181$ & $9.944 \pm 0.209$ & 0.062
      & $9.965 \pm 0.338$ & $9.956 \pm 0.228$ & 0.070 \\
    $Z_\star$ (light-weighted)
      & $0.018 \pm 0.004$ & $0.018 \pm 0.004$ & 0.034
      & $0.017 \pm 0.004$ & $0.018 \pm 0.004$ & 0.071 \\
    $Z_\star$ (mass-weighted)
      & $0.020 \pm 0.005$ & $0.021 \pm 0.006$ & 0.084
      & $0.018 \pm 0.005$ & $0.019 \pm 0.005$ & 0.042 \\
    $\sigma_\star$ (km\,s$^{-1}$)
      & $88.0 \pm 33.1$ & $83.2 \pm 28.7$ & 0.060
      & $125.8 \pm 50.1$ & $117.7 \pm 42.7$ & 0.093 \\
    $A_V$ (mag)
      & $0.486 \pm 0.318$ & $0.645 \pm 0.456$ & 0.172
      & $0.336 \pm 0.310$ & $0.319 \pm 0.258$ & 0.045 \\
    $\log(M_\star)$ ($M_\odot$)
      & $7.735 \pm 0.366$ & $7.749 \pm 0.421$ & 0.051
      & $8.084 \pm 0.555$ & $7.998 \pm 0.546$ & 0.080 \\
    $\log\,{\rm SFR}_{20}$
      & $-2.669 \pm 0.656$ & $-2.622 \pm 0.736$ & 0.074
      & $-2.598 \pm 0.738$ & $-2.790 \pm 0.660$ & 0.134 \\
    $\sigma_{\rm gas}$ (km\,s$^{-1}$)
      & $82.3 \pm 13.7$ & $82.1 \pm 10.7$ & 0.038
      & $138.3 \pm 51.9$ & $120.5 \pm 35.5$ & 0.162 \\
    $\log\,L_{\rm [\ion{O}{iii}]}$ (erg\,s$^{-1}$)
      & $36.77 \pm 0.46$ & $36.60 \pm 0.42$ & 0.167
      & $37.47 \pm 0.86$ & $36.76 \pm 0.58$ & 0.396 \\
    $\log\,\Sigma_{\rm [\ion{O}{iii}]}$ (erg\,s$^{-1}$\,kpc$^{-2}$)$^{a}$
      & $37.69 \pm 0.37$ & $37.65 \pm 0.39$ & 0.051
      & $38.38 \pm 0.73$ & $37.78 \pm 0.49$ & 0.364 \\
    $\log$([\ion{O}{iii}]/H$\beta$)
      & $-0.424 \pm 0.288$ & $-0.428 \pm 0.237$ & 0.058
      & $0.503 \pm 0.314$ & $0.291 \pm 0.340$ & 0.324 \\
    $12+\log(\mathrm{O/H})^{b}$
      & $8.744 \pm 0.061$ & $8.758 \pm 0.054$ & 0.099
      & $8.630 \pm 0.096$ & $8.650 \pm 0.093$ & 0.129 \\
    \hline
  \end{tabular}
  \begin{tablenotes}[flushleft]
    \small
    \item $^a$ Surface brightness per spaxel, $\Sigma = L/A$.
    \item $^b$ Gas-phase metallicity: SF spaxels calibrated with the N2 index from \citet{Curti2017}; AGN spaxels (Seyfert + LINER-like) with the N2 index from \citet{Carvalho2020}.
  \end{tablenotes}
  \end{threeparttable}
\end{table*}

\subsubsection{Gas-Phase Metallicity}
\label{sec:metallicity_result}

We compared the gas-phase oxygen abundance, $12+\log(\mathrm{O/H})$, between nuclear and off-nuclear hosts. Because emission-line ratios depend on the dominant ionizing source, using a single \ion{H}{ii}-region calibration for AGN-ionized spaxels can yield unphysical abundances. We therefore adopted BPT-class-dependent calibrations, using the N2 index
for both to ensure a consistent abundance scale:

\begin{itemize}
    \item \textit{SF spaxels:} the N2 calibration of \citet{Curti2017},
          derived empirically via the direct-$T_\mathrm{e}$ method.
    \item \textit{AGN spaxels (Seyfert + LINER-like):} the N2 calibration of
          \citet{Carvalho2020}, developed for AGN using
          \textsc{cloudy} photoionization models.
\end{itemize}

Adopting the N2 index across all spatial classifications provides a methodological advantage: because the $[\ion{N}{ii}]/\mathrm{H}\alpha$ emission lines are so closely spaced in wavelength, the N2 index is virtually unaffected by uncertainties produced by absolute flux calibration and internal dust reddening \citep{Pettini2004, Kewley2008}.

\begin{figure}
    \centering
    \includegraphics[trim=0cm 0cm 0cm 1.4cm, clip, width=\columnwidth]{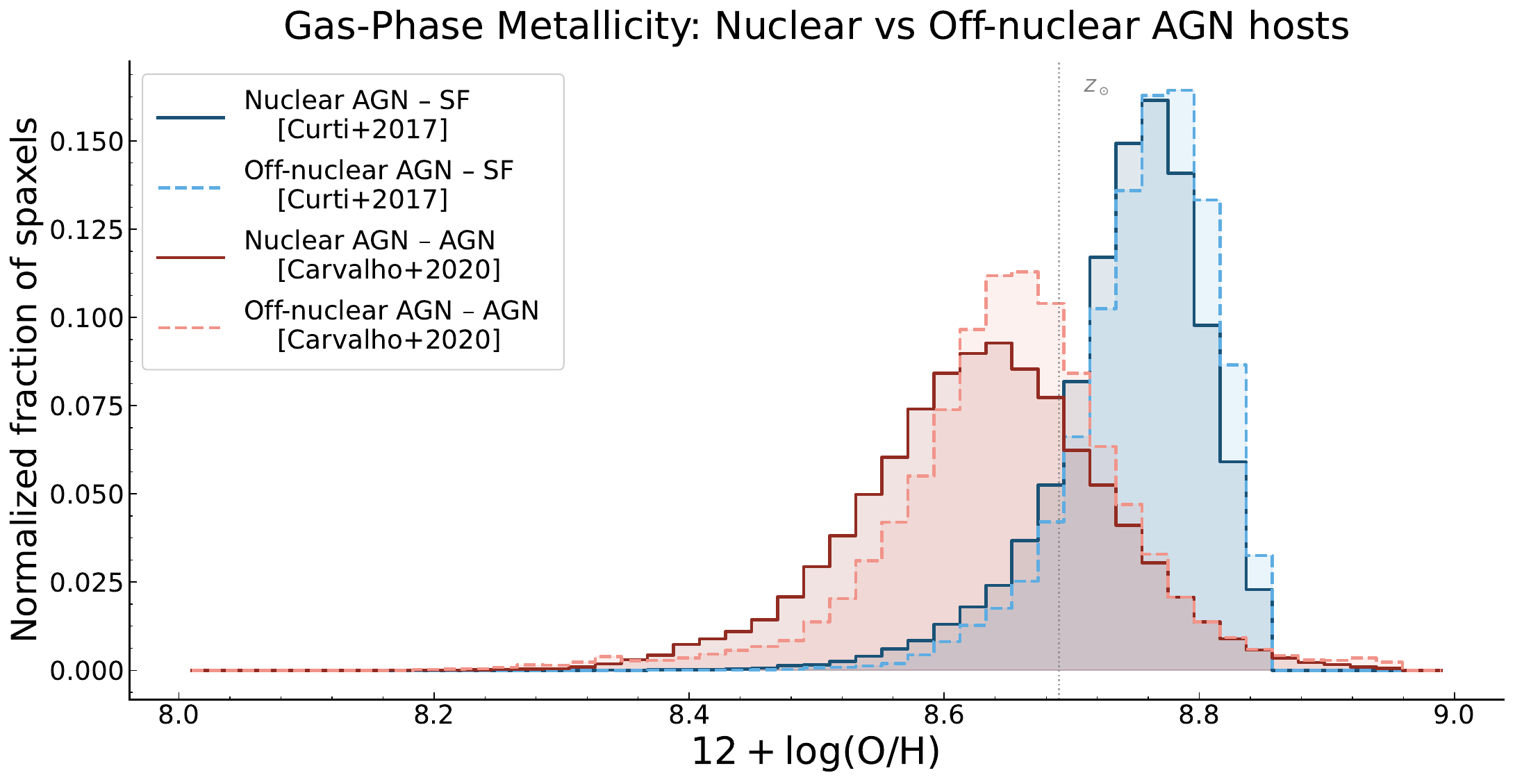}
    \caption{Normalized distributions of $12+\log(\mathrm{O/H})$ for nuclear (solid) and off-nuclear (dashed) AGN hosts, separated into SF-ionized spaxels (blue) and AGN-ionized spaxels (red).}
    \label{fig:gas-phase}
\end{figure}

We find that the gas-phase metallicities of nuclear and off-nuclear hosts are indistinguishable within the calibration uncertainty (Figure~\ref{fig:gas-phase}). For SF spaxels, the mean metallicity in nuclear hosts ($\langle 12+\log(\mathrm{O/H}) \rangle = 8.74$) is nearly identical to that in off-nuclear hosts ($8.76$). For AGN-ionized spaxels, the samples yield $\langle 12+\log(\mathrm{O/H}) \rangle = 8.63$ for nuclear and $8.65$ for off-nuclear hosts. These values are close to the solar abundance ($12+\log(\mathrm{O/H})_\odot = 8.69$).

In both the SF- and AGN-ionized regimes, the nuclear-off-nuclear offset is $\lesssim$ 0.02 dex. Because the intrinsic scatter of the adopted calibrations is on the order of 0.10 dex, this 0.02 dex difference is within the statistical uncertainty of the indices. We therefore find no chemical divergence between the two host classes; any intrinsic metallicity offset between nuclear and off-nuclear environments is too small to resolve robustly given the uncertainties of current abundance diagnostics.

\subsection{Host Galaxy Orientation}
\label{sec:orientation}

A key concern in identifying off-nuclear AGN is whether the inferred spatial offset is intrinsic or driven by projection effects. In highly inclined systems, outflow cones or extended emission projected along the disk plane could produce apparent off-nuclear signatures. We assessed this possibility using the host-galaxy axial ratio ($b/a$) as an inclination proxy.

\subsubsection{Axial Ratio Distributions}
\label{sec:axial_ratio}

\begin{figure}
    \centering
    \includegraphics[trim=0cm 0cm 0cm 1cm, clip,
                     width=\columnwidth]{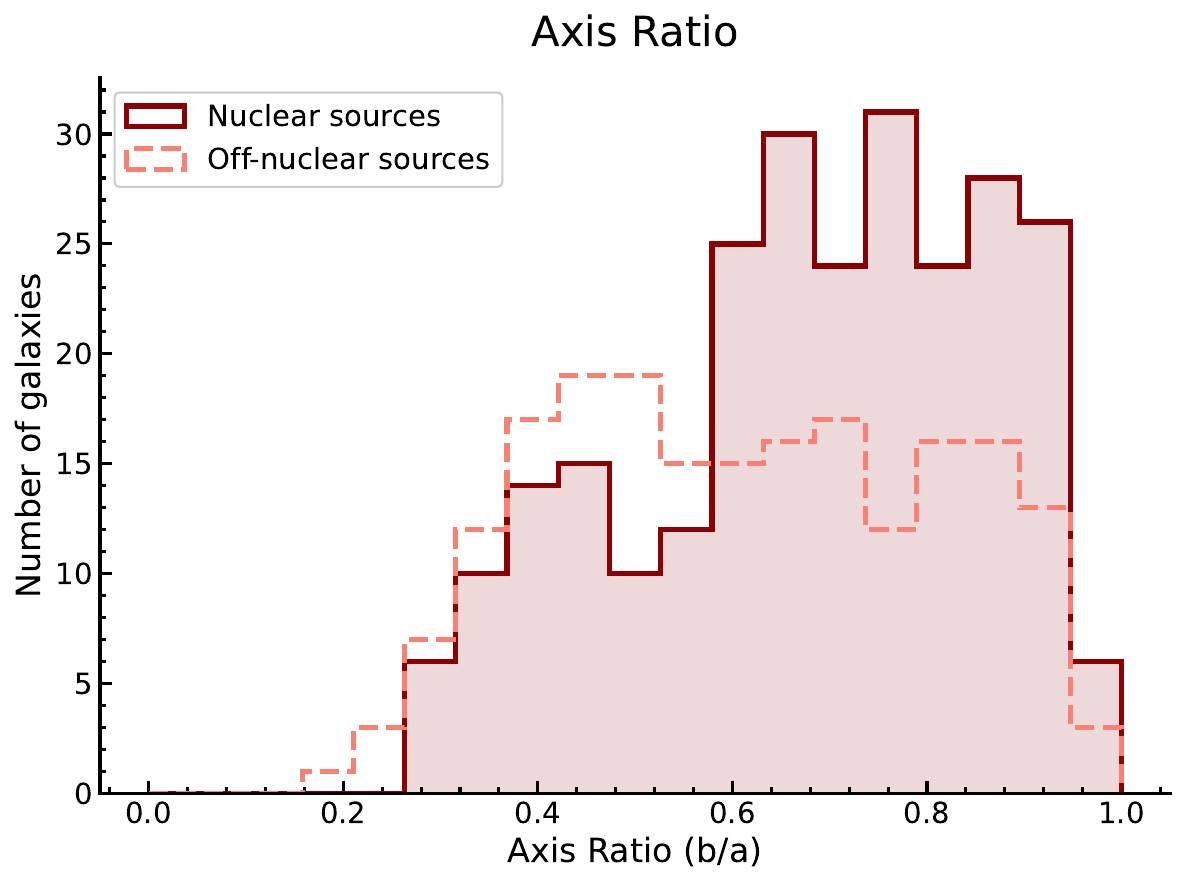}
    \caption{Distribution of axial ratios for nuclear AGN (261 galaxies) and off-nuclear AGN (201 galaxies). Nuclear hosts are predominantly face-on, while off-nuclear hosts show a flatter distribution extending to lower $b/a$.}
    \label{fig:axial_ratio}
\end{figure}

Figure~\ref{fig:axial_ratio} presents the $b/a$ distributions for the two host classes. The samples differ: 72\% of nuclear AGN hosts are face-on ($b/a > 0.6$) \citep{nair2010catalog}, compared to 50\% of off-nuclear hosts. Conversely, off-nuclear hosts include larger fractions of edge-on ($b/a < 0.4$; 16\% vs.\ 9\%) and intermediate-inclination ($0.4 \leq b/a \leq 0.6$; 34\% vs.\ 19\%) galaxies. We interpret this orientation dependence as reflecting a scenario in which face-on viewing angles favor direct visibility of the central regions, and hence a nuclear spaxel classification, whereas in edge-on systems the nucleus can be more effectively diluted by the disk, increasing the prominence of extended or off-nuclear emission.

To test whether the physical differences reported in Section~\ref{sec:nuc_vs_offnuc} persist at fixed orientation, we subdivided each AGN class into three inclination bins following \citet{nair2010catalog}:

\begin{itemize}
  \item Face-on: $b/a > 0.6$ \quad (187 nuclear, 100 off-nuclear)
  \item Intermediate: $0.4 \leq b/a \leq 0.6$ \quad (50 nuclear, 68 off-nuclear)
  \item Edge-on: $b/a < 0.4$ \quad (24 nuclear, 33 off-nuclear)
\end{itemize}

We then repeated the spaxel-by-spaxel comparison within each bin (Table~\ref{tab:orientation_full}).

\begin{figure*}
    \centering
    \includegraphics[trim=0cm 0cm 0cm 1.4cm, clip, width=\textwidth]{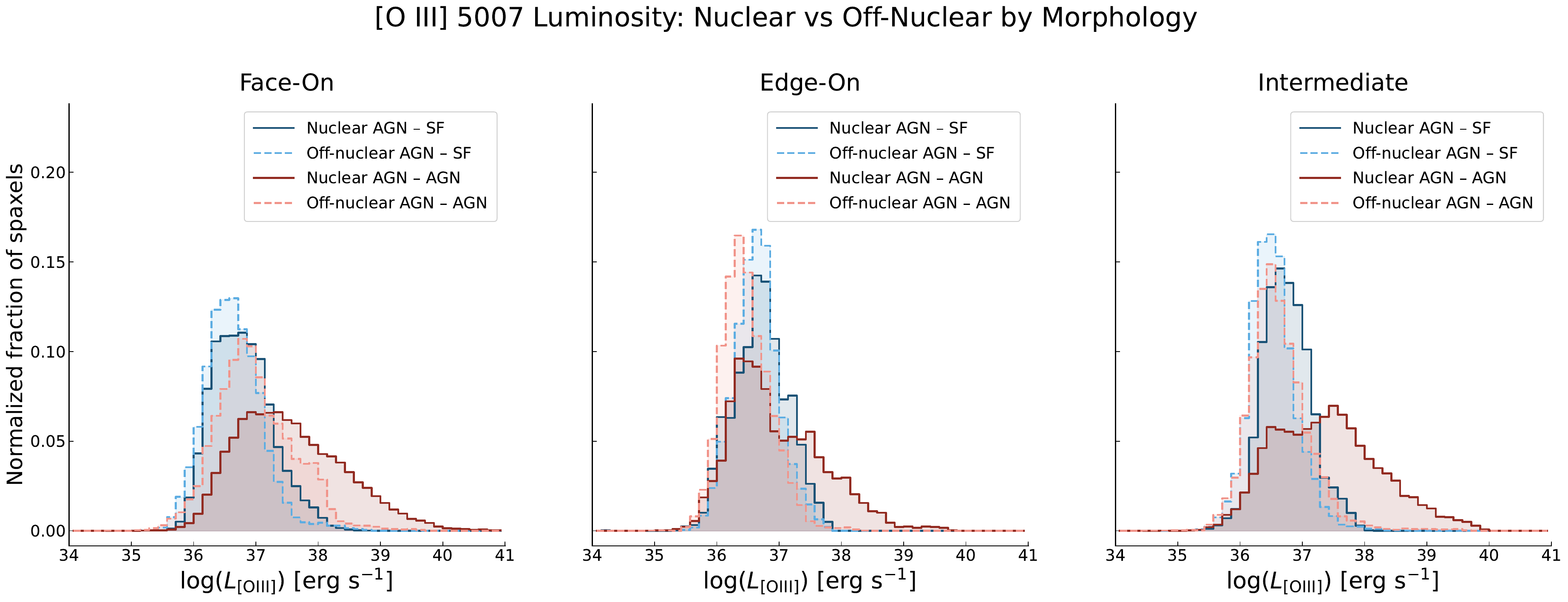}
    \caption{[\ion{O}{iii}]$\lambda$5007 luminosity distributions for nuclear (solid) and off-nuclear (dashed) AGN hosts, split by inclination into face-on, edge-on, and intermediate bins. The luminosity gap between nuclear and off-nuclear AGN spaxels persists across all orientations.}
    \label{fig:loiii_incl}
\end{figure*}

The principal result—the [\ion{O}{iii}] luminosity offset and the corresponding deficit in intrinsic gas excitation ($\log$([\ion{O}{iii}]/H$\beta)$) between nuclear and off-nuclear AGN spaxels—is present in all three inclination bins (Table~\ref{tab:orientation_full}). While the luminosity drops by $\Delta \approx 0.5$-$0.8$~dex (Figure~\ref{fig:loiii_incl}), the excitation ratio simultaneously drops by $\approx 0.13$ to $0.30$~dex in the off-nuclear spaxels across all orientations, indicating that the luminosity gap is not driven by viewing angle. Other trends are less uniform. The $\sigma_{\rm gas}$ difference between nuclear and off-nuclear AGN spaxels ($\sim$14~km\,s$^{-1}$ in face-on systems and $\sim$27~km\,s$^{-1}$ at intermediate inclinations) is largely absent in edge-on galaxies ($\Delta \sim 1$~km\,s$^{-1}$), potentially because disk rotation contributes comparably to line broadening in both classes when viewed edge-on. Dust attenuation increases by approximately a factor of two from face-on to edge-on in both host classes, consistent with longer disk sight lines at higher inclinations. The stellar ages and metallicities of SF spaxels remain comparable across orientation bins, in agreement with Section~\ref{sec:sf_spaxels}. We provide distributions for the inclination-binned properties in Appendix~\ref{sec:incl_properties}.

\subsubsection{Nuclear-region Balmer Decrement}
\label{sec:balmer}

We measured nebular Balmer decrements in the central 2~kpc diameter of every host to test whether off-nuclear classifications in inclined systems are a dust effect. Nuclear attenuation does rise with inclination, reaching $A_V = 2.09$~mag in edge-on off-nuclear hosts, but edge-on nuclear AGN hosts sit at $1.78$~mag, which we cannot distinguish from it (KS~$= 0.312$, $p = 0.11$). Nuclear AGN are thus identified behind the same dust columns, so attenuation is not what separates the two classes. Dust is also unable to alter the classification itself. Because [\ion{O}{iii}]/H$\beta$ and [\ion{N}{ii}]/H$\alpha$ are close in wavelength, BPT line ratios are almost insensitive to reddening, so the only way dust can hide a nucleus is by pushing H$\beta$ below our detection limit. That does not happen: every nuclear spaxel, in both classes, is detected in H$\alpha$, H$\beta$, and [\ion{O}{iii}] at S/N~$\geq 3$, a threshold stricter than the amplitude-to-noise criterion of Section~\ref{sec:diagnostics}, in all three bins. The central kpc is also brighter in total [\ion{O}{iii}] emission (summed over all spaxels, irrespective of BPT class) than the outer disk emission by a median of 1.3~dex, and in off-nuclear hosts most of that light is SF-classified, so there is no missing nuclear flux for dust to explain.
The centers of off-nuclear hosts are therefore dusty but detected. We understand their SF-like excitation as dilution by superposed disk \ion{H}{ii} emission along the inclined sightline rather than extinction. This is consistent with \citet{Mezcua2024}, who found higher central SFRs in off-nuclear than in nuclear AGN dwarf hosts, indicating star-formation dilution of the nuclear line ratios. This does not rule out a compact circumnuclear obscurer, which can hide a broad-line region and soft X-rays while the surrounding kpc-scale gas exhibits disk attenuation, a scenario confirmed by our X-ray detections in sources such as 1-339010 (Section~\ref{sec:chandra}).

\begin{table*}
  \centering
  \caption{Mean physical properties of nuclear and off-nuclear AGN hosts, split by inclination. Top panel: SF spaxels. Bottom panel: AGN spaxels.}
  \label{tab:orientation_full}
  \begin{tabular}{lcccccc}
    \hline
    & \multicolumn{2}{c}{Face-on} &
      \multicolumn{2}{c}{Intermediate} &
      \multicolumn{2}{c}{Edge-on} \\
    \hline
    & Nuc & Off & Nuc & Off & Nuc & Off \\
    \hline
    \multicolumn{7}{c}{\textit{SF Spaxels}} \\
    \hline
    $\log\,t_L$
      & 9.30 & 9.33 & 9.32 & 9.31 & 9.31 & 9.16 \\
    $\log\,t_M$
      & 9.77 & 9.88 & 9.85 & 9.88 & 9.93 & 9.81 \\
    $Z_\star$ (LW)
      & 0.018 & 0.019 & 0.018 & 0.018 & 0.018 & 0.019 \\
    $Z_\star$ (MW)
      & 0.020 & 0.021 & 0.020 & 0.021 & 0.021 & 0.022 \\
    $\sigma_\star$ (km\,s$^{-1}$)
      & 83.5 & 81.1 & 82.0 & 76.9 & 74.5 & 73.3 \\
    $A_V$ (mag)
      & 0.36 & 0.39 & 0.47 & 0.63 & 0.78 & 0.89 \\
    $\log(M_\star)$
      & 7.42 & 7.45 & 7.59 & 7.53 & 7.70 & 7.58 \\
    $\log\,{\rm SFR}_{20}$
      & $-2.89$ & $-2.94$ & $-2.75$ & $-2.82$ & $-2.58$ & $-2.59$ \\
    $\sigma_{\rm gas}$ (km\,s$^{-1}$)
      & 82.0 & 81.2 & 82.6 & 83.0 & 83.6 & 83.1 \\
    $\log\,L_{\rm [\ion{O}{iii}]}$ (erg\,s$^{-1}$)
      & 36.79 & 36.64 & 36.75 & 36.53 & 36.70 & 36.64 \\
    $\log$([\ion{O}{iii}]/H$\beta)$
      & $-0.38$ & $-0.42$ & $-0.50$ & $-0.44$ & $-0.49$ & $-0.41$ \\
    \hline
    \multicolumn{7}{c}{\textit{AGN Spaxels}} \\
    \hline
    $\log\,t_L$
      & 9.59 & 9.49 & 9.34 & 9.30 & 9.25 & 8.91 \\
    $\log\,t_M$
      & 9.85 & 9.78 & 9.75 & 9.60 & 9.59 & 9.36 \\
    $Z_\star$ (LW)
      & 0.017 & 0.018 & 0.018 & 0.018 & 0.016 & 0.019 \\
    $Z_\star$ (MW)
      & 0.018 & 0.019 & 0.019 & 0.018 & 0.016 & 0.021 \\
    $\sigma_\star$ (km\,s$^{-1}$)
      & 119.9 & 114.6 & 117.9 & 104.6 & 88.4 & 93.6 \\
    $A_V$ (mag)
      & 0.29 & 0.28 & 0.44 & 0.45 & 0.59 & 0.68 \\
    $\log(M_\star)$
      & 7.93 & 7.75 & 7.73 & 7.50 & 7.38 & 7.03 \\
    $\log\,{\rm SFR}_{20}$
      & $-2.67$ & $-2.94$ & $-2.76$ & $-2.94$ & $-2.89$ & $-3.15$ \\
    $\sigma_{\rm gas}$ (km\,s$^{-1}$)
      & 139.0 & 124.6 & 147.9 & 120.9 & 113.5 & 112.1 \\
    $\log\,L_{\rm [\ion{O}{iii}]}$ (erg\,s$^{-1}$)
      & 37.56 & 36.98 & 37.40 & 36.63 & 36.97 & 36.48 \\
    $\log$([\ion{O}{iii}]/H$\beta)$
      & 0.51 & 0.38 & 0.52 & 0.22 & 0.41 & 0.20 \\
    \hline
  \end{tabular}
\end{table*}

\subsection{Quiet AGN Re-examination}
\label{sec:results_quiet}
The 159 galaxies labeled quiet by our main pipeline satisfy the global pre-selection of \citet{Wylezalek2018}, but none reach the 40\% AGN spaxel fraction required within the nuclear aperture, nor the 15\% fraction required in any of the outer zones. As discussed in Section~\ref{sec:quiet_class}, we attribute the diagnostic inconsistency to the standard WHaN boundary and therefore adopt a revised threshold of $-0.25$ (Equation~\ref{eq:new_whan}). Here we quantify the impact of this adjustment on the quiet sample.

Shifting the WHaN boundary from $-0.40$ to $-0.25$ reclassifies 48,735 spaxels (across the 159 quiet galaxies) from AGN to SF. The reclassified population is strongly off-nuclear: 45,781 spaxels (94\%) lie outside the 2 kpc nuclear aperture, whereas only 2,954 (6\%) fall within it. This spatial imbalance indicates that the previous boundary preferentially selected extended DIG in the outer disk rather than genuine AGN emission; correspondingly, nuclear spaxels are comparatively insensitive to the revised cut.

The analysis above is restricted to the quiet sample. To verify that the revised boundary does not affect our primary classification, we applied it to the full classified sample. Over 90\% of galaxies retain their original class, and no galaxy moves to a more active one, as expected since the shift can only remove AGN spaxels. The nuclear classification is largely insensitive to the boundary choice, while roughly one in six off-nuclear candidates is reassigned as quiet.

\begin{figure}
    \centering
    \includegraphics[width=\columnwidth]{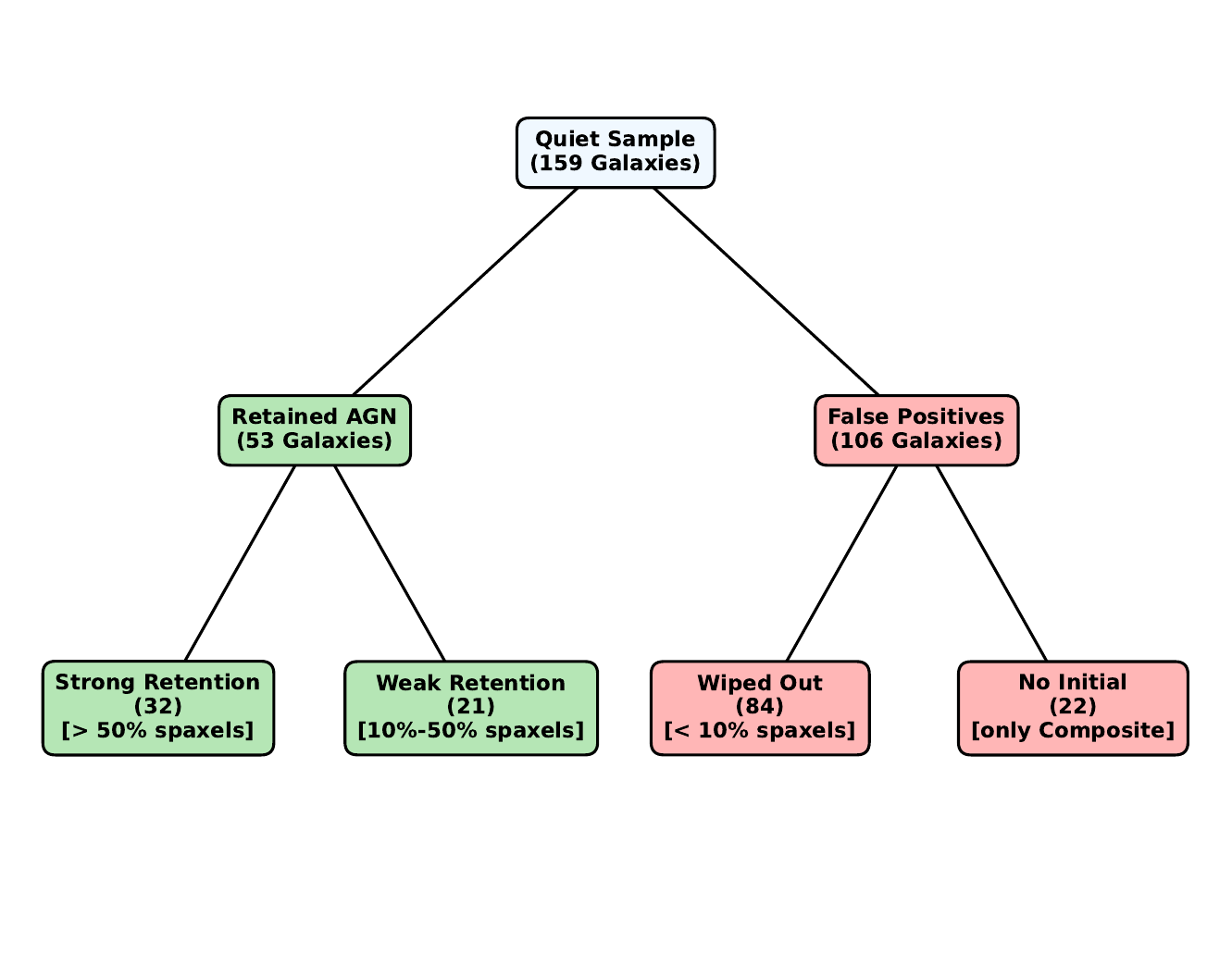}
    \caption{Breakdown of the 159 quiet galaxies after shifting the WHaN boundary to $-0.25$. Of the original sample, 53 retain AGN spaxels, while 106 are reclassified as false positives or inactive.}
    \label{fig:quiet_flowchart}
\end{figure}

To assess the consequences at the galaxy level, we computed an AGN-spaxel retention fraction and partitioned the 159 sources into four categories (Figure~\ref{fig:quiet_flowchart}; Table~\ref{tab:quiet_whan}).

More than half of the sample loses the majority of its AGN spaxels, with the removed emission arising almost exclusively from the off-nuclear disk. A further 22 galaxies contain no AGN spaxels even before the shift, implying that they satisfy the initial \citet{Wylezalek2018} selection because composite spaxels contribute a 20\% weight in the [\ion{N}{ii}] fraction criterion. Consequently, approximately two-thirds of the quiet galaxies are consistent with being inactive or SF under the revised boundary.

The remainder retain AGN spaxels. Strong-retention systems preserve more than half of their original AGN-spaxel counts, indicating that these spaxels lie well above $-0.25$ and are therefore unaffected by the boundary change. Weak-retention systems lose the extended component but continue to host a non-negligible nuclear AGN-spaxel population.

Finally, we examined the location of the reclassified spaxels in the [\ion{N}{ii}]-BPT plane. Spaxels removed by the shift populate the composite region between the \citet{Kauffmann2003} and \citet{Kewley2001} demarcations (Figure~\ref{fig:quiet_before_after}), consistent with DIG contamination and/or high-metallicity SF. In contrast, the AGN spaxels retained after applying the $-0.25$ cut remain above the \citet{Kewley2001} line. The improved consistency between diagnostics supports interpreting the $-0.25$ boundary as a cleaner separator of AGN emission from contamination.

\begin{table}
  \centering
  \caption{Reclassification of the 159 quiet galaxies after shifting the WHaN boundary.}
  \label{tab:quiet_whan}
  \begin{tabular}{lcc}
    \hline
    Category & $N$ & Fraction (\%) \\
    \hline
    Wiped Out (kept $<10\%$ of AGN spaxels)  & 84  & 52.8 \\
    Strong Retention (kept $>50\%$)           & 32  & 20.1 \\
    No Initial AGN Spaxels                    & 22  & 13.8 \\
    Weak Retention (kept 10--50\%)            & 21  & 13.2 \\
    \hline
  \end{tabular}
\end{table}
\subsection{Sensitivity of the classification to the choice of BPT diagnostic}
\label{sec:3dbpt}

As an independent check on our classifications, we repeated the spatial
pipeline, replacing the two standard 2D BPT diagrams with the 3D framework of \citet{ji2020constraining}, which places spaxels in the combined $\log([\ion{O}{iii}]/\mathrm{H}\beta)$, $\log([\ion{N}{ii}]/\mathrm{H}\alpha)$ and $\log([\ion{S}{ii}]/\mathrm{H}\alpha)$ space. The WHaN part of the classification was left unchanged so that only the BPT criterion varies. 

This classification returns 108 nuclear AGN (compared to 261 for the 2D classification), 99 off-nuclear AGN (compared to 201), and 414 quiet sources (compared to 159). The increase in quiet galaxies is expected: the 3D framework resolves the LINER-composite degeneracy present in the 2D projections, reclassifying many borderline LINER-like spaxels as composite and pulling those galaxies below our AGN fraction thresholds. Figure~\ref{fig:3dbpt} shows the effect directly, with the extended LINER-like haloes of the 2D maps reclassified as composite in 3D while the Seyfert spaxels remain intact. Only one galaxy moves to a more active class, indicating that the 3D criterion is uniformly stricter than the standard 2D diagnostics.

We then repeated this using the kinematically calibrated 2D BPT boundaries of \citet{law2021diagnostics} for MaNGA, again holding the WHaN criterion fixed (Figure~\ref{fig:3dbpt}, center column). This returns 279 nuclear, 168 off-nuclear, and 174 quiet sources close to our primary classification. Here, 78\% of galaxies retain their class, and galaxies move in both directions in comparable numbers. The two 2D schemes thus agree with each other, and the larger shift produced by the 3D framework reflects its high purity threshold rather than the extra dimension. 

Of our 201 off-nuclear candidates, 38 are classified as off-nuclear under all three schemes (Table~\ref{tab:scheme_comparison}). We retain the 2D BPT+WHaN classification as our primary result, since it is directly comparable to the existing MaNGA literature.

\begin{table}
  \centering
  \caption{Classification of the 621 galaxies under three diagnostic schemes.
  The WHaN criterion is identical in all three; only the BPT criterion varies.
  The final row gives the number of galaxies assigned the same class by all
  three schemes.}
  \label{tab:scheme_comparison}
  \begin{tabular}{lccc}
    \hline
    Scheme & Nuclear & Off-nuclear & Quiet \\
    \hline
    This work (Traditional 2D-BPT)                     & 261 & 201 & 159 \\
    \citet{law2021diagnostics} (2D)    & 279 & 168 & 174 \\
    \citet{ji2020constraining} (3D)    & 108 &  99 & 414 \\
    \hline
    All three                          & 107 &  38 & 123 \\
    \hline
  \end{tabular}
\end{table}

\section{Discussion}
\label{sec:discussion}

\subsection{What are Off-Nuclear AGN?}
\label{sec:disc_nature}

A central question that arises at this stage is: {\it What physical mechanism is responsible for producing Seyfert/LINER-like emission at projected distances beyond the central kiloparsec?} Previous work has primarily invoked two scenarios: (i) wandering SMBHs that have been displaced following a merger, or (ii) an “offset” AGN in an ongoing interaction, in which one nucleus is active but spatially displaced from the host galaxy’s photometric center \citep[e.g.,][]{Comerford_2014, Steinborn2015}. While wandering SMBHs can certainly persist long after the visible morphological features of a past merger have faded, this scenario alone does not account for the inclination trend we observe across our sample. A third scenario, highlighted by \citet{Casey2026}  and T. Brosi et al.\ (submitted), invokes `fading' AGNs, in which the central black hole has recently shut down or significantly reduced its accretion rate, yet extended emission-line regions and fossil outflows persist due to light-travel-time delays and the longer dynamical timescales of the outflowing gas; we quantify this further in Section \ref{sec:selection}. Nevertheless, the orientation bias discussed in Section~\ref{sec:orientation} indicates that galaxy inclination may provide a conventional explanation for a subset of the off-nuclear detections.

For this inclined subset, superposed disk emission dilutes the nuclear line ratios, while an extended ionization cone or outflow remains observable. In this picture, the off-nuclear classification reflects geometric and line-of-sight effects rather than the presence of a physically displaced black hole \citep{Munoz2022, Tutterow_2025}. However, orientation alone does not account for all of the observed trends. The systematically lower [\ion{O}{iii}]$\lambda$5007 luminosities and reduced $\sigma_{\rm gas}$ in off-nuclear spaxels persist even after we subdivide the sample by inclination (Table~\ref{tab:orientation_full}). If the off-nuclear detections were simply scattered or reprocessed emission from an obscured central engine, then controlling for viewing angle should largely remove these discrepancies; this is not observed.

Instead, the emitting gas responsible for the off-nuclear signal appears to reside in a distinct physical environment characterized by lower density, a weaker ionizing radiation field, and reduced kinematic disturbance, irrespective of whether the ultimate power source is a wandering SMBH or a centrally located AGN whose line ratios are diluted. The off-nuclear AGN class may therefore encompass a composite population, including both extended ionization cones and genuinely offset black holes, but generally tracing a lower-excitation regime than classical nuclear AGNs, as reflected in the lower $\log$([\ion{O}{iii}]/H$\beta)$ measured in Section \ref{sec:agn_spaxels}.

The preference for face-on orientations among nuclear AGN hosts also has important implications for survey completeness. Inclined systems are intrinsically more challenging to classify, as projection effects systematically hinder the identification of nuclear activity. Consequently, any AGN census based on IFU data is likely to miss a non-negligible fraction of AGN in galaxies viewed at higher inclinations, precisely where projection is most significant. The true AGN incidence in MaNGA-like samples is therefore plausibly higher than what is inferred from resolved emission-line selections alone.

\subsection{Do the host galaxy properties dictate the AGN-like spaxel location?}
\label{sec:disc_host}

To check whether the physical properties of a host galaxy dictate the spatial distribution of its AGN-like spaxels, we carried out a spaxel-by-spaxel comparison, which yielded a striking result: the host galaxies of nuclear and off-nuclear AGN are essentially the same. Stellar mass, light- and mass-weighted ages, recent SFR, and stellar kinematics are indistinguishable between the two populations (Table \ref{tab:ks_results}). 
The gas-phase metallicity supports this: both within the SF disk and within the AGN-ionized gas, the oxygen abundance is consistent across host types, even after applying ionization-specific calibrations to each region (Section \ref{sec:metallicity_result}). The chemical enrichment history of the galaxy carries no clear specific signs of whether its AGN ended up central or offset. The one difference is dust attenuation, which is higher in off-nuclear hosts, but this follows from the inclination bias we already discussed, making this a geometric effect rather than a physical one. Everything points to the idea that nuclear vs. off-nuclear classification is of the ionized gas, not of the galaxy. The stellar body beneath is indifferent to where the ionizing source is located.

\subsection{Does the standard WHaN boundary work for IFU data?}
\label{sec:disc_whan}

The standard WHaN diagram places the SF-AGN boundary at $\log([\ion{N}{ii}]/\mathrm{H}\alpha) = -0.4$ \citep{stasinska2006, CidFernandes2010}. This was calibrated on single-fiber spectra, where the emission from the galaxy's central region is blended into one measurement. In that context, $-0.4$ works well as a lower limit for pure star formation.

On the other hand, IFU data classify individual spaxels. The outer disk is no longer averaged out; instead, each spaxel stands on its own. As a result, DIG-contaminated regions, high-metallicity \ion{H}{ii} complexes, and old stellar populations in the extended disk produce $\log([\ion{N}{ii}]/\mathrm{H}\alpha)$ values that exceed $-0.4$ without any AGN contribution. The \citet{Kewley2001} theoretical maximum starburst line corresponds to $\log([\ion{N}{ii}]/\mathrm{H}\alpha)= -0.1$ \citep{CidFernandes2010}, meaning the true SF/AGN transition in resolved data lies somewhere between $-0.4$ and $-0.1$. Our empirically determined value of $-0.25$ falls in the middle of this range and successfully brings the WHaN and BPT classifications into agreement.

Our results show this directly: shifting the boundary to $-0.25$ reclassifies 48,735 spaxels from AGN to SF across the 159 quiet galaxies, and 94\% of those sat beyond the 2 kpc nuclear aperture. The [\ion{N}{ii}] BPT classification of these same spaxels places them in the composite region, confirming they were never purely AGN-ionized.

\subsection{Implications of Off-Nuclear Detections for AGN Surveys}

IFS surveys have opened an observational regime that is inaccessible to single-fiber surveys, enabling the detection of phenomena such as extended ionization cones and emission components that fall outside any fixed aperture. However, this enhanced sensitivity also introduces systematic challenges, which are emphasized by our analysis.

The first issue concerns the sensitivity of the reduction pipeline. As shown in Section \ref{sec:validation}, changes between MaNGA DRP versions alter local flux gradients sufficiently to move thousands of spaxels across the SF/composite boundary and, in some cases, to exclude entire galaxies from the AGN sample. Any study that utilizes the fraction of AGN-classified spaxels is therefore intrinsically dependent on the adopted DRP version. Robust cross-survey and cross-pipeline comparisons are required to track and quantify these systematics.

The second concerns the choice of emission-line diagnostic. To address this, we compared our primary classifications against both the kinematically calibrated 2D BPT boundaries for MaNGA and a 3D BPT diagnostic framework (Section \ref{sec:3dbpt}). The 3D approach evaluates the three principal line ratios jointly rather than as two independent projections, separating the highest-excitation Seyfert spaxels from the LINER-like/composite degeneracy that affects conventional 2D diagrams. Our most secure nuclear detections survive all three schemes. The divergence between the schemes arises in the extended, low-excitation emission component, which directly underpins the off-nuclear classification. Off-nuclear samples are consequently far more sensitive to the diagnostic adopted than nuclear ones, and this should be stated wherever such a census is built on a single-line ratio cut.

It is worth noting that, first, although the observed orientation trend is clear, the distributions of physical properties within the edge-on bin should be interpreted cautiously until they can be confirmed with a larger sample. Second, our DIG correction is purely empirical; it is anchored to BPT classifications rather than derived from a physically motivated decomposition of the ionized gas. Finally, comparing gas-phase metallicities across distinct ionization regimes is intrinsically challenging. By applying separate, ionization-matched calibrations for SF- and AGN-ionized spaxels within the same galaxy \citep{Curti2017, Carvalho2020}, we ensured that relative comparisons within each ionization class are internally consistent. However, because these calibrations are anchored to fundamentally different methodologies (empirical $T_e$-based techniques vs. \textsc{cloudy}-based photoionization grids), reconciling the absolute metallicity scale across these regimes remains an open problem in spatially resolved spectroscopy. To mitigate this discrepancy, we are currently developing a model-independent, ionization-source-dependent metallicity calibration as part of an ongoing follow-up study.

\subsection{Comparison with Independent Fading-AGN and Fossil Outflow Works}
\label{sec:selection}

Despite these challenges, off-nuclear detections offer a direct window into the evolution of nuclear activity. All three studies considered in this discussion—namely, this work, \citet{Casey2026}, and T. Brosi et al.\ (submitted)—are based on data from the MaNGA IFU survey, but they adopt distinct sample selection methodologies. \citet{Casey2026} identify fading-AGN candidates through emission-line diagnostics, quantifying the fractional AGN contribution on the BPT diagram as a function of galactocentric radius; our work applies a multi-diagnostic scheme (BPT + WHaN + WHaD) paired with a radial spaxel-fraction algorithm; and T. Brosi et al.\ detect fossil outflow candidates via kinematic signatures, specifically elevated [\ion{O}{iii}] $W_{80}$ line widths relative to $\sigma_\star$. 

Following the analytical framework of \citet{Casey2026}, we evaluate our sample of 201 off-nuclear AGN as potential fading-AGN candidates for comparative analysis. Of the 33 candidates identified by \citet{Casey2026}, 22 are recovered within our off-nuclear list; the discrepancies are due to differences in diagnostic boundaries—for instance, we classify one overlapping source (1-3327) as a quiet AGN and exclude another (1-591997) as a merging system—while the remaining 9 sources do not satisfy our $d_{\text{BPT}}$ criteria. By measuring the light-travel time from the galactic center to the mean physical radius of our active spaxels, we obtain a mean fading timescale of $t_{\mathrm{faded}} = (1.16 \pm 0.04) \times 10^{4}$~yr,  where the uncertainties represent the standard error of the mean across the sample distribution. This is consistent with the fading timescale of ${\sim}\,(1.20 \pm 0.08) \times 10^{4}$~yr reported by \citet{Casey2026}. Adopting their methodology, we estimate the AGN turn-off rate by dividing the number of our off-nuclear candidates by $t_{\mathrm{faded}}$. Extending this rate to the full parent sample of 776 AGN yields an AGN phase lifetime of $t_{\mathrm{AGN}} = (4.47 \pm 0.15) \times 10^{4}$~yr, shorter by a factor of ${\sim}\,3$ than their reported lower limit of $(1.40 \pm 0.3)\times 10^{5}$~yr, a difference primarily driven by the distinct parent-sample sizes and selection criteria.

Cross-matching the 269 fossil outflow candidates identified by T. Brosi et al.\ with our catalog yields 60 sources in common: 30 classified as off-nuclear, 18 as quiet, and 12 as nuclear AGN. The remaining 209 objects are not identified as AGN according to our selection criteria, which is consistent with expectations, as T. Brosi et al.\ deliberately target non-AGN host galaxies whose kinematic perturbations persist longer than the radiative signatures accessible through standard emission-line diagnostic diagrams. This longer persistence is supported by semianalytical estimates, which show long dynamical timescales for AGN outflows on the order of ${\sim}\,10^5 - 10^6$~yr \citep[][T. Brosi et al.\ (submitted)]{Zubovas2018}. Because outflow dynamical timescales exceed the ${\sim}\,10^5$~yr duration of an AGN flickering episode \citep{Schawinski2015}, up to 60\% of all observed outflows may be fossil outflows \citep{Zubovas2022}. Consequently, while these 269 fossil outflow candidates represent only about one-third of the multi-wavelength AGN sample identified by T. Brosi et al., their strict S/N requirements imply that this figure is likely a conservative lower limit for the true population of fossil AGN outflows in MaNGA.

T. Brosi et al.\ and \citet{Casey2026} have 10 sources in common. Of these, 8 are also included in our off-nuclear sample, while the remaining 2 are excluded by our $d_{\text{BPT}}$ selection criteria. The convergence of three independent studies on this common subset of 8 galaxies provides strong evidence that these systems host genuinely fading nuclear engines, whose outflows remain observable in both their ionization and kinematics. Furthermore, the similarity between the fading timescales inferred from our sample and from \citet{Casey2026}, together with the overlap in candidates across all three catalogs, supports a scenario in which the central SMBHs in these systems undergo rapid, episodic accretion rather than long-lived, steady activity. This picture of a flickering AGN duty cycle has recently gained attention from studies combining IFU data with radio observations, which have revealed a connection between radio detections and perturbed gas in AGN \citep{Alban2024, Escott2025, Nandi_2025}. In this context, \citet{Albán2026} has found a similar behavior in galaxies without current AGN activity that have radio detections and discusses past or recurrent AGN activity as a possible explanation. Our findings support this and underscore the importance of spatially resolved IFU diagnostics—both in emission lines and in gas kinematics—for constraining the duty cycle of AGN episodes.

We next investigated whether kinematically misaligned gas accretion, commonly implicated in the triggering of nuclear activity \citep{Raimundo2023, Raimundo2025}, is associated with the presence of off-nuclear ionization structures. We cross-matched our samples with the $\Delta \mathrm{PA}$ measurements for ${\sim}\,4600$ systems from \citet{Raimundo2025}, derived from the same MaNGA {\sc megacubes}. Adopting the criterion for misaligned gas of $\Delta \mathrm{PA} \geq 45^{\circ}$, we find misalignment fractions of $17.8 \pm 3.1$\% for nuclear, $11.3 \pm 2.8$\% for off-nuclear, and $14.6 \pm 3.5$\% for quiet AGN. These values are consistent within the uncertainties across the three samples, although the off-nuclear subset exhibits the lowest absolute fraction. Overall, gas–stellar misalignment appears compatible with facilitating nuclear activity, but it does not preferentially characterize any particular evolutionary phase, as anticipated for two reasons: (i) misaligned gas can persist for ${\sim}\,10^{8}-10^{9}$~yr \citep{vandeVoort2015, Baker2025}, substantially exceeding our inferred $t_{\mathrm{faded}}$; and (ii) \citet{Tous2026} show that misaligned galaxies host comparable central ionized-gas reservoirs irrespective of whether an AGN is currently detected. Taken together, these results suggest that the fuel availability is quasi-steady, whereas the nuclear activity is periodic, consistent with the flickering duty cycle discussed above.

\subsection{Chandra Cross-Match: X-ray Confirmation of Off-Nuclear Scenarios}
\label{sec:chandra}
Additionally, to validate the physical nature of our off-nuclear detections, we cross-matched the full parent sample of 776 AGN with observations from the general Chandra archive, an X-ray observatory capable of spatially resolving multiple point sources within the MaNGA FoV. Of these, 48 galaxies have Chandra coverage, and 6 are classified as off-nuclear AGN in our catalog. Notably, four of these six sources overlap with the sample of \citet{Comerford2022}, who similarly used Chandra data to confirm X-ray AGN in off-nuclear Seyfert regions. Although this subsample is relatively small, the results are informative: the majority of these 6 systems are consistent with the orientation-based scenario discussed in Section~\ref{sec:orientation}, in which a nuclear AGN is obscured on circumnuclear scales, while the extended emission dominates the optical classification. (e.g., in 1-339010, Chandra reveals a compact nuclear X-ray source in an edge-on system; see Figure~\ref{fig:chandra_edgeon}). However, one source, 1-258535, which also appears among the fossil outflow candidates of T. Brosi et al.\, supports the fading-AGN interpretation: its X-ray emission is spatially extended (Figure~\ref{fig:chandra_outflow}). The Chandra data therefore independently confirm the presence of two of the physical scenarios proposed for off-nuclear AGN: (1) geometric dilution of an otherwise nuclear AGN and (2) genuinely fading nuclear activity.

Future high-resolution radio continuum observations could further validate this population.
\section{Conclusions}
\label{sec:conclusions}

We identified 261 nuclear AGNs, 201 off-nuclear AGNs, and 159 `quiet' AGNs in the MaNGA IFS survey. We then performed a spaxel-by-spaxel analysis of the nuclear and off-nuclear AGN hosts, comparing the physical properties of their SF disks and AGN-ionized gas while explicitly controlling for host-galaxy orientation. The central question was whether spatially offset AGNs occupy a distinct class of galaxy or whether the offset is purely a local phenomenon in the ionized gas. Our observations consistently support the latter scenario.

Our main conclusions are as follows:
\begin{enumerate}[leftmargin=*]
\item \textbf{Host galaxy properties are broadly similar.} Stellar mass, light- and mass-weighted stellar ages, recent SFR, stellar kinematics, and gas-phase metallicities are consistent between nuclear and off-nuclear AGN hosts. The evolutionary state of the galaxy does not determine whether the AGN emission appears nuclear or off-nuclear.

\item \textbf{The ionized gas environments differ significantly.} Off-nuclear spaxels are fainter in [\ion{O}{iii}] $\lambda$5007 by $\sim 0.7$ dex than nuclear ones and show both lower excitation, traced by the $\log$([\ion{O}{iii}]/H$\beta$) line ratio, and narrower $\sigma_{\rm gas}$. The drop in excitation shows that the luminosity deficit is not simply a normalization effect but reflects conditions expected farther from the central SMBH: lower gas density and a weaker radiation field.

\item \textbf{Orientation and intrinsic physical differences are both at play.} These two effects are not mutually exclusive; rather, they combine to produce the off-nuclear population. Host galaxy orientation contributes to the spatial classification: nuclear AGN hosts are mostly face-on, while off-nuclear hosts show a broader inclination distribution, with the nuclear line ratios more often diluted by superposed disk emission in inclined systems. However, orientation is not the sole driver—half of the off-nuclear hosts are themselves face-on, and classifying by inclination reveals that the $\sim 0.7$ dex [\ion{O}{iii}] gap and the lower $\log$([\ion{O}{iii}]/H$\beta)$ excitation ratios persist across all viewing angles. Thus, the off-nuclear gas represents a genuinely lower-excitation environment, not merely attenuated nuclear emission seen at an inclined angle.

\item \textbf{Off-nuclear AGN signatures constitute a heterogeneous population.} Cross-matching our sample with independent fossil outflow and fading-AGN catalogs, and validated through archival Chandra X-ray observations, demonstrates that off-nuclear detections do not originate from a single physical mechanism. Instead, they comprise compact circumnuclear obscured engines (supported by compact X-ray cores in edge-on galaxies) and genuinely fading AGNs characterized by episodic accretion cycles with durations of $\sim 10^4$ years, supported by the kinematic misalignment test, in which extended ionization signatures persist even after the central engine has shut down.

\item \textbf{Standard diagnostic thresholds need recalibration for IFU data.} The standard WHaN boundary at $\log([\ion{N}{ii}]/\text{H}\alpha) = -0.4$ systematically misclassifies high-metallicity DIG in the outer disk as AGN emission. Shifting the cut to $-0.25$ reassigned the WHaN classifications and removed over 48,000 spurious AGN-like spaxels. Any resolved AGN study using WHaN on MaNGA-like data should empirically verify the boundary before trusting the results.
\end{enumerate}

The broader point is that IFU surveys can detect AGN activity that single-fiber spectroscopy misses entirely, but that spatial reach introduces classification degeneracies such as DIG contamination, pipeline sensitivity, and viewing-angle effects that no single 2D diagnostic can handle cleanly. A reliable census requires multi-diagnostic, spaxel-level checking. Otherwise, the resulting sample will be shaped more by viewing angles and data-reduction choices than by actual astrophysical activity.

\section*{Acknowledgements}
We thank the anonymous referee for their helpful comments and suggestions. SJ acknowledges financial support from the Coordena\c{c}\~ao de Aperfei\c{c}oamento de Pessoal de N\'ivel Superior (CAPES; Finance Code 001). RR acknowledges support from  Conselho Nacional de Desenvolvimento Cient\'{i}fico e Tecnol\'ogico  ( CNPq, Proj. CNPq-445231/2024-6,311223/2020-6, 404238/2021-1, and 310413/2025-7), Funda\c{c}\~ao de amparo \`{a} pesquisa do Rio Grande do Sul (FAPERGS, Proj. 19/1750-2 and 24/2551-0001282-6) and Coordena\c{c}\~ao de Aperfei\c{c}oamento de Pessoal de N\'{i}vel Superior (CAPES, 88881.109987/2025-01). RAR acknowledges the support from the Conselho Nacional de Desenvolvimento Científico e Tecnológico (CNPq; Projects 303450/2022-3, and 403398/2023-1), the Coordenação de Aperfeiçoamento de Pessoal de Nível Superior (CAPES; Project 88887.894973/2023-00), and Fundação de Amparo à Pesquisa do Estado do Rio Grande do Sul (FAPERGS; Project 25/2551-0002765-9). SBR  acknowledges support from the Brazilian research agencies Conselho Nacional de Desenvolvimento Científico e Tecnológico (CNPq) and Fundação de Amparo à Pesquisa do Estado do Rio Grande do Sul (FAPERGS). RCH acknowledges support from NASA via ADAP grant 80NSSC23K0485. QOC acknowledges support from the Dartmouth Fellowship. THSW acknowledges financial support from CAPES (88887.002352/2024-00). MA gratefully acknowledges support from Fundação de Amparo à Pesquisa do Estado de São Paulo (FAPESP, Processo: 2024/03727-3). SIR acknowledges the support from the Science and Technology Facilities Council (STFC) of the UK Research and Innovation via grant reference ST/Y002644/1 and from the Royal Society via grant reference RGS/R2/252734. We acknowledge the use of AI-assisted tools for grammatical corrections and language homogenization.

\section*{Data Availability}
The data used in this study are fully available at \href{https://manga.if.ufrgs.br/}{https://manga.if.ufrgs.br/} and \href{https://manga.linea.org.br/}{https://manga.linea.org.br/}.

\bibliographystyle{mnras}
\bibliography{references} 

\vspace*{\fill}
\bsp

\clearpage
\appendix

\begin{figure*}
    
    \section{Supplementary Figures}
    \label{sec:Supplementary}
    \vspace{0.4cm} 

    \centering
    \includegraphics[width=0.95\textwidth]{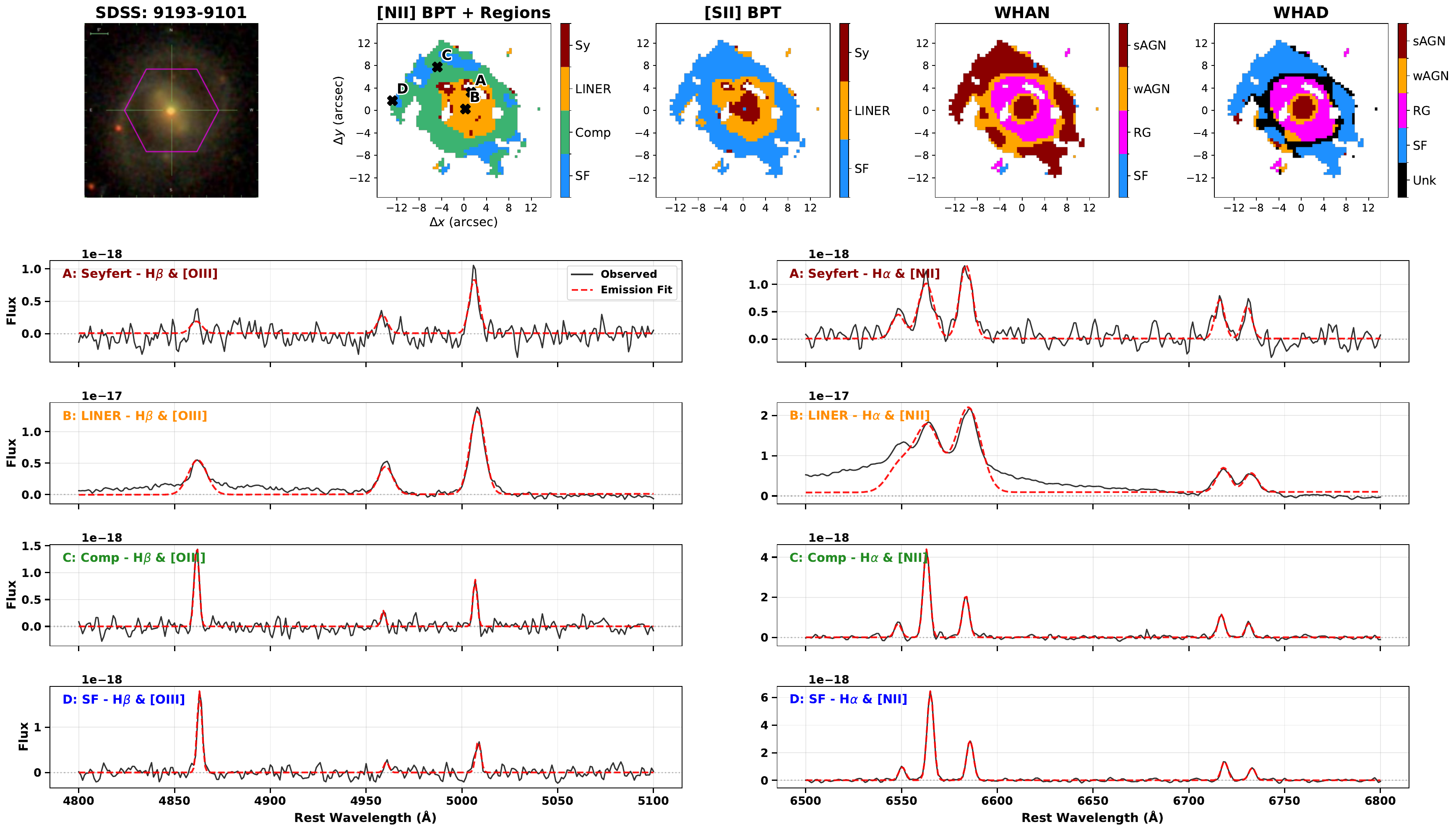}
    \caption{Example of diagnostic maps and the position of the emission-line measurements per spaxel in the [\ion{N}{ii}]-BPT classification for the target MaNGA:1-37385 ($z = 0.045$), which exhibits broad Balmer emission lines.}
    \label{fig:bcomp_example}
\end{figure*}

\begin{figure*} 
    \centering
    \includegraphics[width=0.95\textwidth]{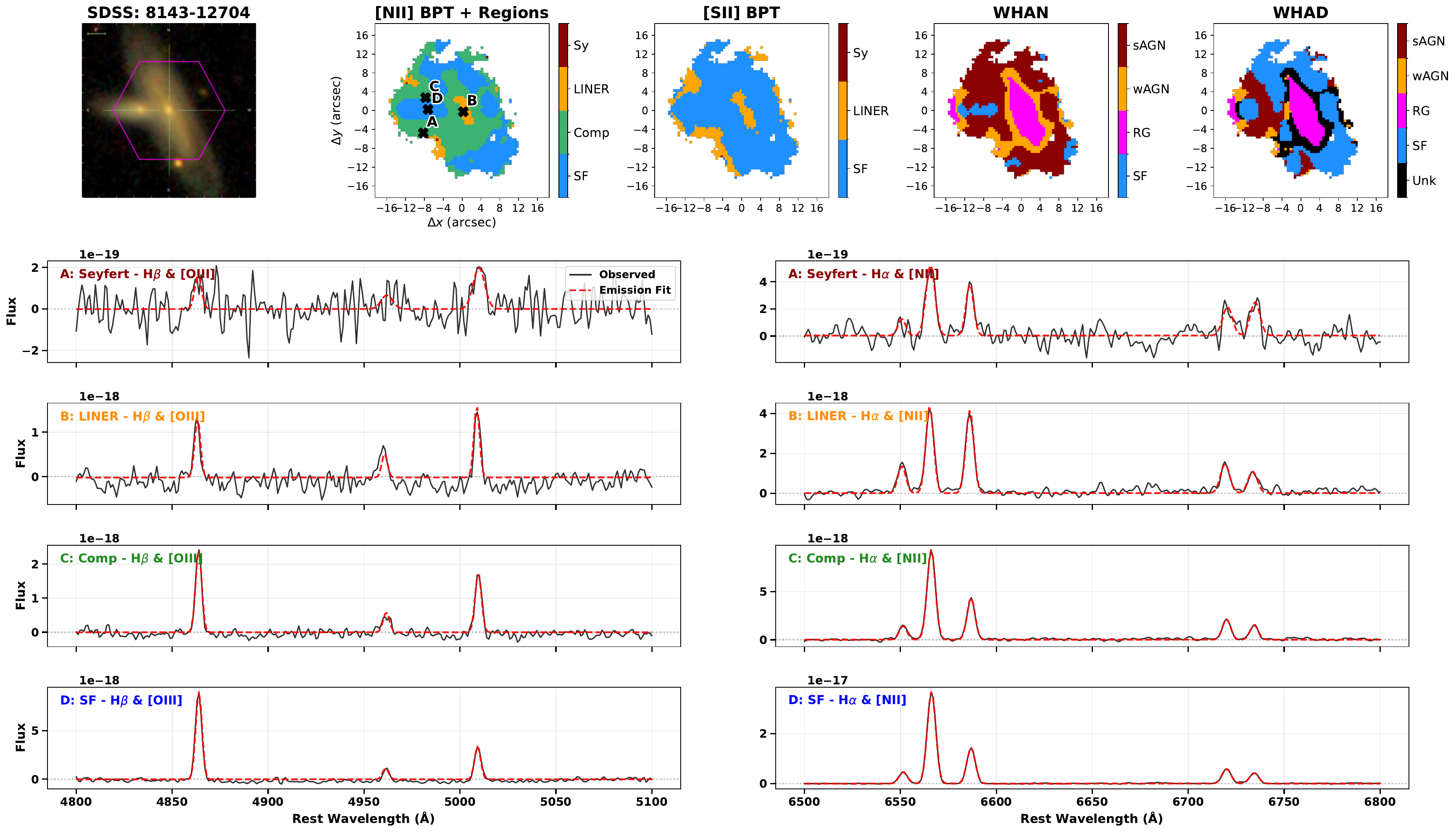}
    \caption{Example of diagnostic maps and the position of the emission-line measurements per spaxel in the [\ion{N}{ii}]-BPT classification for the major merger target MaNGA:1-44487 ($z = 0.032$). Because of the highly disturbed morphology, radial profiling is unreliable, and the source was removed from the sample.}
    \label{fig:merger_example}
\end{figure*}

\begin{figure*} 
    \centering
    \includegraphics[width=0.95\textwidth]{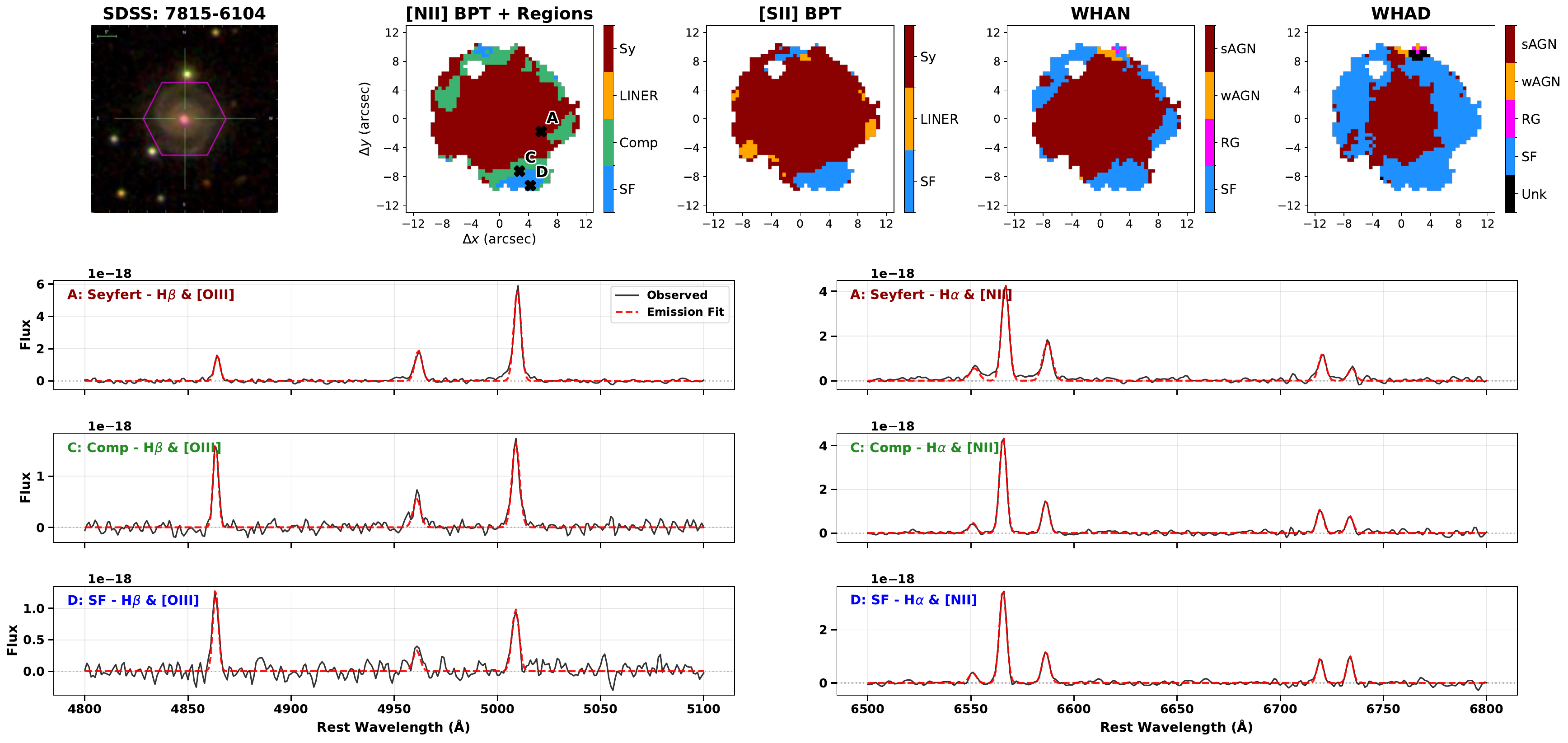}
    \caption{Example of diagnostic maps and the position of the emission-line measurements per spaxel in the [\ion{N}{ii}]-BPT classification for the target MaNGA:1-113712 ($z = 0.081$), which was successfully classified as a nuclear AGN by our pipeline.}
    \label{fig:nucl_example}
\end{figure*}
\begin{figure*} 
    \centering
    \includegraphics[width=0.95\textwidth]{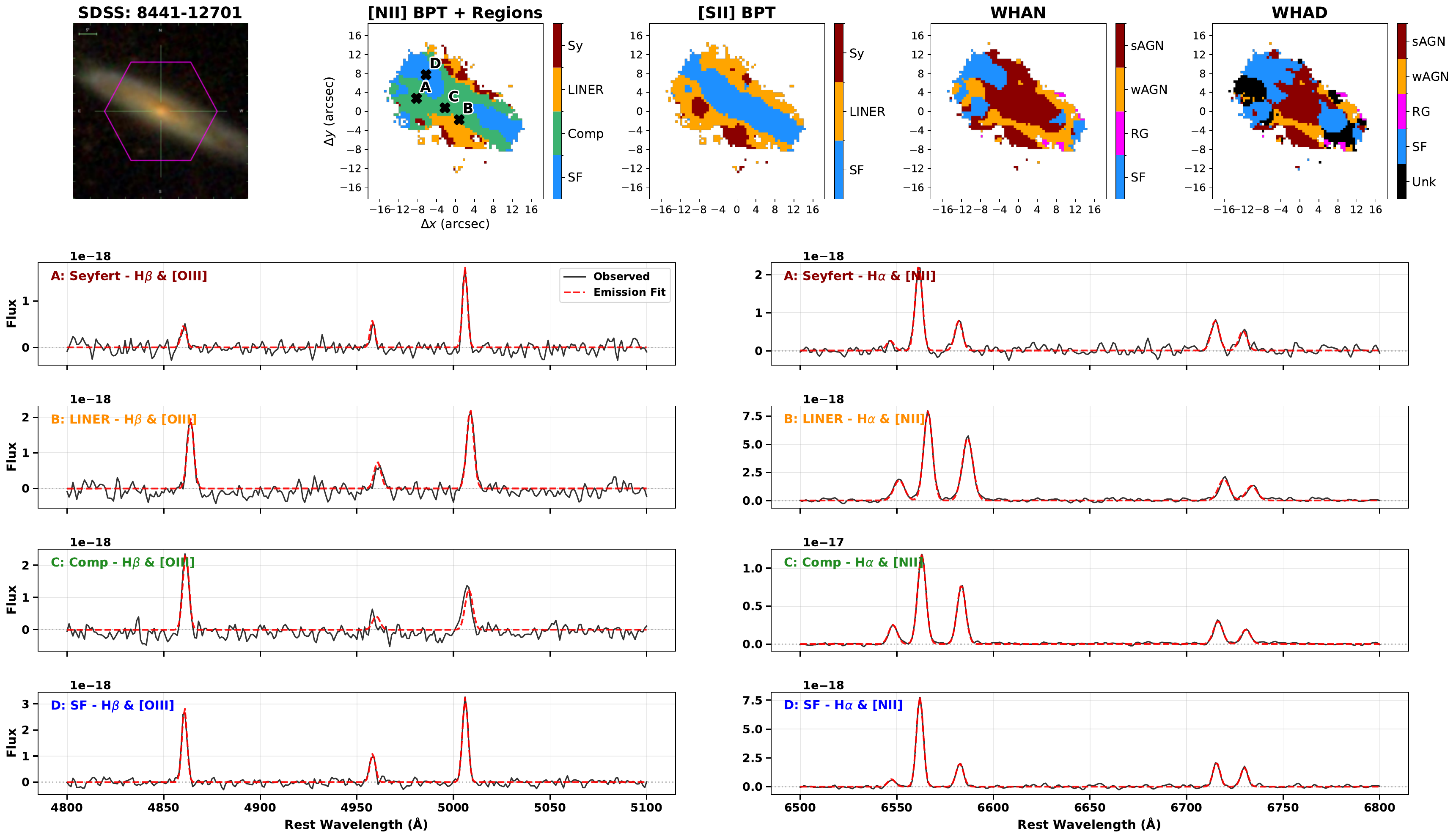}
    \caption{Example of diagnostic maps and the position of the emission-line measurements per spaxel in the [\ion{N}{ii}]-BPT classification for the target MaNGA:1-252202 ($z = 0.0317$), which was successfully classified as an off-nuclear AGN by our pipeline.}
    \label{fig:offnucl_example}
\end{figure*}

\begin{figure*} 
    \centering
    \includegraphics[width=0.95\textwidth]{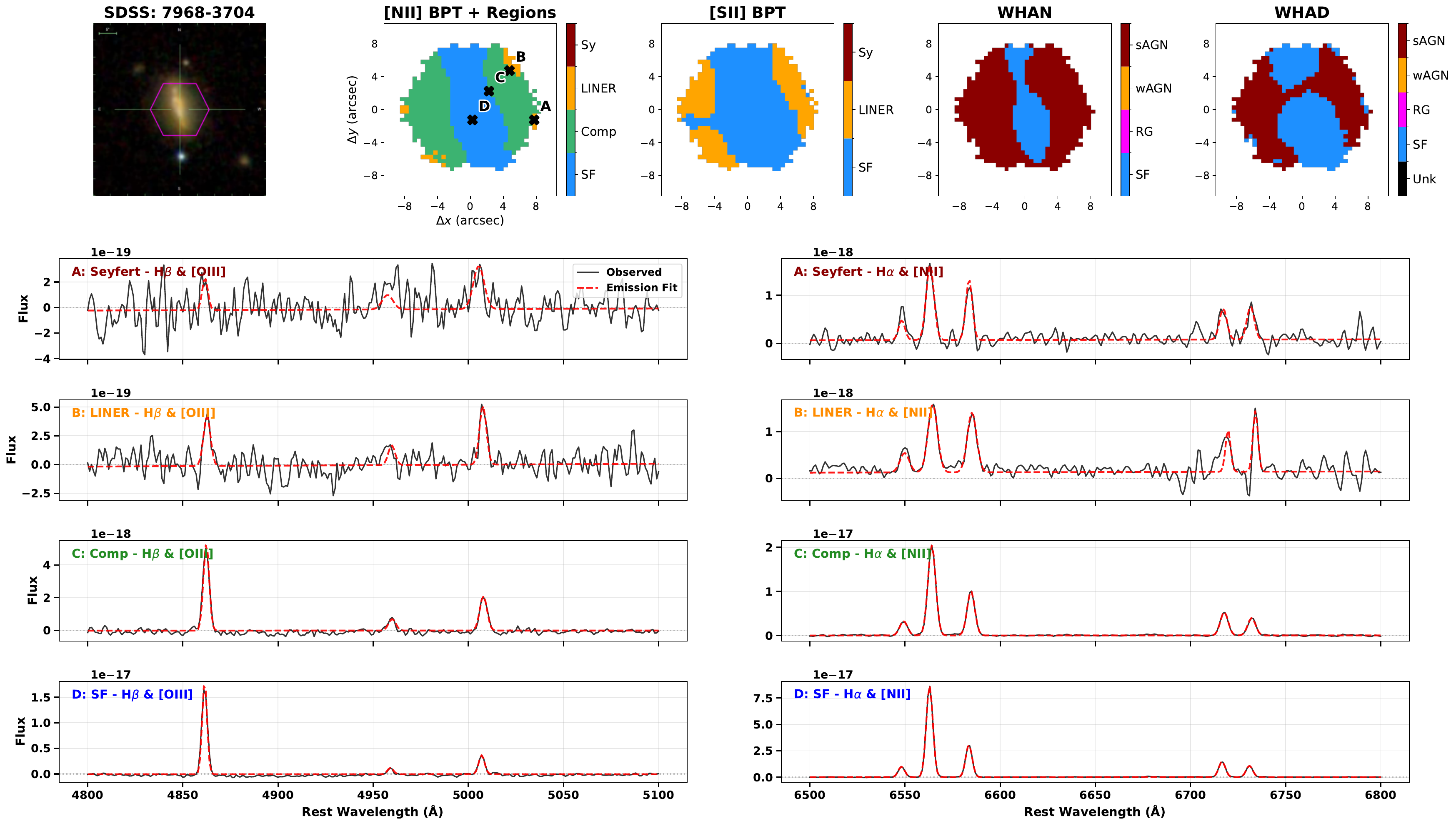}
    \caption{Example of diagnostic maps and the position of the emission-line measurements per spaxel in the [\ion{N}{ii}]-BPT classification for the target MaNGA:1-635629 ($z = 0.020$), which was successfully classified as a quiet AGN by our pipeline.}
    \label{fig:quiet_example}
\end{figure*}

\begin{figure*}
    \centering
    \includegraphics[width=\textwidth]{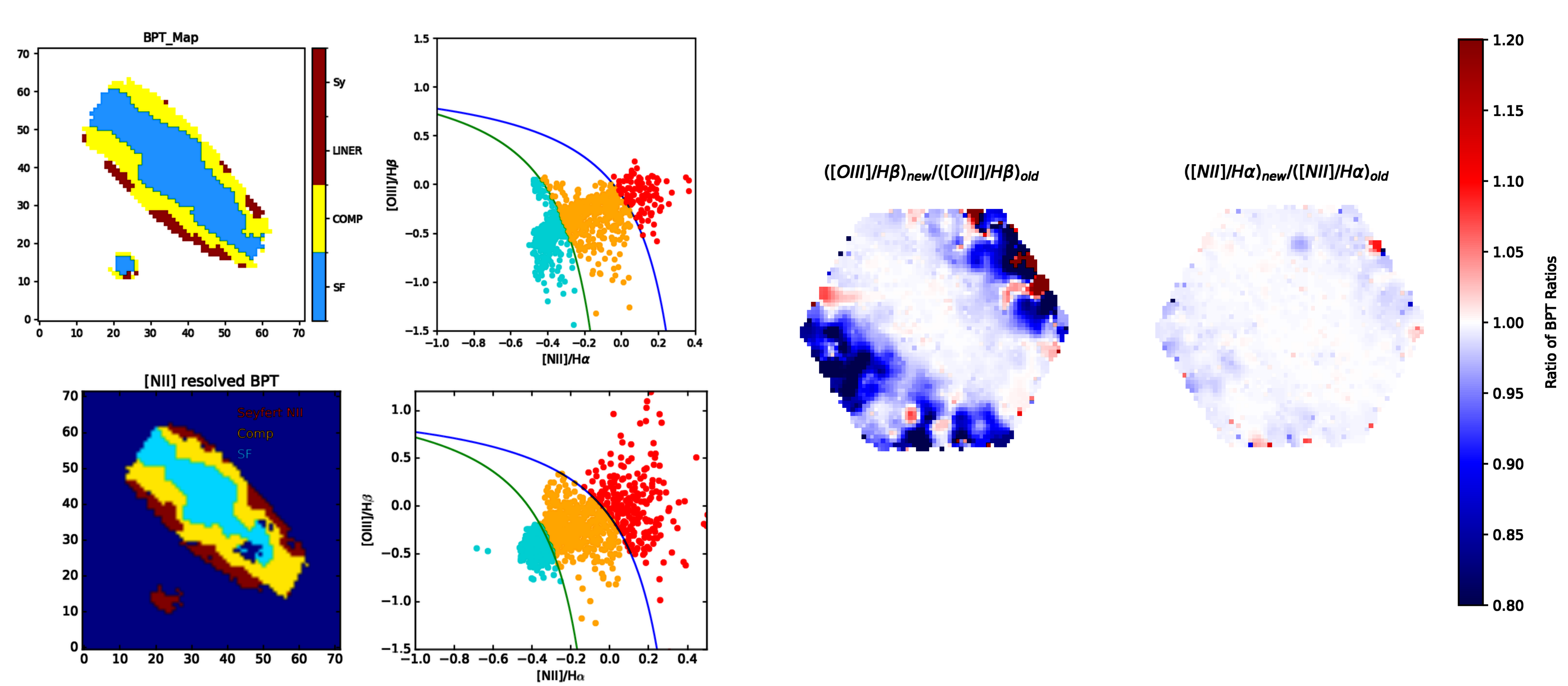}
    \caption{Data reduction impact on BPT classification for a source 
    (MaNGA:1-38042) present in the \citet{Wylezalek2018} sample but absent 
    from our DR17 selection.
    \textit{Top left:} [\ion{N}{ii}]-BPT map and diagram from the DR17 
    {\sc megacubes}.
    \textit{Bottom left:} The same galaxy from the MPL-5.
    \textit{Right:} Ratio of BPT line ratios (new/old),
    $([\ion{O}{iii}]/H\beta)_{\rm new}/([\ion{O}{iii}]/H\beta)_{\rm old}$ (left) and
    $([\ion{N}{ii}]/H\alpha)_{\rm new}/([\ion{N}{ii}]/H\alpha)_{\rm old}$ (right).
    Blue means the new pipeline measured less; red means more. The drop in 
    [\ion{O}{iii}]/H$\beta$ shifts spaxels downward on the BPT, moving 
    borderline cases out of the AGN region.}
    \label{fig:drp_comparison}
\end{figure*}

\begin{figure}
    \centering
    
    \includegraphics[width=0.45\textwidth]{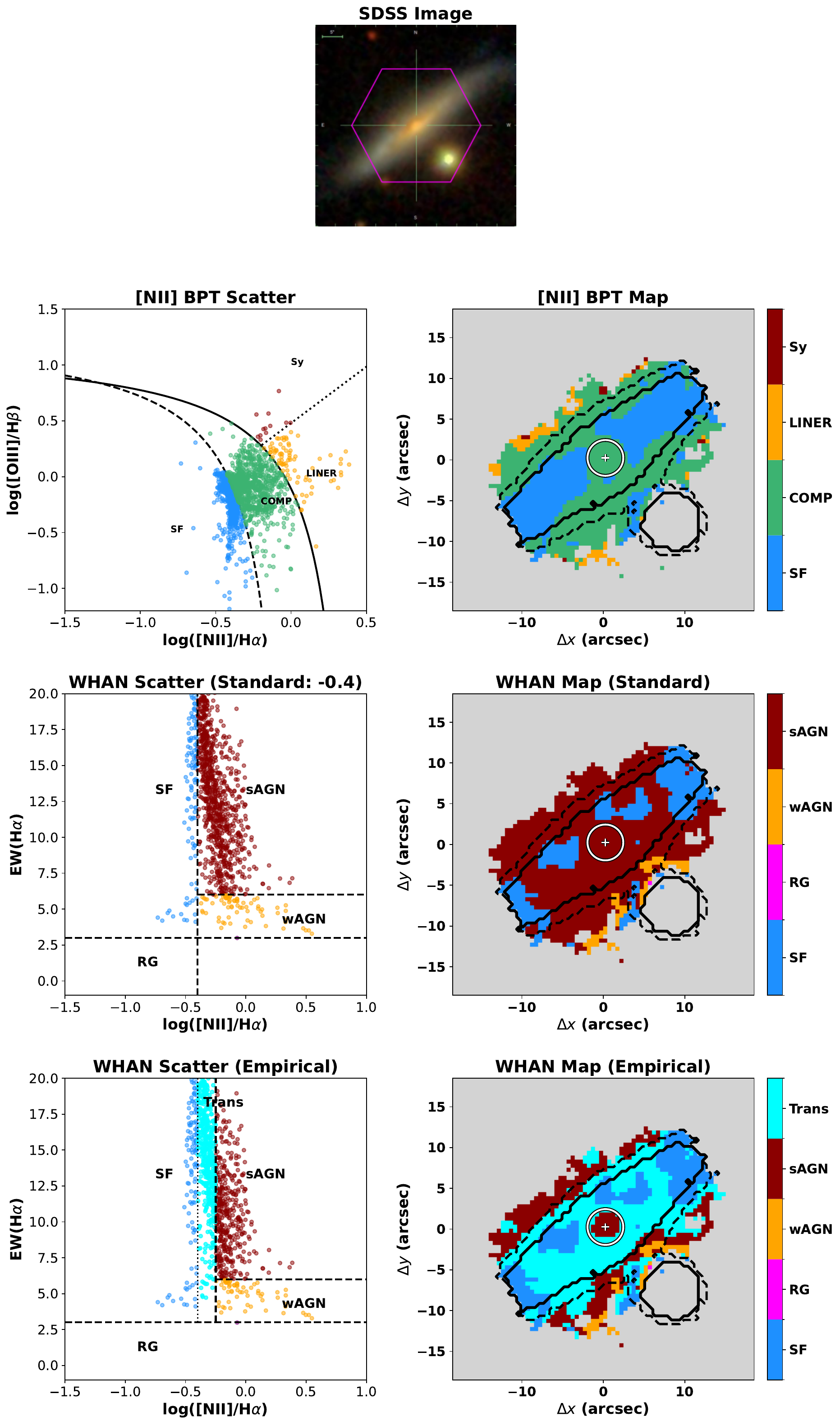}
    \caption{Effect of the WHaN boundary shift on a representative Quiet source (MaNGA:1-136027, $z = 0.0232$). \textit{Top row:} SDSS image of the target. \textit{Second row:} [$\ion{N}{ii}$]-BPT scatter diagram (left) and spatial classification map (right), showing predominantly composite and star-forming (SF) spaxels. \textit{Third row:} WHaN scatter diagram and spatial map using the standard boundary at $\log([\ion{N}{ii}]/\text{H}\alpha) = -0.4$; the galaxy is largely classified as sAGN, in clear disagreement with the BPT. \textit{Bottom row:} Same, after introducing the revised empirical boundary at $-0.25$. Spaxels in the intermediate zone ($-0.4 < \log([\ion{N}{ii}]/\text{H}\alpha) < -0.25$) are now classified as Transitional (cyan), bringing the WHaN map into agreement with the BPT.}
    \label{fig:quiet_before_after}
\end{figure}

\begin{figure}
    \centering
    \includegraphics[trim=0cm 0cm 0cm 0cm, clip,
                     width=0.45\textwidth]{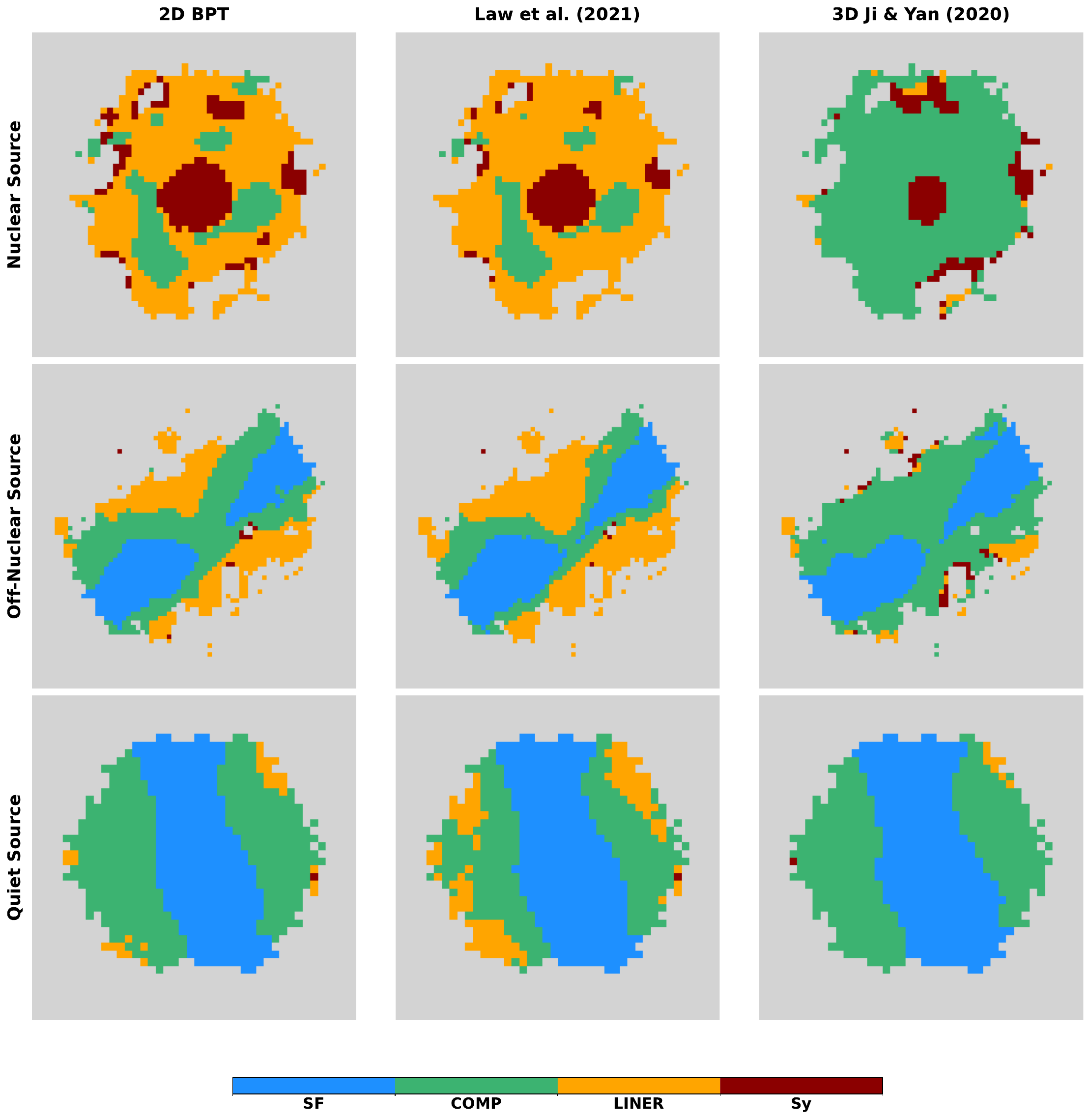}
    \caption{Spatially resolved classification maps for a representative Nuclear AGN (top), Off-Nuclear AGN (middle) and Quiet galaxy (bottom) under three diagnostic schemes: the standard 2D BPT (left), the kinematically calibrated 2D boundaries of \citet{law2021diagnostics} (centre) and the 3D reprojection of \citet{ji2020constraining} (right). The COMP class denotes the Kauffmann--Kewley composite region, the Law Intermediate region, and the Ji \& Yan gap region, respectively. The 3D classification reclassifies much of the extended LINER-like emission (orange) as composite (green), while the Seyfert cores (red) remain unchanged.}
    \label{fig:3dbpt}
\end{figure}

\begin{figure*}
    \centering
    \includegraphics[trim=0cm 0cm 0cm 1.3cm, clip, width=0.75\textwidth]{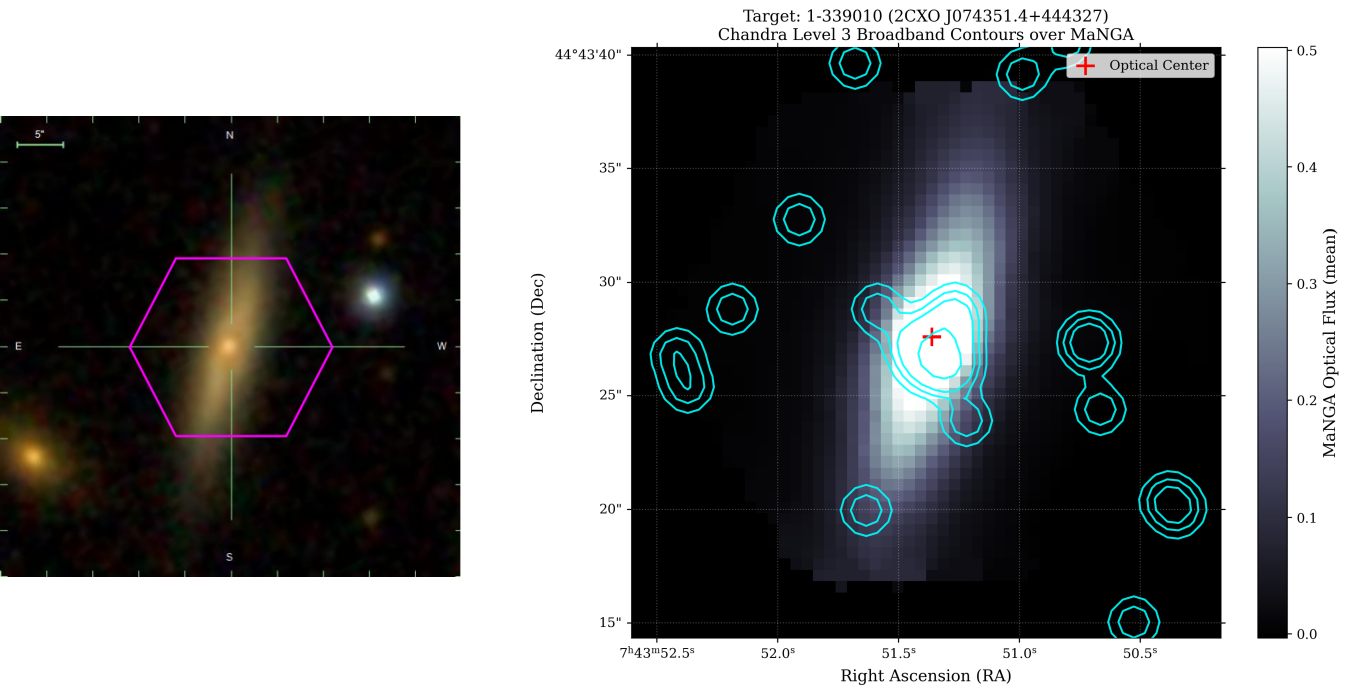}
    \caption{\textit{Left:} SDSS $gri$ composite image of MaNGA:1-339010
    ($z = 0.0313$), an edge-on system classified as an off-nuclear AGN in
    our catalog. The magenta hexagon indicates the MaNGA IFU field of
    view. \textit{Right:} Chandra broadband X-ray
    contours (cyan) overlaid on the MaNGA optical flux map. The red cross
    marks the optical center of the galaxy.}
    \label{fig:chandra_edgeon}
\end{figure*}

\begin{figure*}
    \centering
    \includegraphics[trim=0cm 0cm 0cm 1.3cm, clip, width=0.75\textwidth]{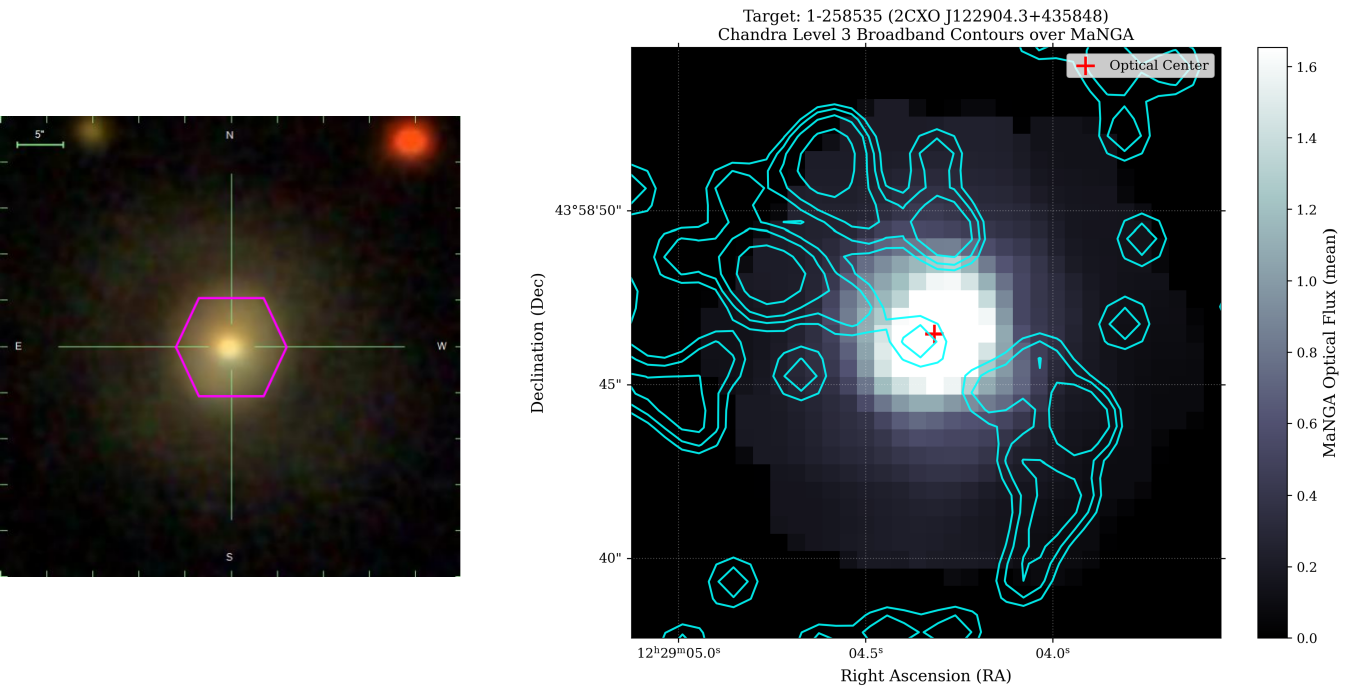}
    \caption{\textit{Left:} SDSS $gri$ composite image of MaNGA:1-258535
    ($z = 0.0237$), a face-on system classified as an off-nuclear AGN in
    our catalog. The magenta hexagon indicates the MaNGA IFU field of
    view. \textit{Right:} Chandra broadband X-ray
    contours (cyan) overlaid on the MaNGA optical flux map. The red cross
    marks the optical center of the galaxy. The X-ray contours reveal the
    fossil outflows.}
    \label{fig:chandra_outflow}
\end{figure*}

\clearpage

\begin{figure*}

    \section{Nuclear vs. Off-Nuclear: Physical Properties}
    \label{sec:other_properties}
    \vspace{0.2cm} 

    \centering
    \begin{minipage}{0.50\textwidth}
        \centering
        \includegraphics[trim=0cm 0cm 0cm 1.2cm, clip, width=0.85\linewidth]{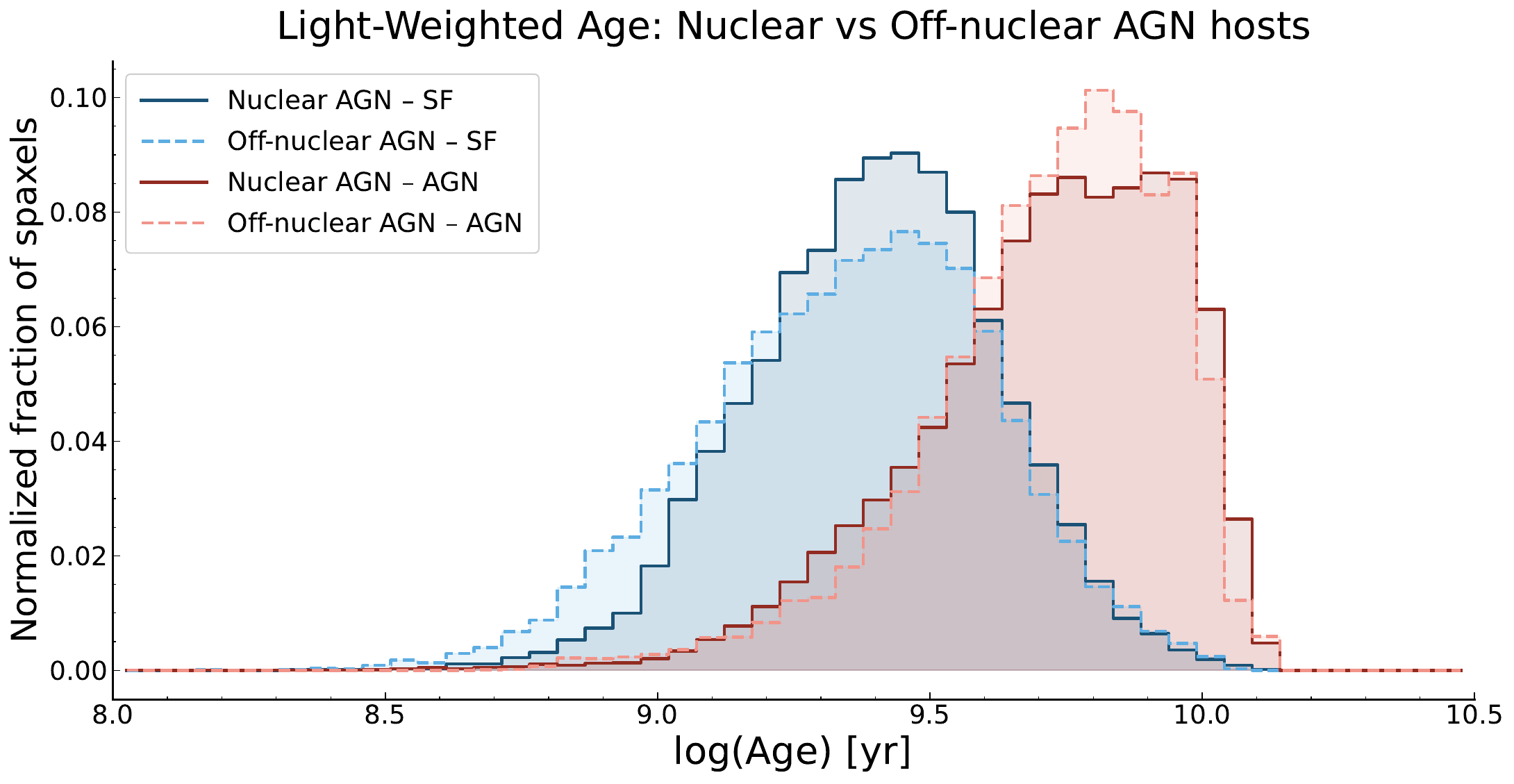}
        \caption{Light-weighted stellar age.}
        \label{fig:app_age_l}
    \end{minipage}\hfill
    \begin{minipage}{0.50\textwidth}
        \centering
        \includegraphics[trim=0cm 0cm 0cm 1.2cm, clip, width=0.85\linewidth]{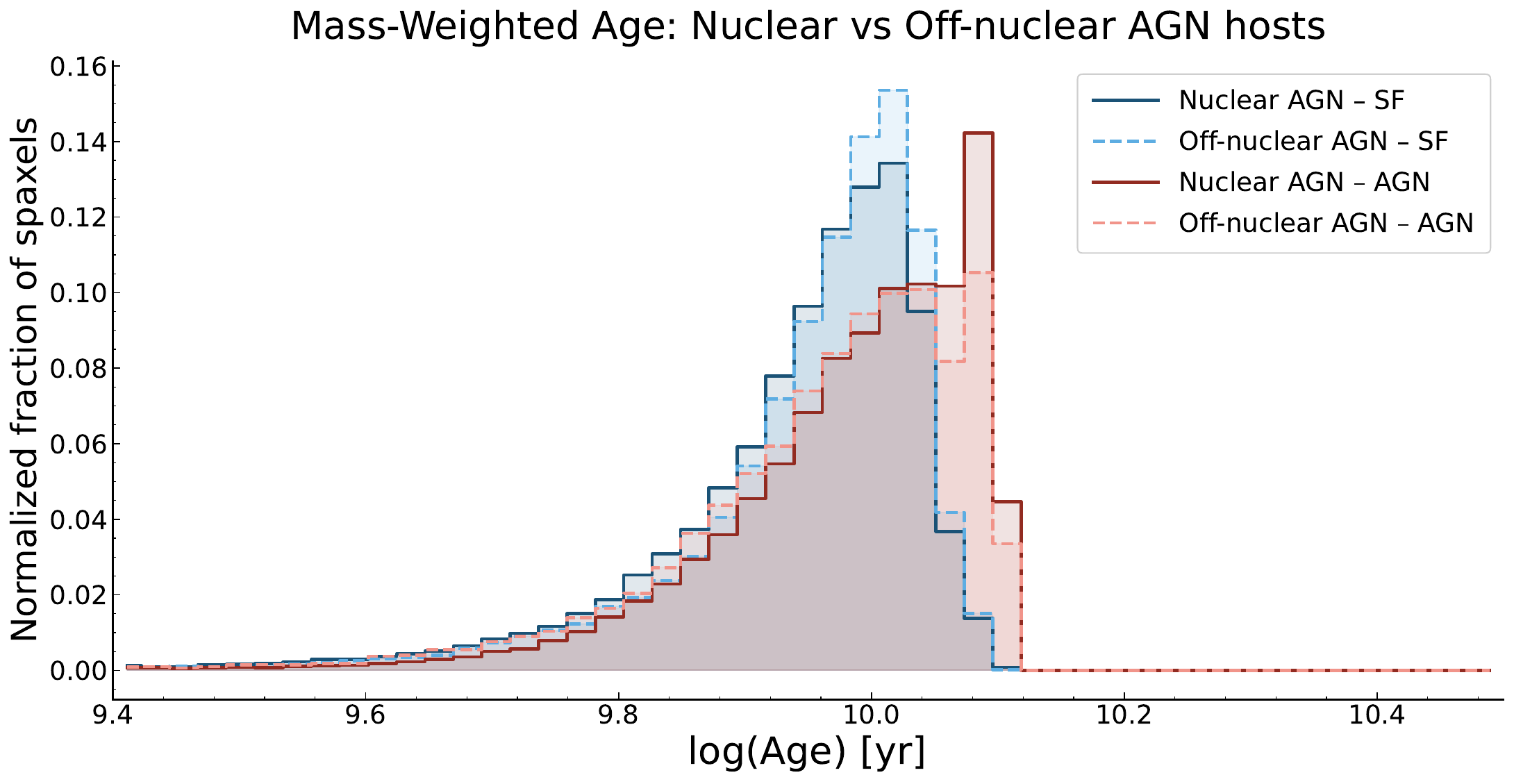}
        \caption{Mass-weighted stellar age.}
        \label{fig:app_age_m}
    \end{minipage}

    \vspace{0.06cm} 

    \begin{minipage}{0.50\textwidth}
        \centering
        \includegraphics[trim=0cm 0cm 0cm 1.0cm, clip, width=0.85\linewidth]{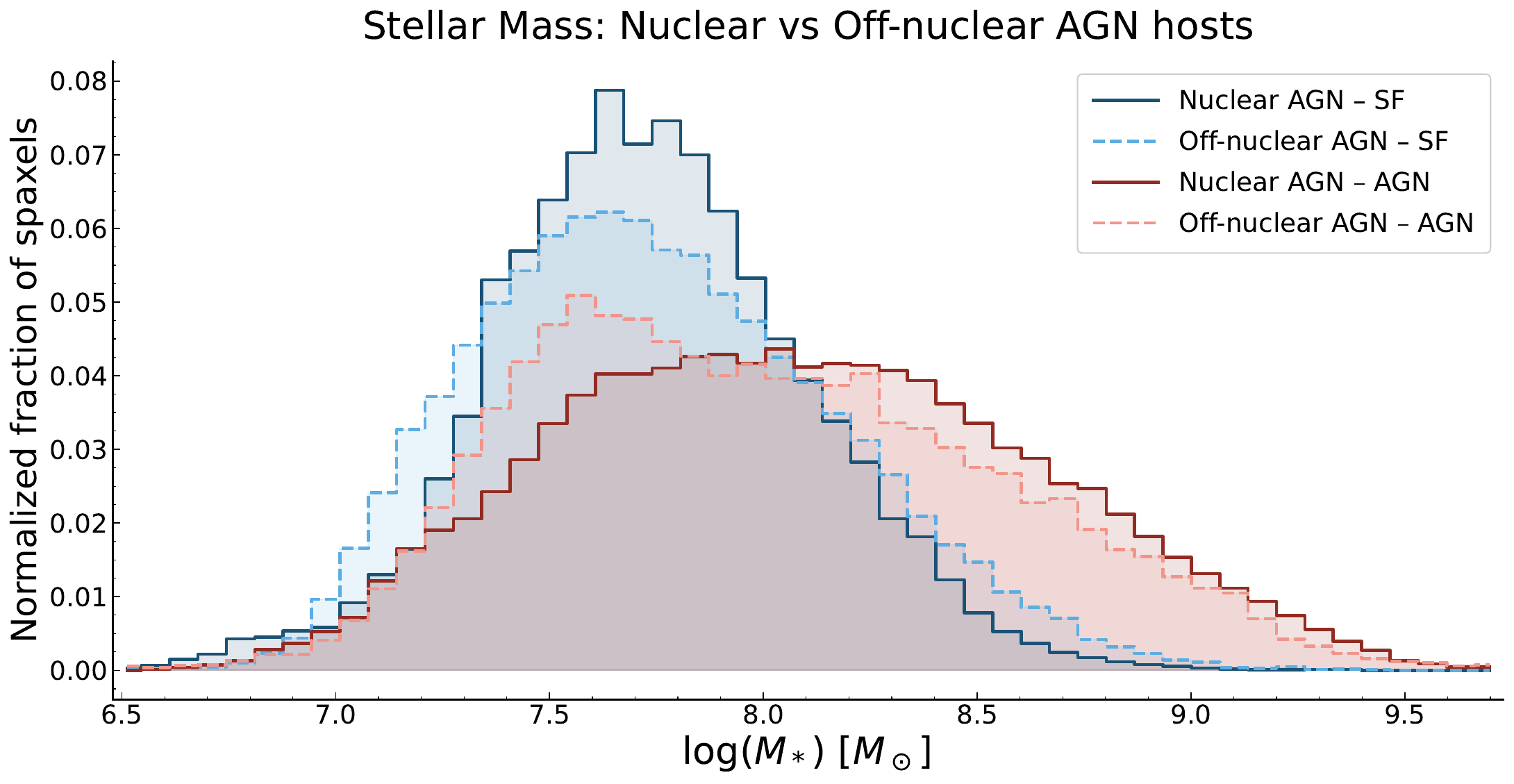}
        \caption{Stellar mass $\log(M_\star)$.}
        \label{fig:app_mass}
    \end{minipage}\hfill
    \begin{minipage}{0.50\textwidth}
        \centering
        \includegraphics[trim=0cm 0cm 0cm 1.2cm, clip, width=0.85\linewidth]{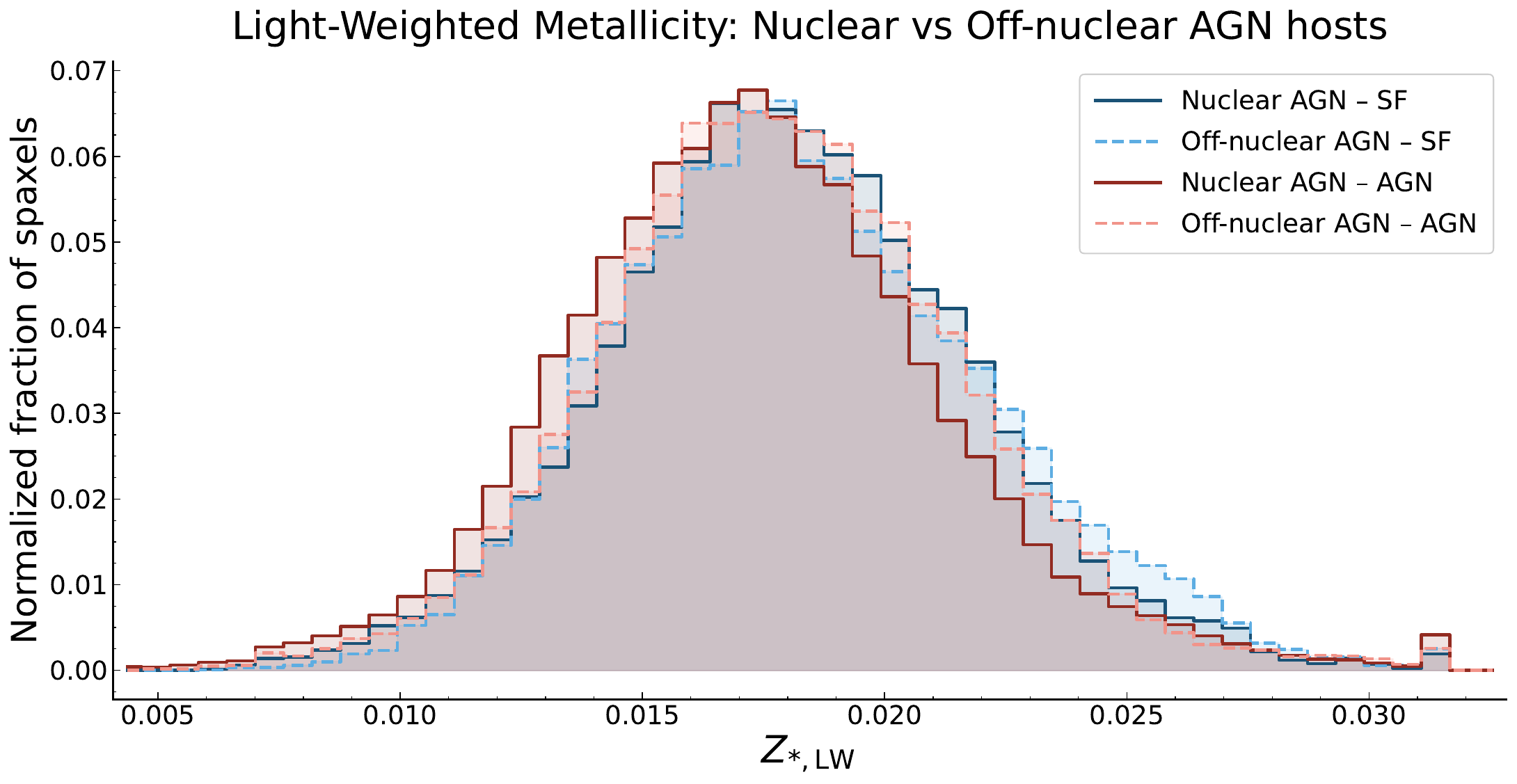}
        \caption{Light-weighted stellar metallicity.}
        \label{fig:app_meta_lw}
    \end{minipage}

    \vspace{0.06cm}

    \begin{minipage}{0.50\textwidth}
        \centering
        \includegraphics[trim=0cm 0cm 0cm 1.2cm, clip, width=0.85\linewidth]{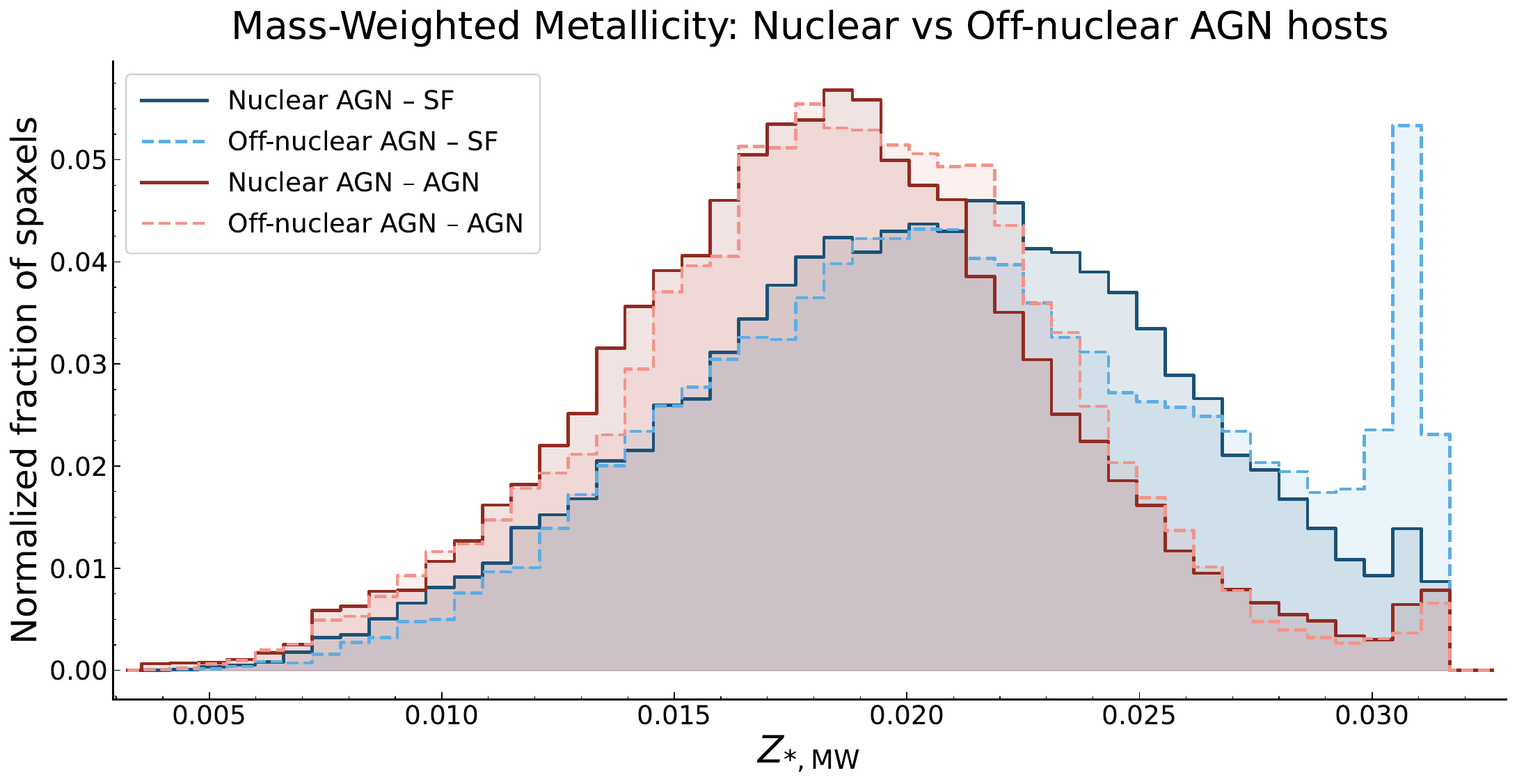}
        \caption{Mass-weighted stellar metallicity.}
        \label{fig:app_meta_mw}
    \end{minipage}\hfill
    \begin{minipage}{0.50\textwidth}
        \centering
        \includegraphics[trim=0cm 0cm 0cm 1.2cm, clip, width=0.85\linewidth]{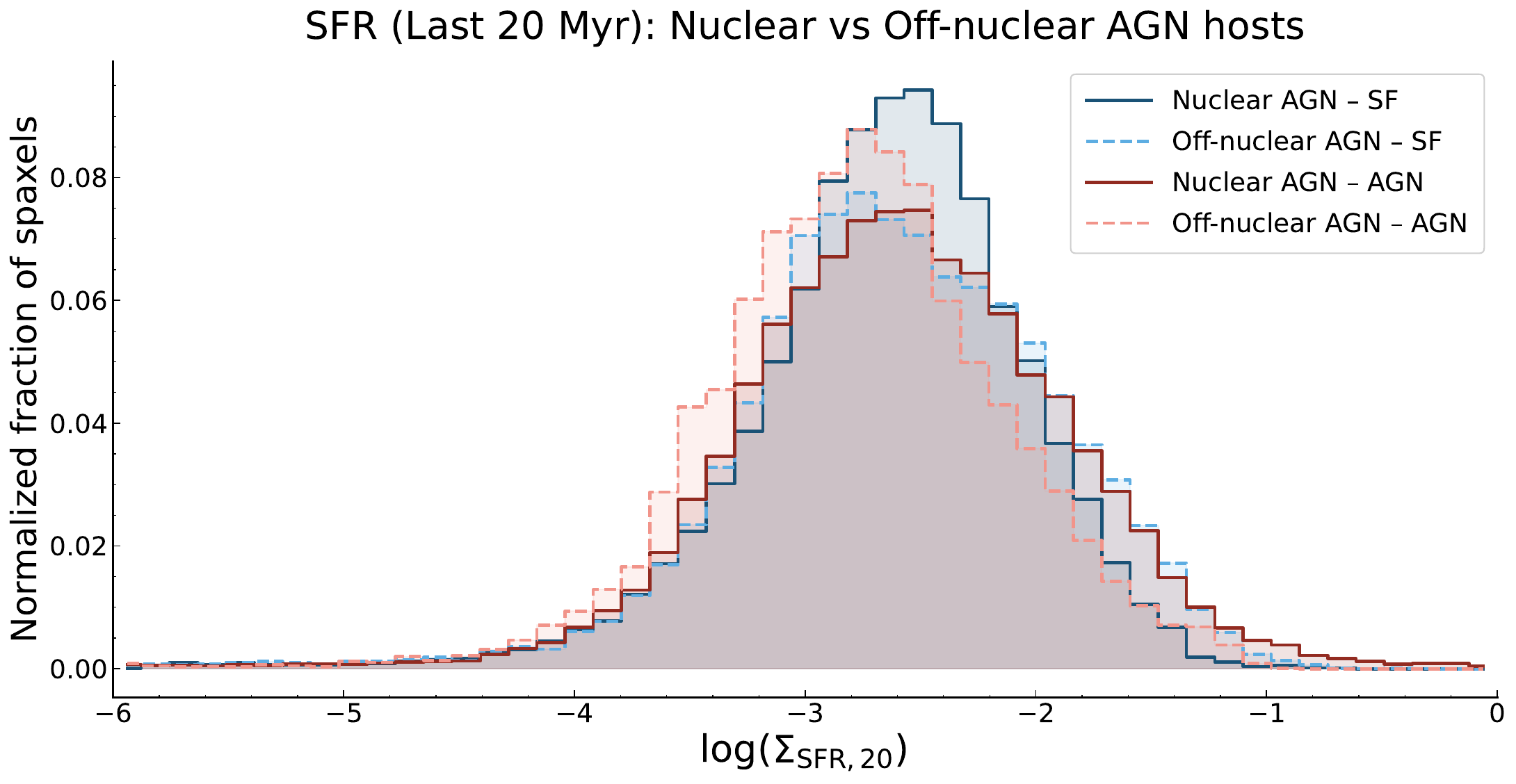}
        \caption{Recent star formation rate (SFR over the last 20 Myr).}
        \label{fig:app_sfr}
    \end{minipage}

        \vspace{0.06cm}
    \begin{minipage}{0.50\textwidth}
        \centering
        \includegraphics[trim=0cm 0cm 0cm 1.2cm, clip, width=0.85\linewidth]{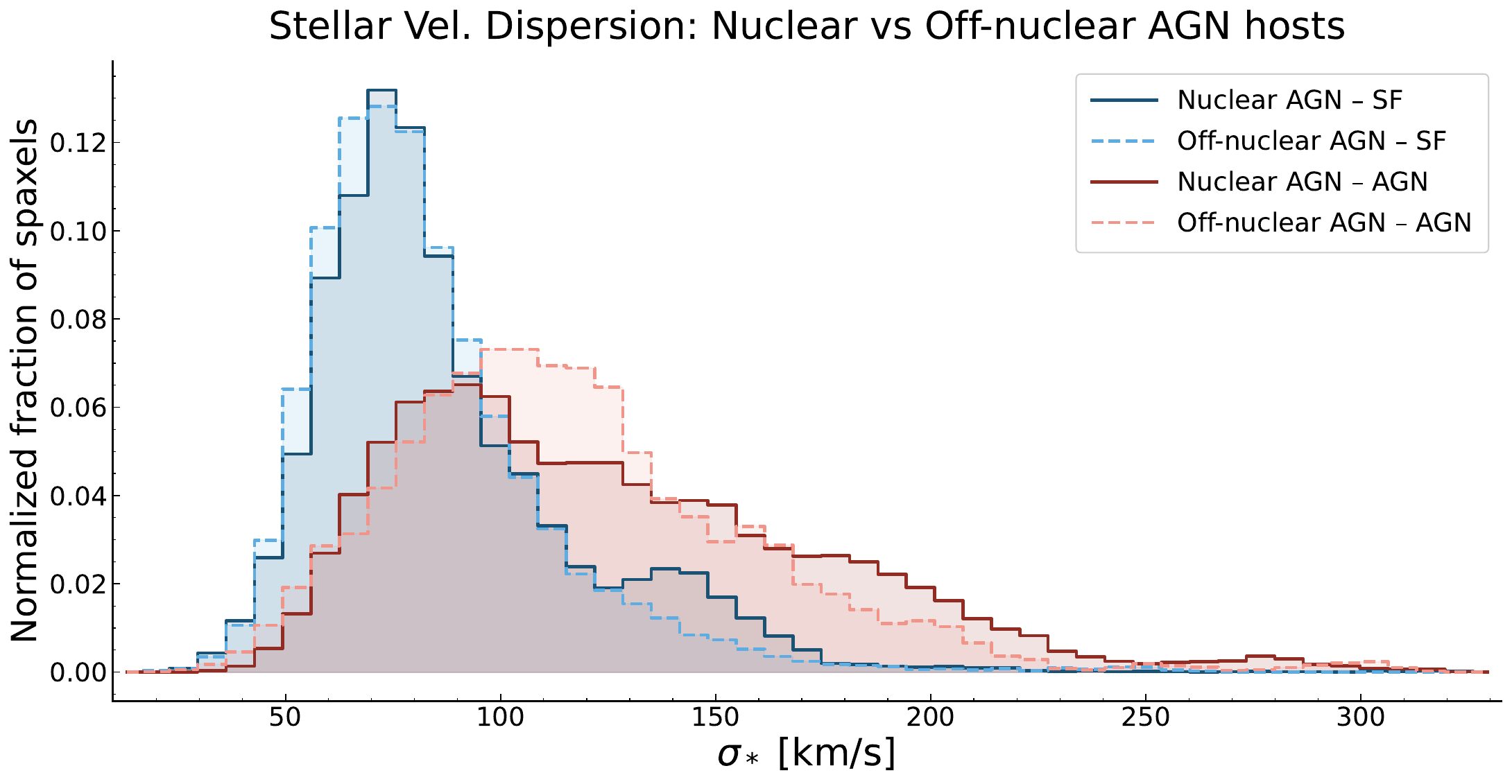}
        \caption{Stellar velocity dispersion ($\sigma_\star$).}
        \label{fig:app_sig_star}
    \end{minipage}\hfill
    \begin{minipage}{0.50\textwidth}
        \centering
        \includegraphics[trim=0cm 0cm 0cm 1.3cm, clip, width=0.85\linewidth]{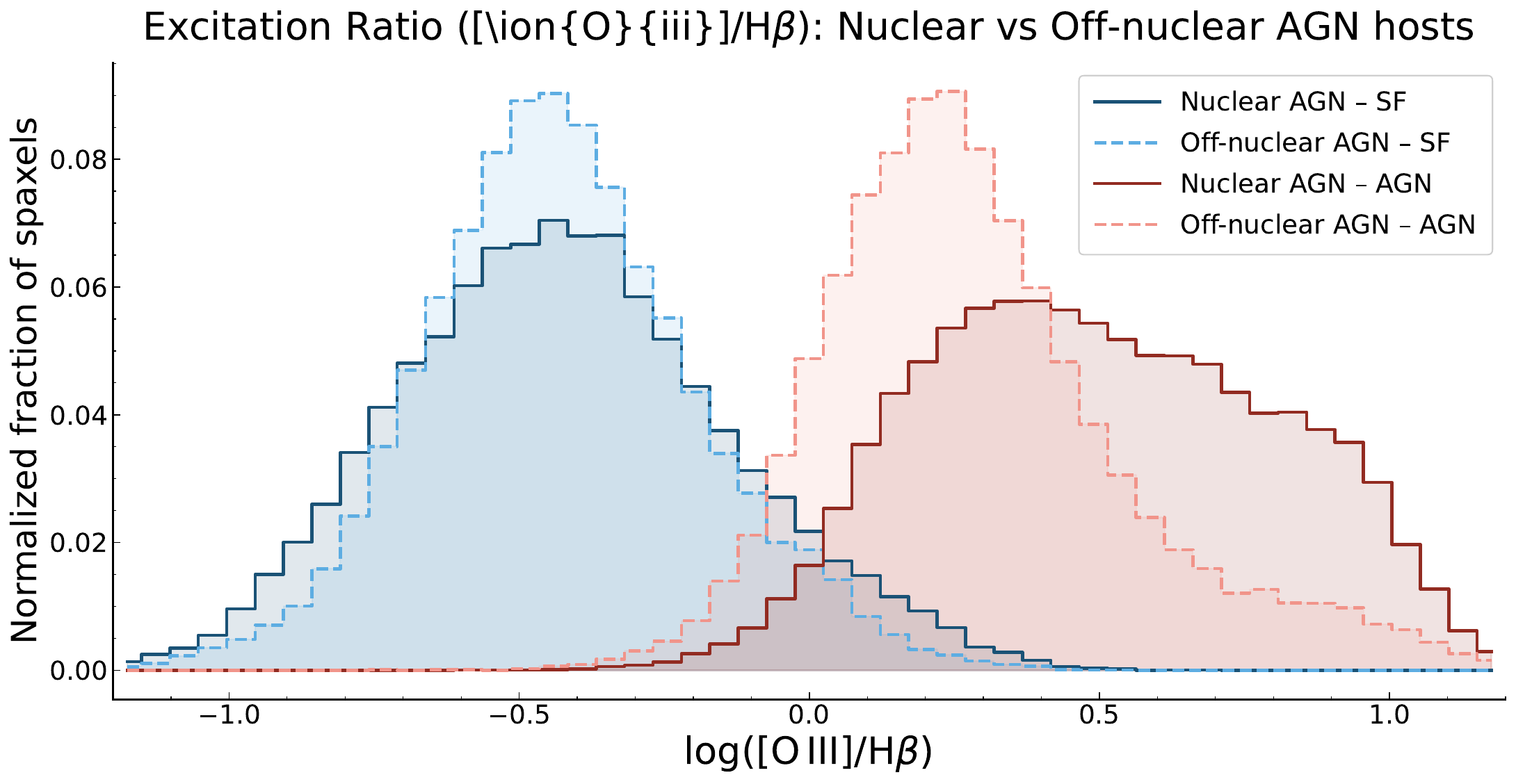}
        \caption{Gas excitation ratio, $\log$([\ion{O}{iii}]/H$\beta$).}
        \label{fig:loiii_and_excitation}
    \end{minipage}

    \vspace{0.06cm}
    \begin{minipage}{0.50\textwidth}
        \centering
        \includegraphics[trim=0cm 0cm 0cm 1.3cm, clip, width=0.85\linewidth]{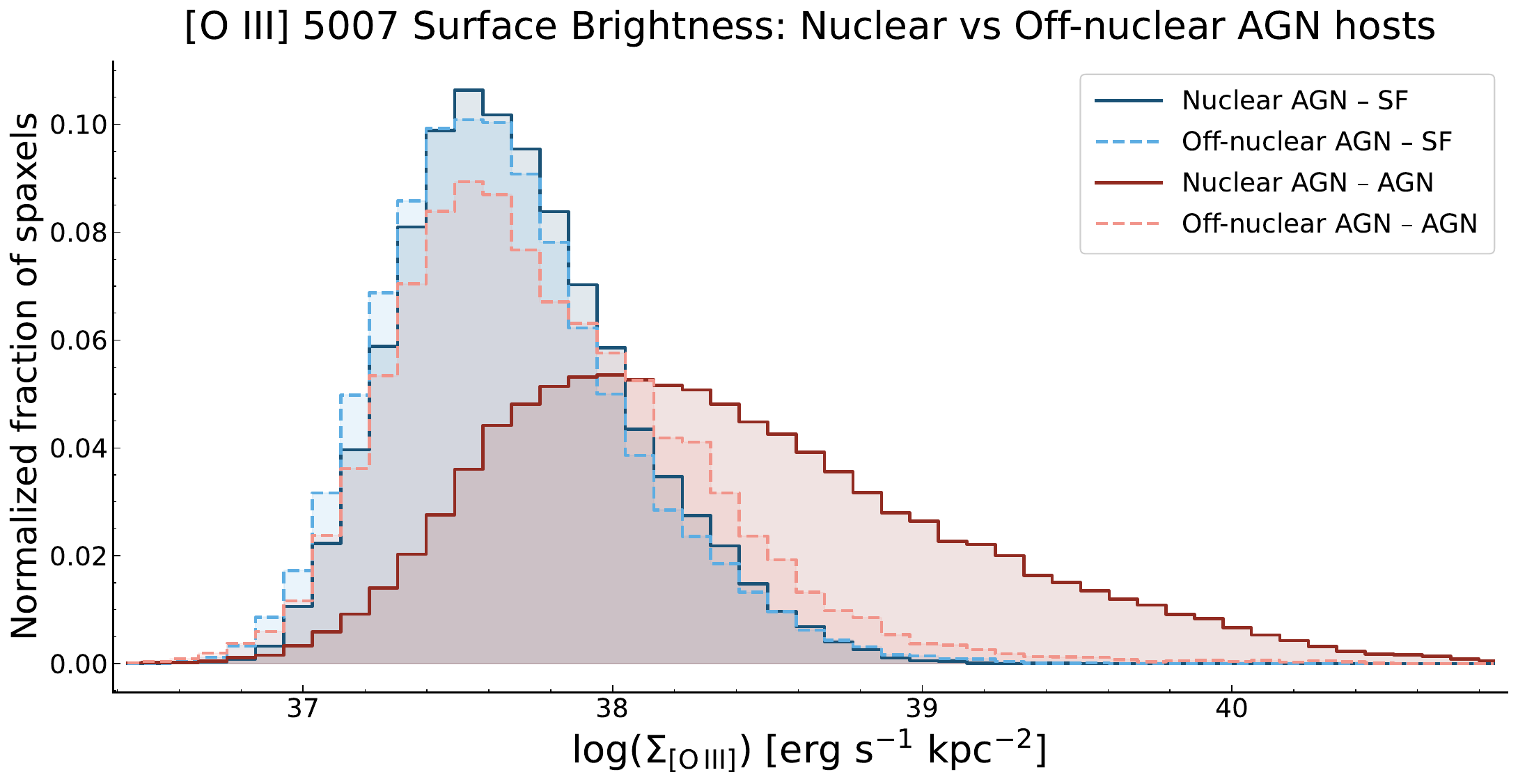}
        \caption{[\ion{O}{iii}]$\lambda$5007 surface brightness, $\log\,\Sigma_{\rm [\ion{O}{iii}]}$, in erg\,s$^{-1}$\,kpc$^{-2}$.}
        \label{fig:sb_oiii}
    \end{minipage}\hfill
    \begin{minipage}{0.50\textwidth}
        ~
    \end{minipage}
\end{figure*}

\clearpage
\begin{figure*}
    \section{Inclination - Property Distributions}
    \label{sec:incl_properties}
    \vspace{0.2cm}

    \centering
    \begin{minipage}{\textwidth}
        \centering
        \includegraphics[trim=0cm 0cm 0cm 1.3cm, clip, width=0.85\linewidth]{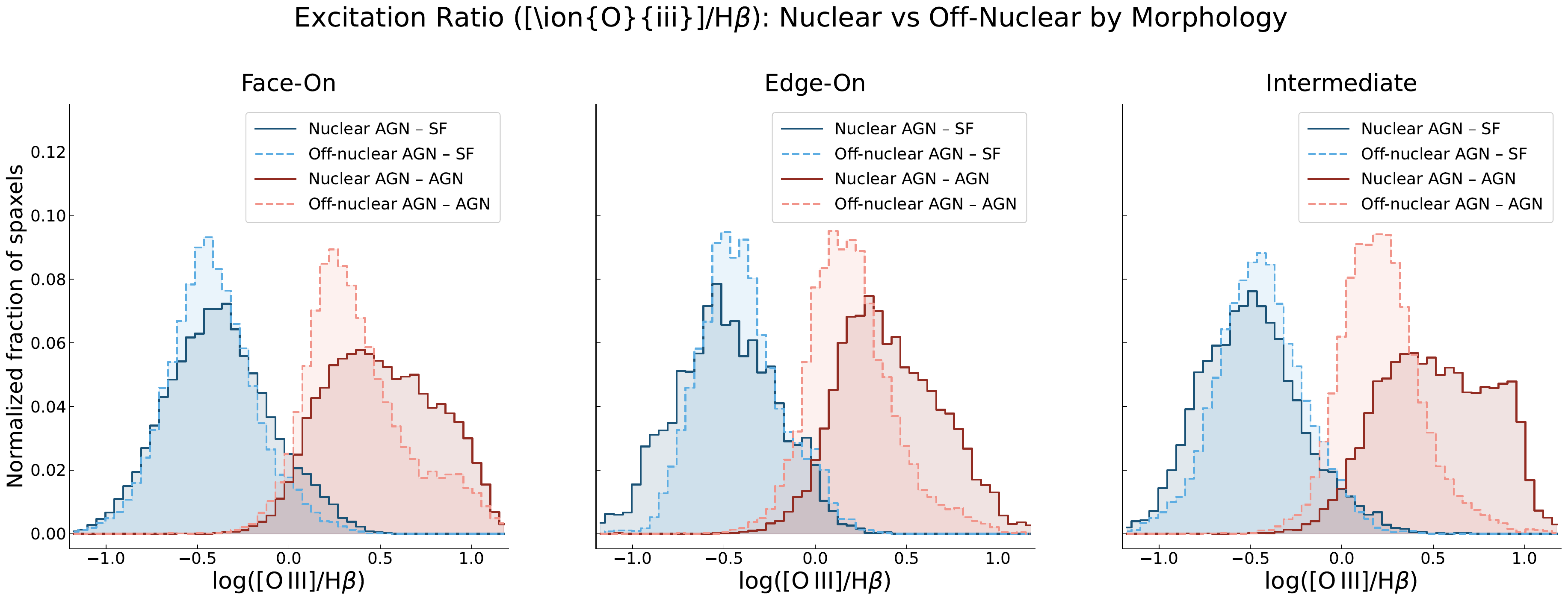}
        \caption{Excitation ratio $\log$([\ion{O}{iii}]/H$\beta$) by inclination.}
        \label{fig:app_o3hb_incl}
    \end{minipage}
    \vspace{0.2cm}
    \begin{minipage}{\textwidth}
        \centering
        \includegraphics[trim=0cm 0cm 0cm 1.3cm, clip, width=0.85\linewidth]{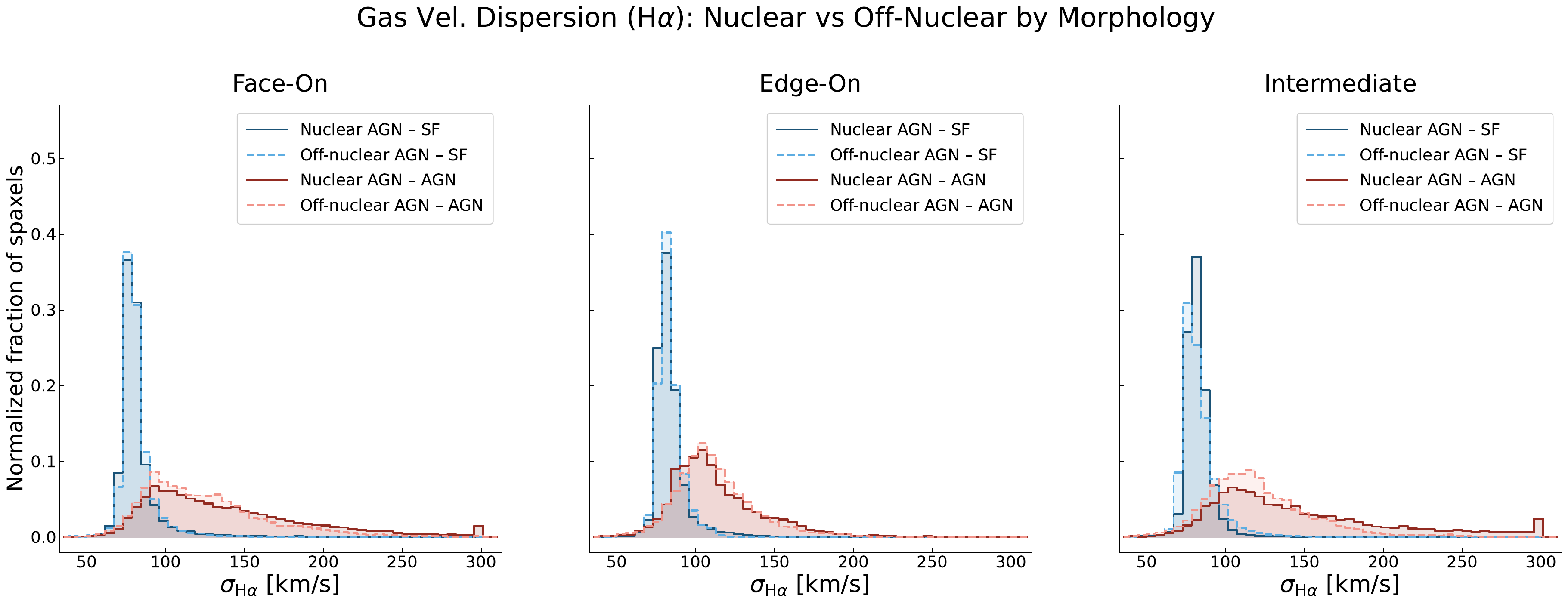}
        \caption{Gas velocity dispersion by inclination.}
        \label{fig:app_siggas_incl}
    \end{minipage}
    \vspace{0.2cm}
    \begin{minipage}{\textwidth}
        \centering
        \includegraphics[trim=0cm 0cm 0cm 1.3cm, clip, width=0.85\linewidth]{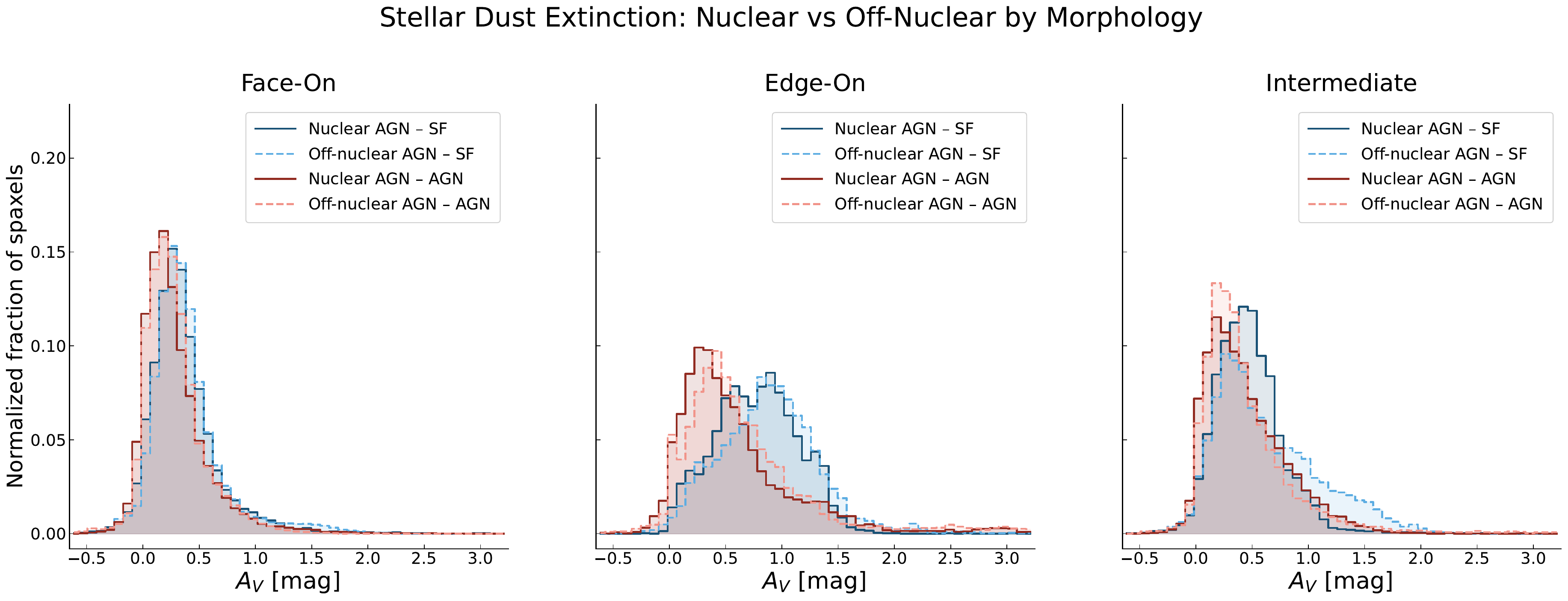}
        \caption{Dust attenuation by inclination.}
        \label{fig:app_dust_incl}
    \end{minipage}
\end{figure*}

\begin{figure*}
    \centering
    \begin{minipage}{\textwidth}
        \centering
        \includegraphics[trim=0cm 0cm 0cm 1.4cm, clip, width=0.85\linewidth]{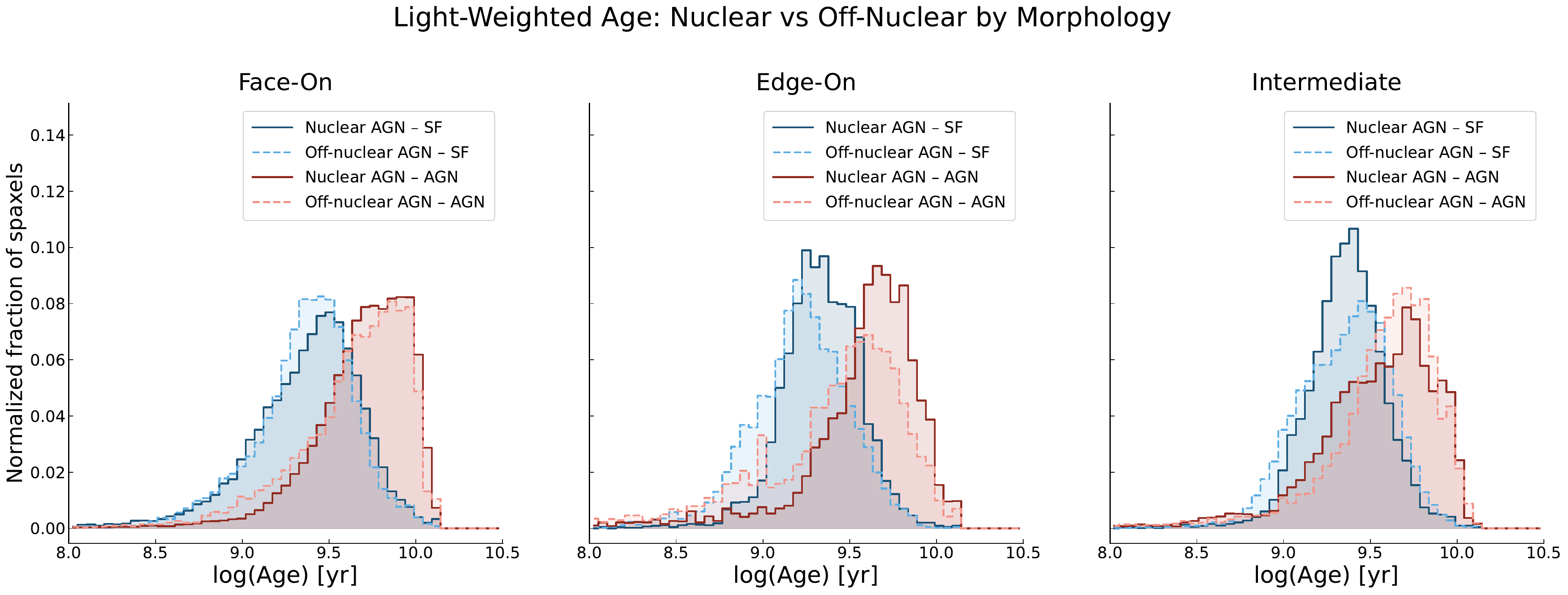}
        \caption{Light-weighted stellar age by inclination.}
        \label{fig:app_agel_incl}
    \end{minipage}
    \vspace{0.2cm}
    \begin{minipage}{\textwidth}
        \centering
        \includegraphics[trim=0cm 0cm 0cm 1.3cm, clip, width=0.85\linewidth]{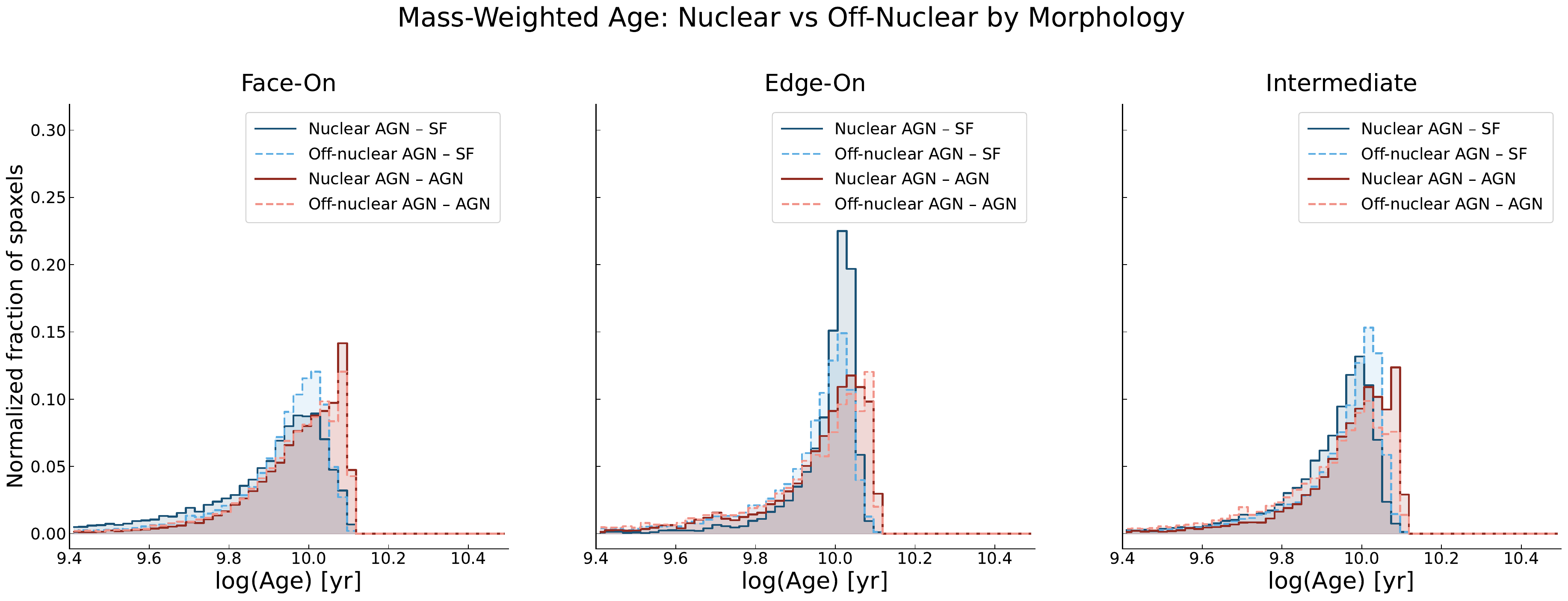}
        \caption{Mass-weighted stellar age by inclination.}
        \label{fig:app_agem_incl}
    \end{minipage}
    \vspace{0.2cm}
    \begin{minipage}{\textwidth}
        \centering
        \includegraphics[trim=0cm 0cm 0cm 1.3cm, clip, width=0.85\linewidth]{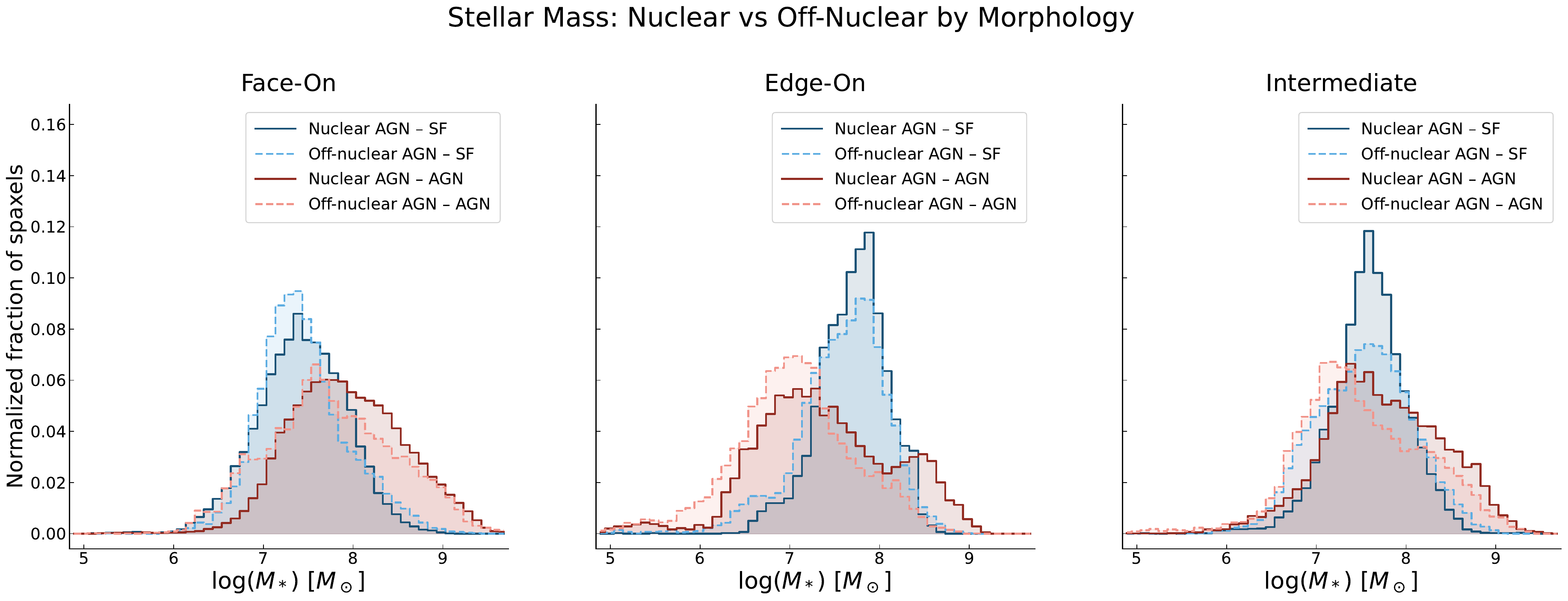}
        \caption{Stellar mass by inclination.}
        \label{fig:app_mass_incl}
    \end{minipage}
\end{figure*}

\begin{figure*}
    \centering
    \begin{minipage}{\textwidth}
        \centering
        \includegraphics[trim=0cm 0cm 0cm 1.55cm, clip, width=0.75\linewidth]{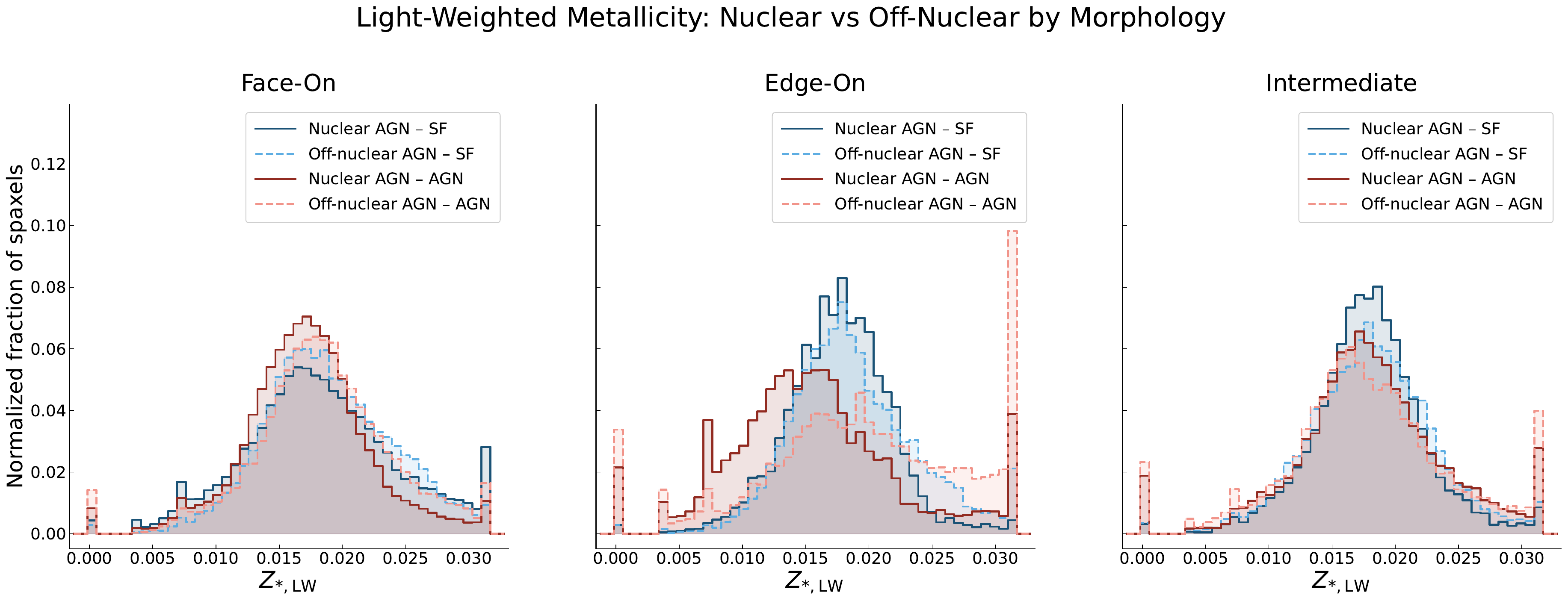}
        \caption{Light-weighted stellar metallicity by inclination.}
        \label{fig:app_metalw_incl}
    \end{minipage}
    
    \vspace{0.05cm} 
    
    \begin{minipage}{\textwidth}
        \centering
        \includegraphics[trim=0cm 0cm 0cm 1.45cm, clip, width=0.75\linewidth]{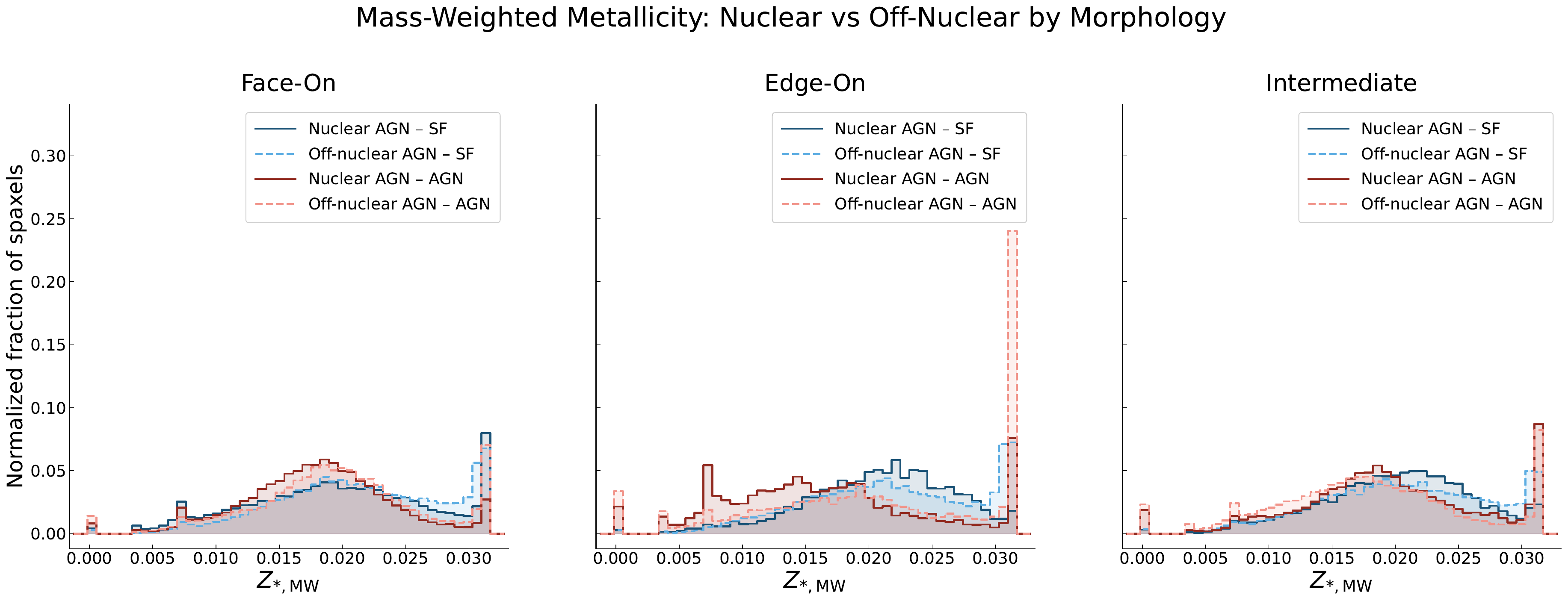}
        \caption{Mass-weighted stellar metallicity by inclination.}
        \label{fig:app_metamw_incl}
    \end{minipage}
    
    \vspace{0.05cm}
    
    \begin{minipage}{\textwidth}
        \centering
        \includegraphics[trim=0cm 0cm 0cm 1.45cm, clip, width=0.75\linewidth]{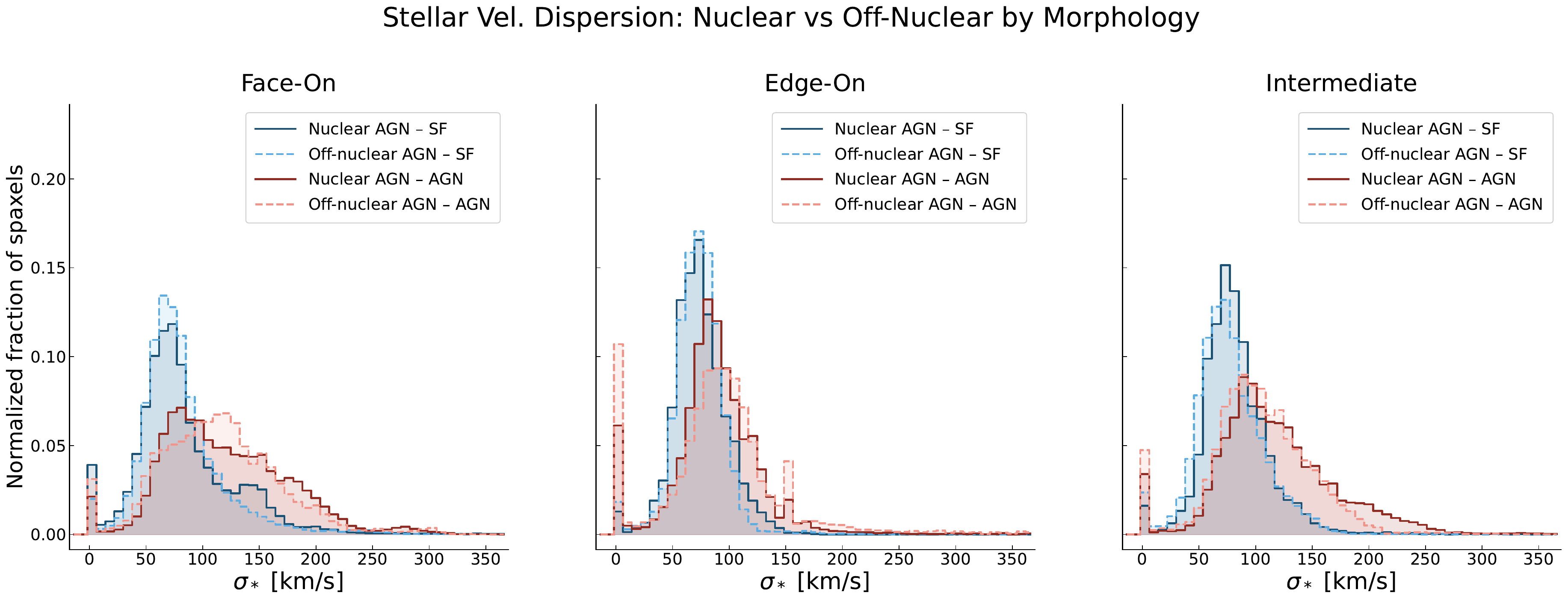}
        \caption{Stellar velocity dispersion by inclination.}
        \label{fig:app_sigstar_incl}
    \end{minipage}
    
    \vspace{0.05cm}
    
    \begin{minipage}{\textwidth}
        \centering
        \includegraphics[trim=0cm 0cm 0cm 1.45cm, clip, width=0.75\linewidth]{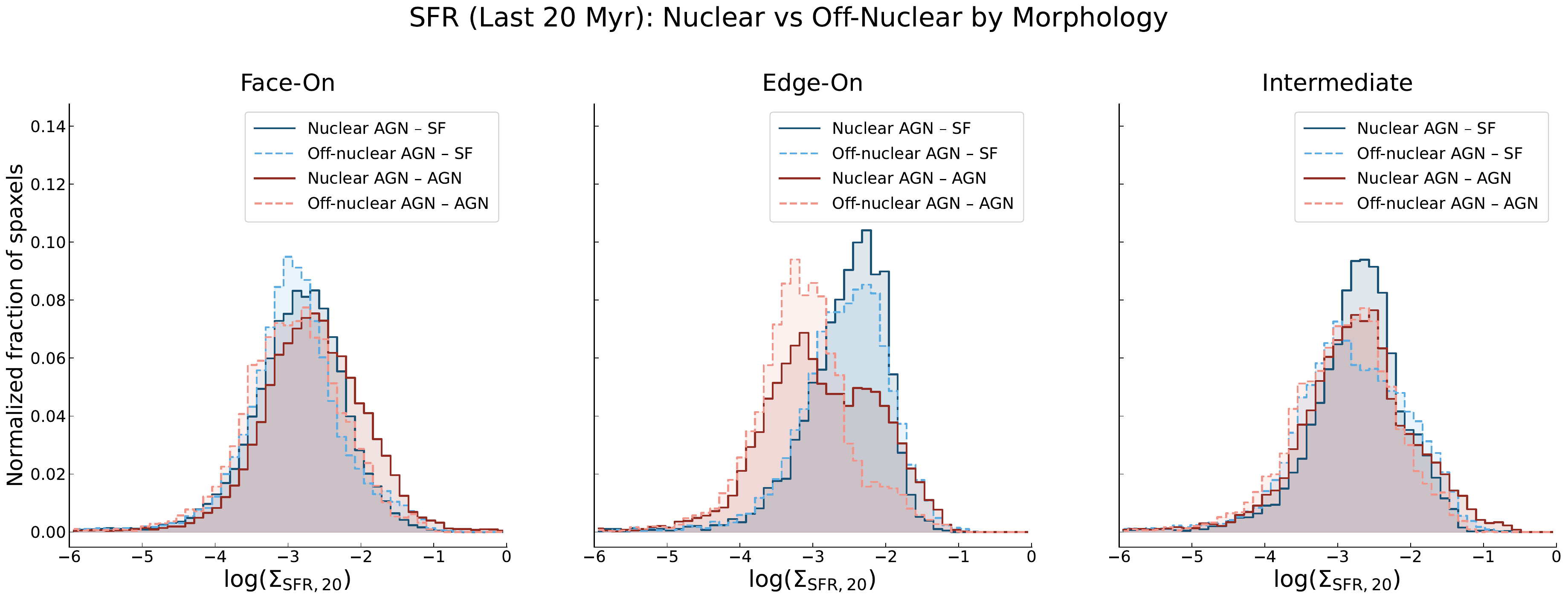}
        \caption{SFR (last 20 Myr) by inclination.}
        \label{fig:app_sfr20_incl}
    \end{minipage}

    \label{lastpage} 
\end{figure*}

\end{document}